ICRC 2013

# The IceCube Neutrino Observatory Part VI: Ice Properties, Reconstruction and Future Developments

THE ICECUBE COLLABORATION

## Contents

## IceCube Collaboration Member List

M. G. Aartsen[2], R. Abbasi[27], Y. Abdou[22], M. Ackermann[41], J. Adams[15], J. A. Aguilar[21], M. Ahlers[27], D. Altmann[9], J. Auffenberg[27], X. Bai[31,a], M. Baker[27], S. W. Barwick[23], V. Baum[28], R. Bay[7], J. J. Beatty[17,18], S. Bechet[12], J. Becker Tjus[10], K.-H. Becker[40], M. Bell[38], M. L. Benabderrahmane[41], S. BenZvi[27], P. Berghaus[41], D. Berley[16], E. Bernardini[41], A. Bernhard[30], D. Bertrand[12], D. Z. Besson[25], G. Binder[8,7], D. Bindig[40], M. Bissok[1], E. Blaufuss[16], J. Blumenthal[1], D. J. Boersma[39], S. Bohaichuk[20], C. Bohm[34], D. Bose[13], S. Böser[11], O. Botner[39], L. Brayeur[13], H.-P. Bretz[41], A. M. Brown[15], R. Bruijn[24], J. Brunner[41], M. Carson[22], J. Casey[5], M. Casier[13], D. Chirkin[27], A. Christov[21], B. Christy[16], K. Clark[38], F. Clevermann[19], S. Coenders[1], S. Cohen[24], D. F. Cowen[38,37], A. H. Cruz Silva[41], M. Danninger[34], J. Daughhetee[5], J. C. Davis[17], C. De Clercq[13], S. De Ridder[22], P. Desiati[27], K. D. de Vries[13], M. de With[9], T. DeYoung[38], J. C. Díaz-Vélez[27], M. Dunkman[38], R. Eagan[38], B. Eberhardt[28], J. Eisch[27], R. W. Ellsworth[16], S. Euler[1], P. A. Evenson[31], O. Fadiran[27], A. R. Fazely[6], A. Fedynitch[10], J. Feintzeig[27], T. Feusels[22], K. Filimonov[7], C. Finley[34], T. Fischer-Wasels[40], S. Flis[34], A. Franckowiak[11], K. Frantzen[19], T. Fuchs[19], T. K. Gaisser[31], J. Gallagher[26], L. Gerhardt[8,7], L. Gladstone[27], T. Glüsenkamp[41], A. Goldschmidt[8], G. Golup[13], J. G. Gonzalez[31], J. A. Goodman[16], D. Góra[41], D. T. Grandmont[20], D. Grant[20], A. Groß[30], C. Ha[8,7], A. Haj Ismail[22], P. Hallen[1], A. Hallgren[39], F. Halzen[27], K. Hanson[12], D. Heereman[12], D. Heinen[1], K. Helbing[40], R. Hellauer[16], S. Hickford[15], G. C. Hill[2], K. D. Hoffman[16], R. Hoffmann[40], A. Homeier[11], K. Hoshina[27], W. Huelsnitz[16,b], P. O. Hulth[34], K. Hultqvist[34], S. Hussain[31], A. Ishihara[14], E. Jacobi[41], J. Jacobsen[27], K. Jagielski[1], G. S. Japaridze[4], K. Jero[27], O. Jlelati[22], B. Kaminsky[41], A. Kappes[9], T. Karg[41], A. Karle[27], J. L. Kelley[27], J. Kiryluk[35], J. Kläs[40], S. R. Klein[8,7], J.-H. Köhne[19], G. Kohnen[29], H. Kolanoski[9], L. Köpke[28], C. Kopper[27], S. Kopper[40], D. J. Koskinen[38], M. Kowalski[11], M. Krasberg[27], K. Krings[1], G. Kroll[28], J. Kunnen[13], N. Kurahashi[27], T. Kuwabara[31], M. Labare[22], H. Landsman[27], M. J. Larson[36], M. Lesiak-Bzdak[35], M. Leuermann[1], J. Leute[30], J. Lünemann[28], J. Madsen[33], G. Maggi[13], R. Maruyama[27], K. Mase[14], H. S. Matis[8], F. McNally[27], K. Meagher[16], M. Merck[27], P. Mészáros[37,38], T. Meures[12], S. Miarecki[8,7], E. Middell[41], N. Milke[19], J. Miller[13], L. Mohrmann[41], T. Montaruli[21,c], R. Morse[27], R. Nahnhauer[41], U. Naumann[40], H. Niederhausen[35], S. C. Nowicki[20], D. R. Nygren[8], A. Obertacke[40], S. Odrowski[20], A. Olivas[16], M. Olivo[10], A. O'Murchadha[12], L. Paul[1], J. A. Pepper[36], C. Pérez de los Heros[39], C. Pfendner[17], D. Pieloth[19], E. Pinat[12], J. Posselt[40], P. B. Price[7], G. T. Przybylski[8], L. Rädel[1], M. Rameez[21], K. Rawlins[3], P. Redl[16], R. Reimann[1], E. Resconi[30], W. Rhode[19], M. Ribordy[24], M. Richman[16], B. Riedel[27], J. P. Rodrigues[27], C. Rott[17,d], T. Ruhe[19], B. Ruzybayev[31], D. Ryckbosch[22], S. M. Saba[10], T. Salameh[38], H.-G. Sander[28], M. Santander[27], S. Sarkar[32], K. Schatto[28], M. Scheel[1], F. Scheriau[19], T. Schmidt[16], M. Schmitz[19], S. Schoenen[1], S. Schöneberg[10], A. Schönwald[41], A. Schukraft[1], L. Schulte[11], O. Schulz[30], D. Seckel[31], Y. Sestayo[30], S. Seunarine[33], R. Shanidze[41], C. Sheremata[20], M. W. E. Smith[38], D. Soldin[40], G. M. Spiczak[33], C. Spiering[41], M. Stamatikos[17,e], T. Stanev[31], A. Stasik[11], T. Stezelberger[8], R. G. Stokstad[8], A. Stößl[41], E. A. Strahler[13], R. Ström[39], G. W. Sullivan[16], H. Taavola[39], I. Taboada[5], A. Tamburro[31], A. Tepe[40], S. Ter-Antonyan[6], G. Tešić[38], S. Tilav[31], P. A. Toale[36], S. Toscano[27], M. Usner[11], D. van der Drift[8,7], N. van Eijndhoven[13], A. Van Overloop[22], J. van Santen[27], M. Vehring[1], M. Voge[11], M. Vraeghe[22], C. Walck[34], T. Waldenmaier[9], M. Wallraff[1], R. Wasserman[38], Ch. Weaver[27], M. Wellons[27], C. Wendt[27], S. Westerhoff[27], N. Whitehorn[27], K. Wiebe[28], C. H. Wiebusch[1], D. R. Williams[36], H. Wissing[16], M. Wolf[34], T. R. Wood[20], K. Woschnagg[7], D. L. Xu[36], X. W. Xu[6], J. P. Yanez[41], G. Yodh[23], S. Yoshida[14], P. Zarzhitsky[36], J. Ziemann[19], S. Zierke[1], M. Zoll[34]

[1]III. Physikalisches Institut, RWTH Aachen University, D-52056 Aachen, Germany
[2]School of Chemistry & Physics, University of Adelaide, Adelaide SA, 5005 Australia
[3]Dept. of Physics and Astronomy, University of Alaska Anchorage, 3211 Providence Dr., Anchorage, AK 99508, USA
[4]CTSPS, Clark-Atlanta University, Atlanta, GA 30314, USA
[5]School of Physics and Center for Relativistic Astrophysics, Georgia Institute of Technology, Atlanta, GA 30332, USA
[6]Dept. of Physics, Southern University, Baton Rouge, LA 70813, USA
[7]Dept. of Physics, University of California, Berkeley, CA 94720, USA
[8]Lawrence Berkeley National Laboratory, Berkeley, CA 94720, USA
[9]Institut für Physik, Humboldt-Universität zu Berlin, D-12489 Berlin, Germany
[10]Fakultät für Physik & Astronomie, Ruhr-Universität Bochum, D-44780 Bochum, Germany
[11]Physikalisches Institut, Universität Bonn, Nussallee 12, D-53115 Bonn, Germany
[12]Université Libre de Bruxelles, Science Faculty CP230, B-1050 Brussels, Belgium
[13]Vrije Universiteit Brussel, Dienst ELEM, B-1050 Brussels, Belgium
[14]Dept. of Physics, Chiba University, Chiba 263-8522, Japan
[15]Dept. of Physics and Astronomy, University of Canterbury, Private Bag 4800, Christchurch, New Zealand
[16]Dept. of Physics, University of Maryland, College Park, MD 20742, USA
[17]Dept. of Physics and Center for Cosmology and Astro-Particle Physics, Ohio State University, Columbus, OH 43210, USA
[18]Dept. of Astronomy, Ohio State University, Columbus, OH 43210, USA
[19]Dept. of Physics, TU Dortmund University, D-44221 Dortmund, Germany
[20]Dept. of Physics, University of Alberta, Edmonton, Alberta, Canada T6G 2E1
[21]Département de physique nucléaire et corpusculaire, Université de Genève, CH-1211 Genève, Switzerland
[22]Dept. of Physics and Astronomy, University of Gent, B-9000 Gent, Belgium
[23]Dept. of Physics and Astronomy, University of California, Irvine, CA 92697, USA
[24]Laboratory for High Energy Physics, École Polytechnique Fédérale, CH-1015 Lausanne, Switzerland
[25]Dept. of Physics and Astronomy, University of Kansas, Lawrence, KS 66045, USA
[26]Dept. of Astronomy, University of Wisconsin, Madison, WI 53706, USA
[27]Dept. of Physics and Wisconsin IceCube Particle Astrophysics Center, University of Wisconsin, Madison, WI 53706, USA
[28]Institute of Physics, University of Mainz, Staudinger Weg 7, D-55099 Mainz, Germany
[29]Université de Mons, 7000 Mons, Belgium
[30]T.U. Munich, D-85748 Garching, Germany
[31]Bartol Research Institute and Department of Physics and Astronomy, University of Delaware, Newark, DE 19716, USA
[32]Dept. of Physics, University of Oxford, 1 Keble Road, Oxford OX1 3NP, UK
[33]Dept. of Physics, University of Wisconsin, River Falls, WI 54022, USA
[34]Oskar Klein Centre and Dept. of Physics, Stockholm University, SE-10691 Stockholm, Sweden
[35]Department of Physics and Astronomy, Stony Brook University, Stony Brook, NY 11794-3800, USA
[36]Dept. of Physics and Astronomy, University of Alabama, Tuscaloosa, AL 35487, USA
[37]Dept. of Astronomy and Astrophysics, Pennsylvania State University, University Park, PA 16802, USA
[38]Dept. of Physics, Pennsylvania State University, University Park, PA 16802, USA
[39]Dept. of Physics and Astronomy, Uppsala University, Box 516, S-75120 Uppsala, Sweden
[40]Dept. of Physics, University of Wuppertal, D-42119 Wuppertal, Germany
[41]DESY, D-15735 Zeuthen, Germany
[a]Physics Department, South Dakota School of Mines and Technology, Rapid City, SD 57701, USA
[b]Los Alamos National Laboratory, Los Alamos, NM 87545, USA
[c]also Sezione INFN, Dipartimento di Fisica, I-70126, Bari, Italy
[d]Department of Physics, Sungkyunkwan University, Suwon 440-746, Korea
[e]NASA Goddard Space Flight Center, Greenbelt, MD 20771, USA

**Acknowledgements**

We acknowledge the support from the following agencies: U.S. National Science Foundation-Office of Polar Programs, U.S. National Science Foundation-Physics Division, University of Wisconsin Alumni Research Foundation, the Grid Laboratory Of Wisconsin (GLOW) grid infrastructure at the University of Wisconsin - Madison, the Open Science Grid (OSG) grid infrastructure; U.S. Department of Energy, and National Energy Research Scientific Computing Center, the Louisiana Optical Network Initiative (LONI) grid computing resources; Natural Sciences and Engineering Research Council of Canada, WestGrid and Compute/Calcul Canada; Swedish Research Council, Swedish Polar Research Secretariat, Swedish National Infrastructure for Computing (SNIC), and Knut and Alice Wallenberg Foundation, Sweden; German Ministry for Education and Research (BMBF), Deutsche Forschungsgemeinschaft (DFG), Helmholtz Alliance for Astroparticle Physics (HAP), Research Department of Plasmas with Complex Interactions (Bochum), Germany; Fund for Scientific Research (FNRS-FWO), FWO Odysseus programme, Flanders Institute to encourage scientific and technological research in industry (IWT), Belgian Federal Science Policy Office (Belspo); University of Oxford, United Kingdom; Marsden Fund, New Zealand; Australian Research Council; Japan Society for Promotion of Science (JSPS); the Swiss National Science Foundation (SNSF), Switzerland.

# IceTop as a veto in astrophysical neutrino searches for IceCube

THE ICECUBE COLLABORATION[1]

[1] *See special section in these proceedings*

*jauffenb@icecube.wisc.edu*

**Abstract:** IceCube, the world's largest high-energy neutrino observatory, was built at the South Pole. It consists of photomultipliers deployed 1.5-2.5 km deep into the Antarctic ice cap and detects the trajectory of charged leptons produced during high-energy neutrino interactions in the surrounding ice. The surface air shower detector IceTop located above IceCube can be used to veto the cosmic ray induced background in IceCube to measure astrophysical neutrinos from the southern sky. The implementation of the IceTop veto technique and the impact on different IceCube analyses are presented.

**Corresponding authors:**
Jan Auffenberg [1]
[1] *University of Wisconsin Madison*

## 1 Introduction and motivation

A recent search for high-energy neutrinos in IceCube finds significant evidence for neutrinos of astrophysical origin in the region of several hundreds of TeV neutrino primary energy [1]. A statistically sufficient number of events and high reconstruction quality are essential to be able to do astronomy and measure the spectrum of those neutrinos. The main backgrounds for extraterrestrial neutrino detection are cosmic ray (CR) induced particles. One way to suppress these backgrounds in IceCube is to look at particles that traversed the whole Earth to shield CR induced muons. The disadvantage is that one measures not only astrophysical but also atmospheric neutrinos. At neutrino primary energies above a hundred TeV the Earth starts to get opaque. Thus analyses that look for cosmogenic neutrinos ($E >$ PeV), find the major part of their signal above the horizon. Here, in the southern sky, the Earth cannot shield the charged particles; thus the CR muon background dominates the neutrino spectrum. CR induced air showers produce a large bundle of coincident secondary particles. Those particles can get detected at the surface with IceTop to veto the CR events. This proceedings describe improvements of the analysis to detect cosmogenic neutrinos in the energy range of $10^{8-11}$ GeV by using the IceTop as a veto.

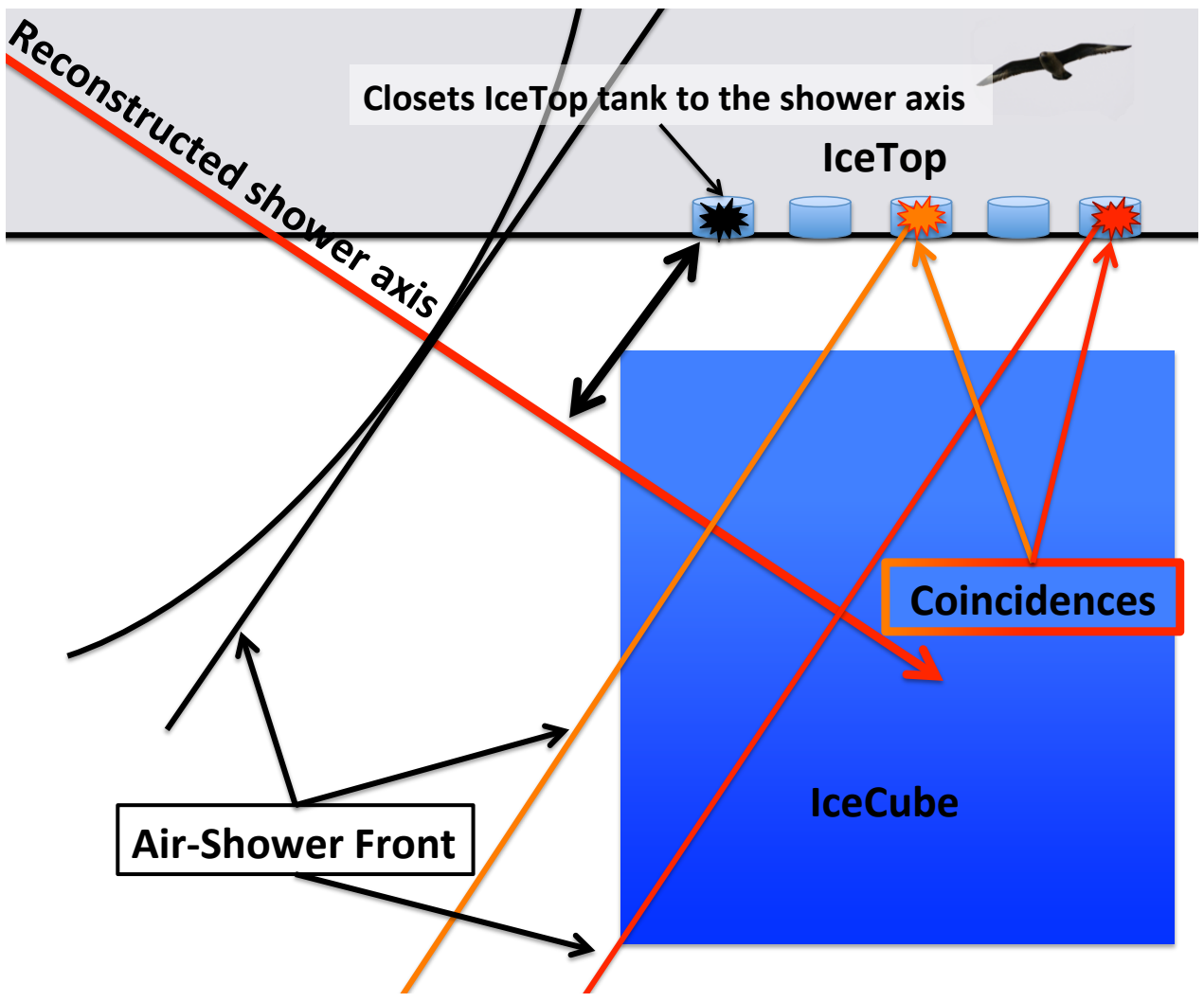

**Fig. 1**: Schematic view of the IceTop veto strategy for the IceCube Observatory. If IceTop records hits consistent in time with a hypothetical shower front the event is considered to be a CR.

## 2 IceCube and IceTop

The search for high energy astrophysical neutrinos, that are believed to get produced in CR sources, and point back to the cosmic ray accelerators is one purpose of the IceCube Observatory [2]. Located at the South Pole, it consists of a surface air shower array IceTop [4], and a deep-ice Cherenkov detector (IceCube) with a more densely instrumented core in the center, called DeepCore. All three detector components use digital optical modules (DOMs) that contain a photomultiplier tube (PMT) in a glass sphere. They detect Cherenkov light emitted by relativistic charged particles passing through the ice. IceCube consists of 86 strings with 60 DOMs deployed between 1450 m and 2450 m depth, encompassing a volume of 1 km$^3$ (see Figure 1).

For the air shower detector IceTop, the optical modules are arranged in pairs of two cylindrical plastic tanks filled with clear ice. Each pair of tanks makes up one station, and is deployed above IceCube strings. The focus of IceTop is the detection of electrons and muons from cosmic ray induced air showers. The main science goal of IceTop is to reconstruct the energy, type, and direction of the cosmic ray primary particle.

IceTop and IceCube can be combined to measure the electromagnetic and muonic component of the air shower on the surface and the high-energy muonic component in the deep ice, improving sensitivity to the composition of the cosmic rays [3]. An event in IceCube with coincident hits in IceTop is a clear evidence for an atmospheric muon bundle or a neutrino generated in a cosmic ray air shower. A charged lepton induced by an astrophysical neutrino should have only random coincident background hits in IceTop. Hence IceTop can be used as a veto for CRs to enhance the sensitivity for neutrino detection in IceCube.

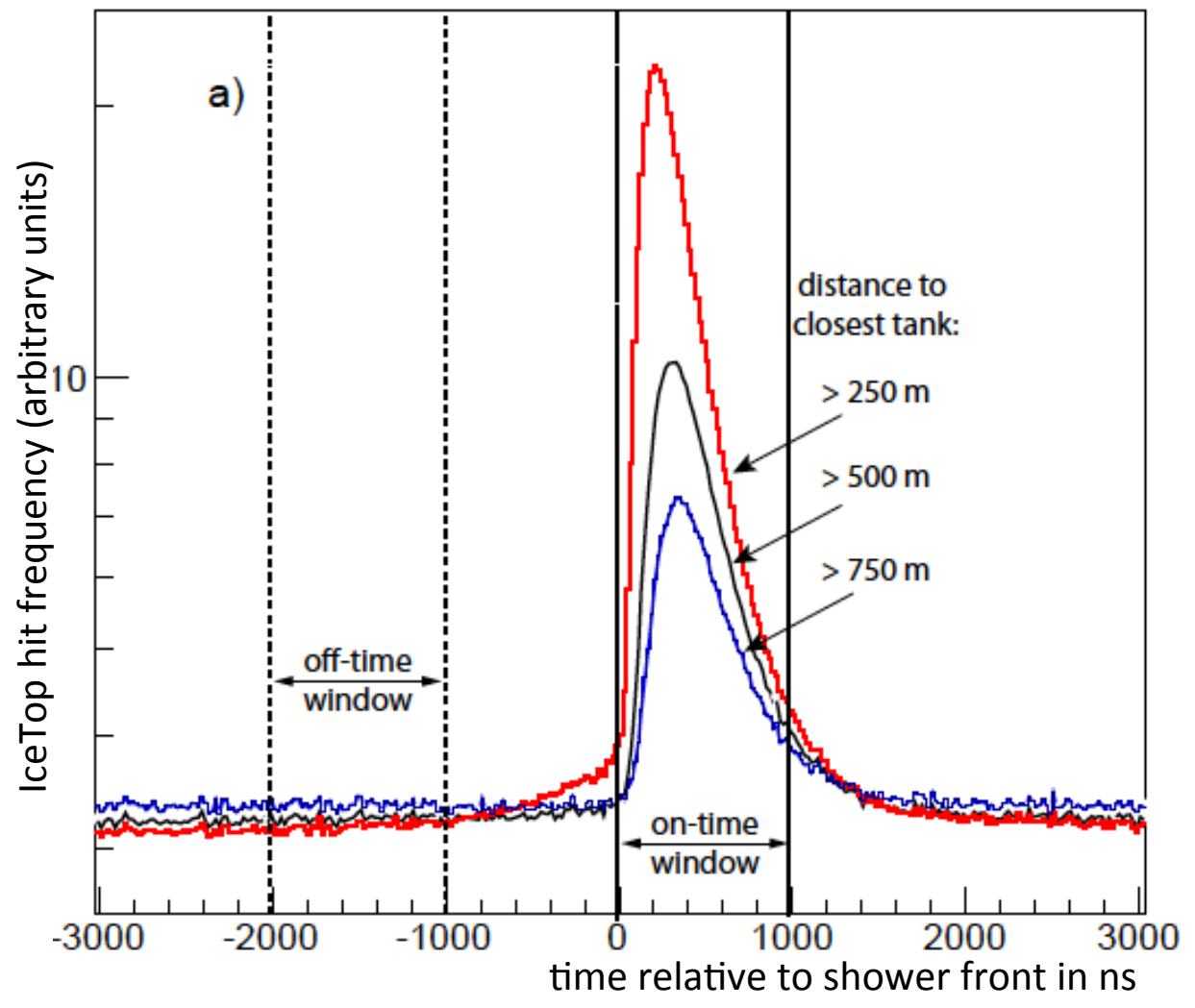

**Fig. 2**: IceTop hit distribution relative to the expected air shower front based on IceCube reconstruction. The closer the IceTop tanks are relative to the shower axis the higher the expected number of hits.

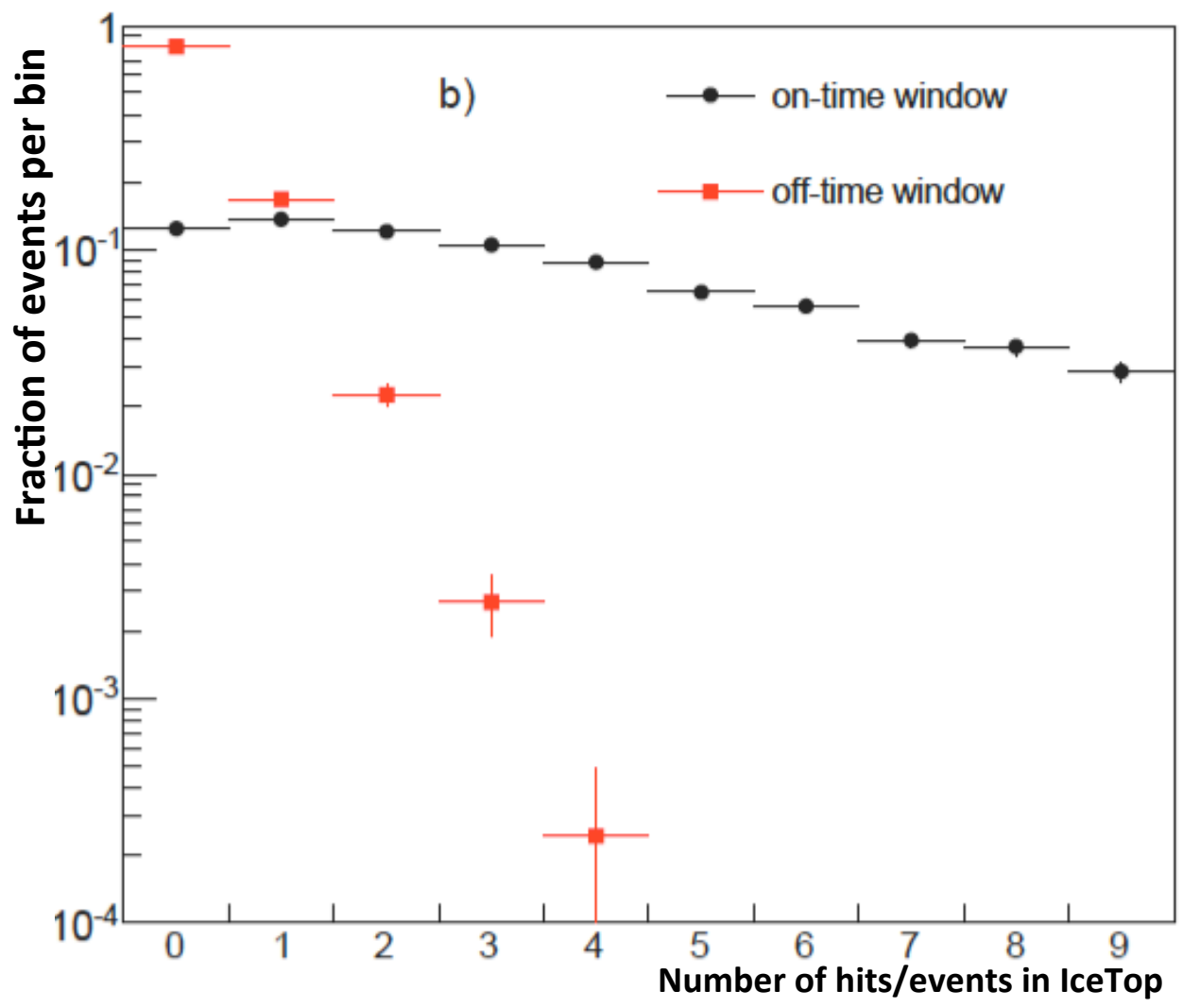

**Fig. 3**: Hit multiplicity of IceTop for events with more than 500 m distance of the closest tank to the shower axis. The multiplicity of hits in the off-time window is steeply falling, reducing the probability of removing a neutrino signal with the veto.

### 2.1 The IceTop Veto

To use IceTop information to veto high-energy CRs in IceCube, a simple algorithm was implemented (Fig. 1). Based on the reconstructed shower axis for each IceCube event a flat hypothetical shower front is constructed. The time difference of the shower front to the hits in IceTop on the surface is calculated. Every hit in IceTop coincident with the predicted shower front, or up to $1\mu$s later, is counted as a possible air shower correlated hit on-time. If the number of hits in IceTop exceeds the threshold, the event is vetoed. Figure 2 shows the hit frequency of events from 33.5 days of real IceTop data relative to the shower front. The veto efficiency weakens as the distance of the IceTop tank from the shower axis increases.

Figure 3 shows the IceTop hit multiplicity at the on-time window of correlated hits (black horizontal points) and off-time window of uncorrelated events (red fast dropping points). The probability of accidentally vetoing an extraterrestrial neutrino-induced event ($P_{av}$) can be determined from real data by dividing the number of events vetoed in the off-time window over the total number of events.

| Number of IceTop Hits | Fraction of Signal events |
|---|---|
| $>3$ | $4\times10^{-4}$ |
| $>2$ | $2.9\times10^{-3}$ |
| $>1$ | $2.65\times10^{-2}$ |
| $>0$ | $1.994\times10^{-1}$ |

**Table 1**: Rejection probability for astrophysical neutrinos given a number of IceTop hits in the off-time window.

$P_{av}$ to veto an extraterrestrial neutrino with more than 2 IceTop hits, for example, is calculated to be smaller than $3\times10^{-3}$ in a 1 $\mu$s time window. For more than 0 hits in IceTop, 20% of the neutrinos are rejected (Tab. 1).

## 3 The IceTop Veto applied to current IceCube EHE analyses

The search for cosmogenic neutrinos with IceCube [7] attempts to extract the neutrino signal from the CR background by cutting in the parameter space of inclination and NPE. This, after some additional quality cuts (reconstruction and size of the event in the detector), leads to a good separation. The CR background dominates the southern sky at lower NPE, while the cosmogenic neutrino signal clusters at high NPE at the horizon. Figure 4 shows that the island of expected cosmogenic neutrino signal from [6] at and above the horizon gets less and less inclined with increasing NPE. This is due to the correlation of NPE with the primary energy of the corresponding neutrino. With increasing energy of the cosmogenic neutrinos, the cross-section increases too, such that neutrinos are reaching IceCube more and more vertically. Figure 5 shows the distribution of the CR background as a function of inclination and NPE. Here we see that the CR background decreases with increasing NPE and inclination due to the absorption in Earth and the steep CR spectrum. The red dotted line indicates the cut in the current analysis without using the IceTop veto, which resulted in a 3 $\sigma$ discovery potential for cosmogentic neutrinos from [6].

IceTop detects a large fraction of the CR background at high NPE (see Fig. 6). This IceTop information can be used to veto cosmic ray induced events in IceCube. The best CR veto efficiency for the extremely high energy neutrino detection analysis is found for a minimum number of IceTop hits in the on-time window lager than 2. It is divided in four inclination regions, $\cos(\theta)>0.75$, $0.75\geq\cos(\theta)>0.5$, $0.5\geq\cos(\theta)>0.25$, and $0.25\geq\cos(\theta)>0$, where, due to limited statistics, only the first three are used in this analysis. Figure 6 shows the veto efficiency in the different inclination regions as a function of the number of photoelectrons (NPE) detected in IceCube. It was found that the veto efficiency gets worse with increasing inclination. This is mostly due to the increasing lateral distance of the air shower axis from the nearest IceTop tank. The statistical error on the veto efficiency is dominating the overall error.

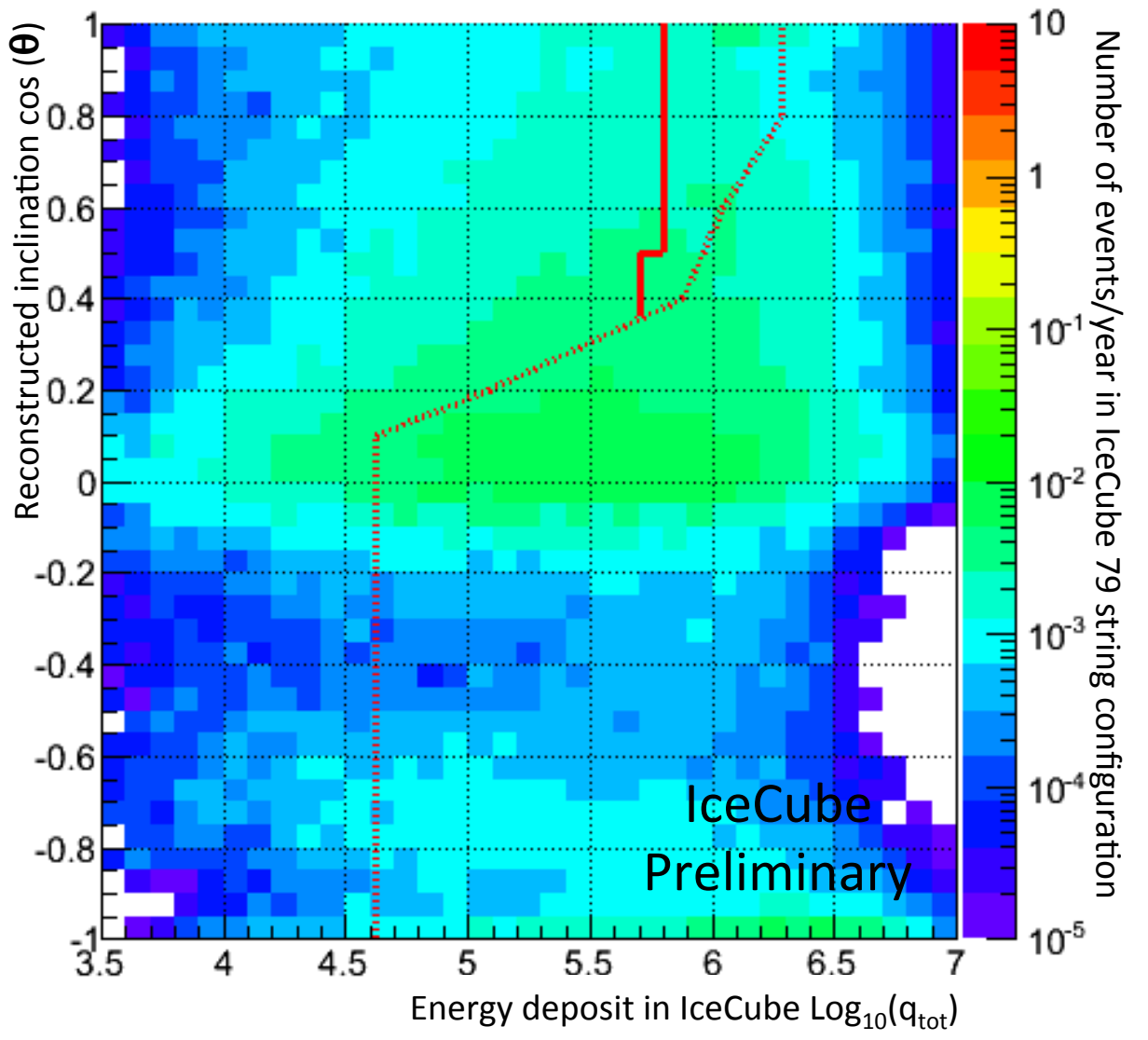

**Fig. 4**: Expected cosmogenic neutrino signal from [6] as a function of reconstructed inclination and the number of NPE detected in IceCube. Most of the signal is expected to cluster at the horizon and get more vertical with increasing NPE. The event rate is normalized to one year of IceCube lifetime. The dashed line indicates the cut region without IceTop veto. The solid line is the new cut including the IceTop veto.

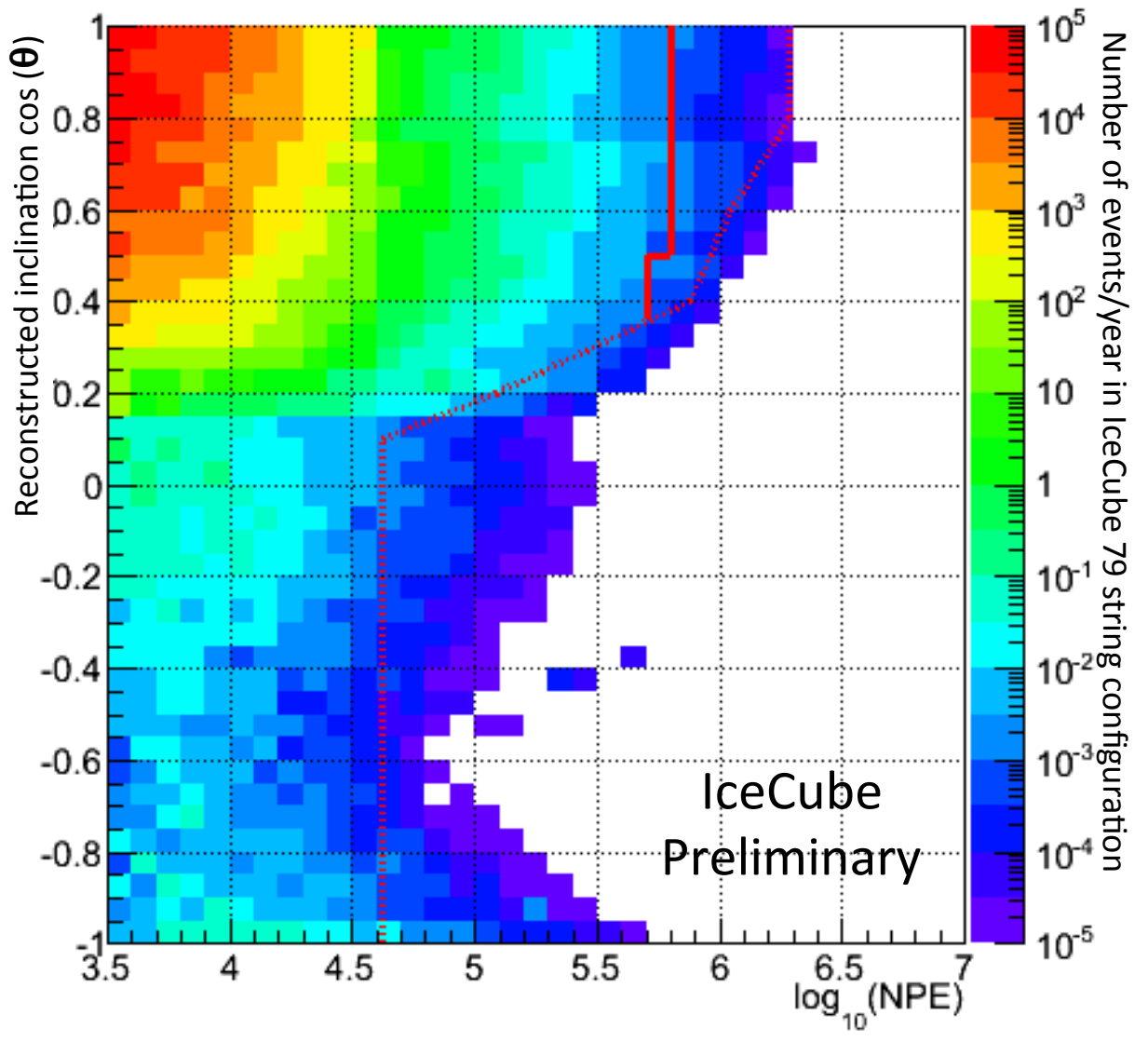

**Fig. 5**: Expected cosmic ray background based on CORSIKA simulations after applying the IceTop Veto. The cosmic ray background is thinning with increasing inclination and an increasing NPE in IceCube. The event rate is normalized to one year of IceCube lifetime. The thin line indicates the cut region without IceTop veto. The solid line is the new cut after IceTop veto. The expected number of events per year are given in Table 3.

| inclination | $V_{best}$ | stat. err. | NPE |
|---|---|---|---|
| $\cos(\theta) \geq 0.75$ | 99.0% | -2.0% +1% | $10^{4.6}$ |
| $0.75 > \cos(\theta) \geq 0.5$ | 88.5% | -6.5% +5% | $10^{4.7}$ |
| $0.5 > \cos(\theta) \geq 0.25$ | 100% | -24.0% +0% | $10^{4.7}$ |
| $0.25 > \cos(\theta) \geq 0.0$ | 100% | -100.0% +0% | $10^{4.5}$ |

**Table 2**: The highest veto efficiency $V_{best}$ for each inclination region including the statistical uncertainties and the corresponding lower NPE limit above which the veto will be applied.

Assuming that, with increasing light deposit in IceCube (higher NPE), the number of air shower particles that can get detected by IceTop will only increase, the veto efficiency determined for a given inclination region and NPE (Table 2) can be applied to the events at higher NPE for a given inclination region. As no coincident IceTop and IceCube high energy cosmic ray CORSIKA [8] simulations are available, the IceTop veto results were applied to the signal and background simulations as a constant reduction factor in each inclination region above the NPE values for which the best Veto efficiency $V_{best}$ was found.

Figure 5 shows the expected CR background from simulation after applying the veto efficiency to the simulated data. The CR background is reduced due to the IceTop veto at high NPE compared to the older analysis [7]. The overall cosmogenic neutrino flux gets only reduced by 0.3% calculated with Table 1. The remaining cosmogenic neutrino flux is shown as a function of NPE and reconstructed inclination in Fig. 4. Based the remaining CR background including the IceTop veto and the remaining signal, we recalculate the cut for a $3\sigma$ discovery potential on cosmogenic neutrinos from [6]. The solid lines in the Figs. 4, and 5 indicate the improved cut. Table 3 summarizes the expected cosmogenic signal and background. The overall improvement is about 10%, mostly in the region above 10 PeV neutrino primary energy.

| inclination band | NPE cut | backgr./a | signal/a |
|---|---|---|---|
| $\cos(\theta) \geq 0.75$ | $> 10^{5.8}$ | 0.00739 | 0.0596 |
| $0.75 > \cos(\theta) \geq 0.5$ | $> 10^{5.8}$ | 0.0128 | 0.0797 |
| $0.5 > (\theta) \geq 0.36$ | $> 10^{5.7}$ | 0.00534 | 0.0613 |
| **Σ all downgoing** | **var** | **0.0255** | **0.201** |
| $0.36 \geq (\theta) \geq -1.0$ | $10^4$.7 | 0.0472 | 0.903 |
| **Σ all inclination** | **var** | **0.0727** | **1.104** |

**Table 3**: Table of the expected cosmic ray background and cosmogenic neutrino induced signal/year after applying the IceTop veto. The overall cosmogenic neutrino signal that IceCube can detect increases from 1.01 without IceTop to 1.104. The IceTop veto is improving the analysis in the down-going region where the detectable signal almost doubles from 0.11 to 0.2.

This leads to an improved all flavor cosmogenic neutrino sensitivity shown in Fig. 7. The systematic uncertainties are the same as those from the older analyses [7]. Figure 8 shows the ratio of sensitivity and effective area before and after using the IceTop veto. The improvement is mostly in the region from $10^{8-11}$ GeV neutrino primary, the region where most of the cosmogenic neutrino flux is expected.

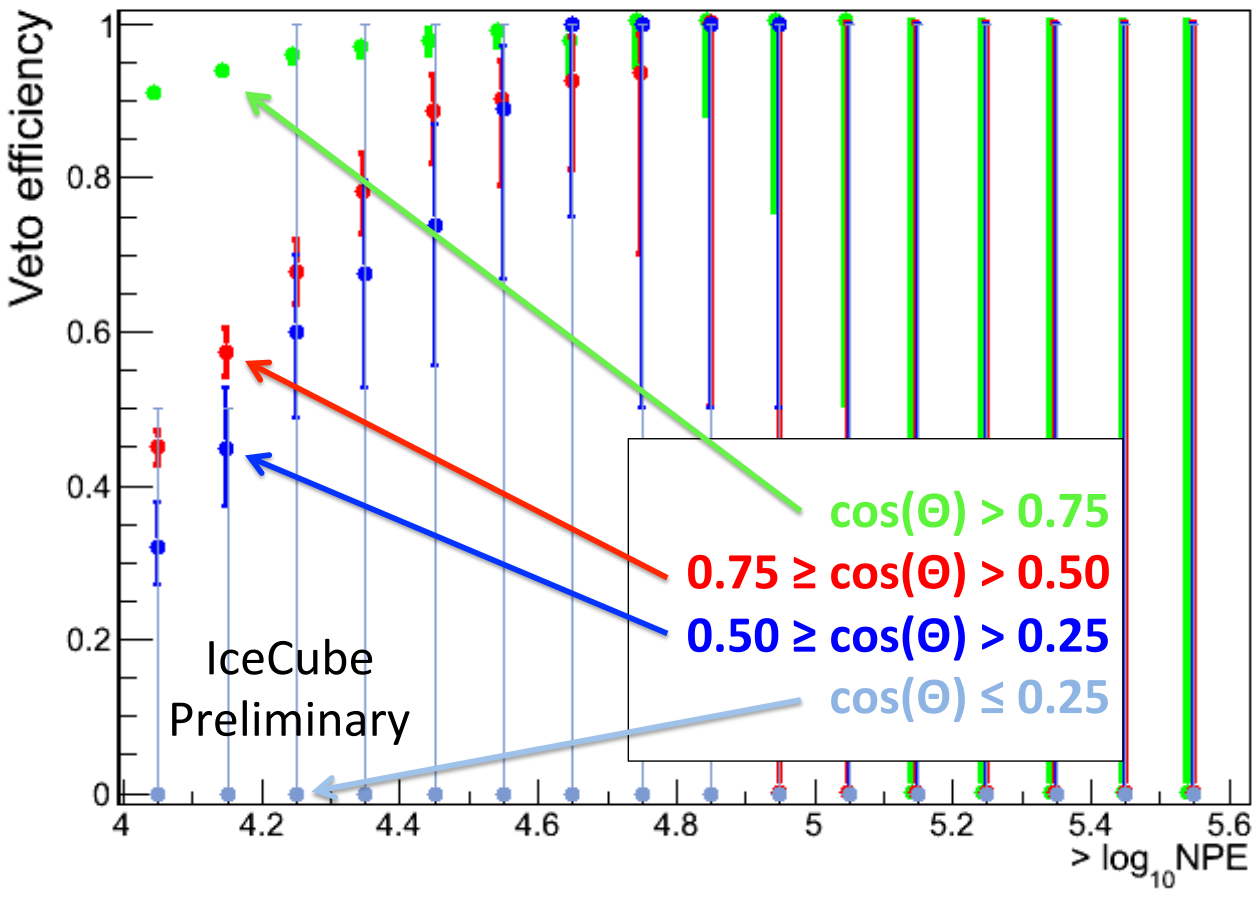

**Fig. 6**: IceTop hit veto efficiency. An event is vetoed if more than 2 Hits are seen in the 1 $\mu$s on-time window. We see the veto efficiency increasing with increasing light deposit in IceCube (NPE). The best veto efficiency is found for the most vertical showers.

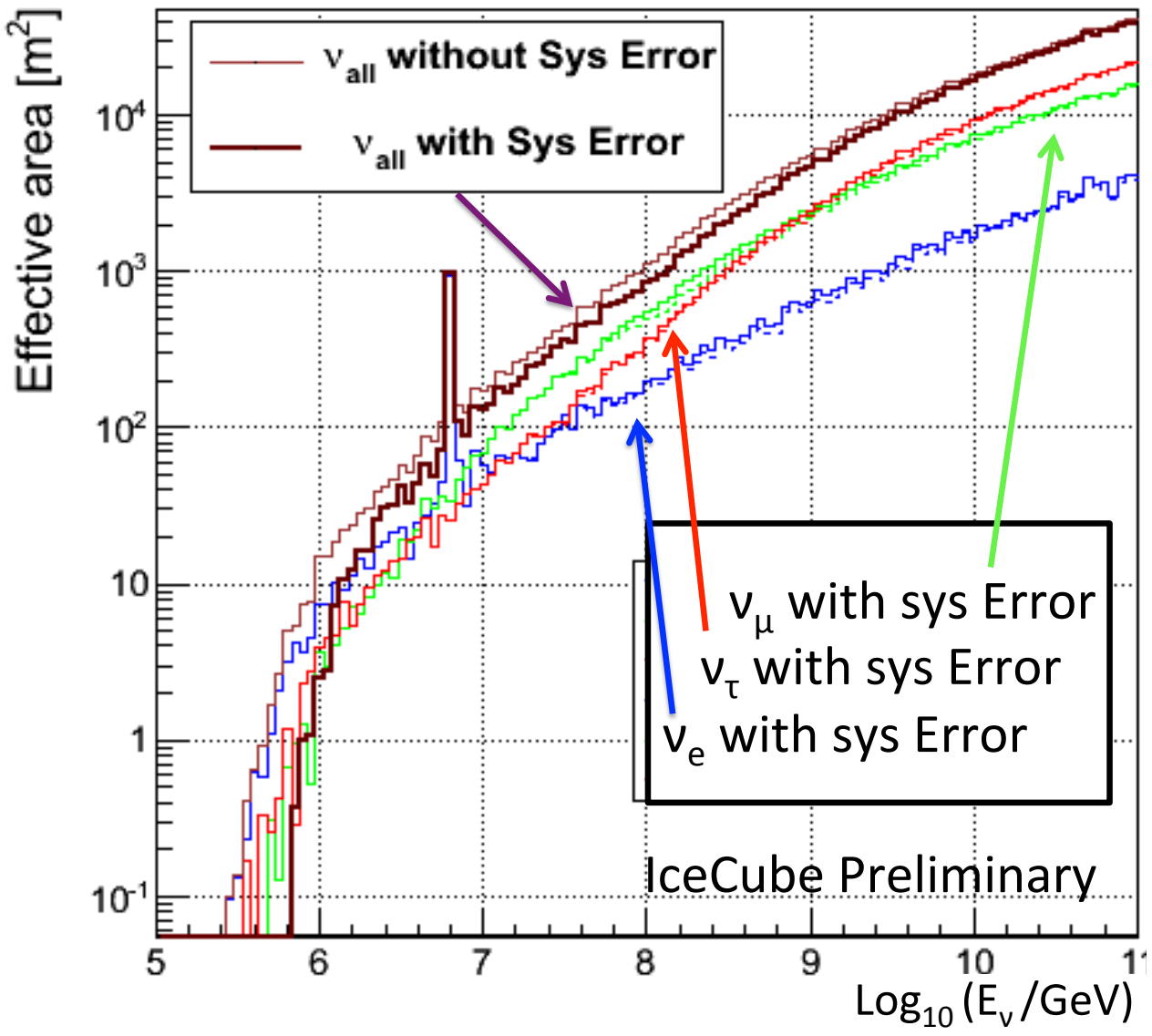

**Fig. 7**: Sensitivity of the cosmogenic neutrino search in IceCube for different neutrino favors and its sum. The sensitivity improved by 10% compared to the same analysis without using the IceTop veto.

## 4 Summary and Conclusion

Looking for coincident time hits in IceTop is a powerful tool to improve the ability to identify and ultimately filter high energy cosmic rays in current IceCube neutrino analyses. In current cosmogenic neutrino searches, the IceTop veto method is most effective in the energy region from $10^{8-11}$ GeV. This technique is also effective in reducing the background of point source analyses in the down going region [9]. The overall improvement for cosmogenic neutrinos searches in IceCube is now about 10% by doubling the observable flux in the down going region. The IceTop veto efficiency was optimized using only 33.5 days of 79-string IceCube configuration. Adding more data will reduce the statistical uncertainties in the veto efficiency calculation and further enhance the veto power.

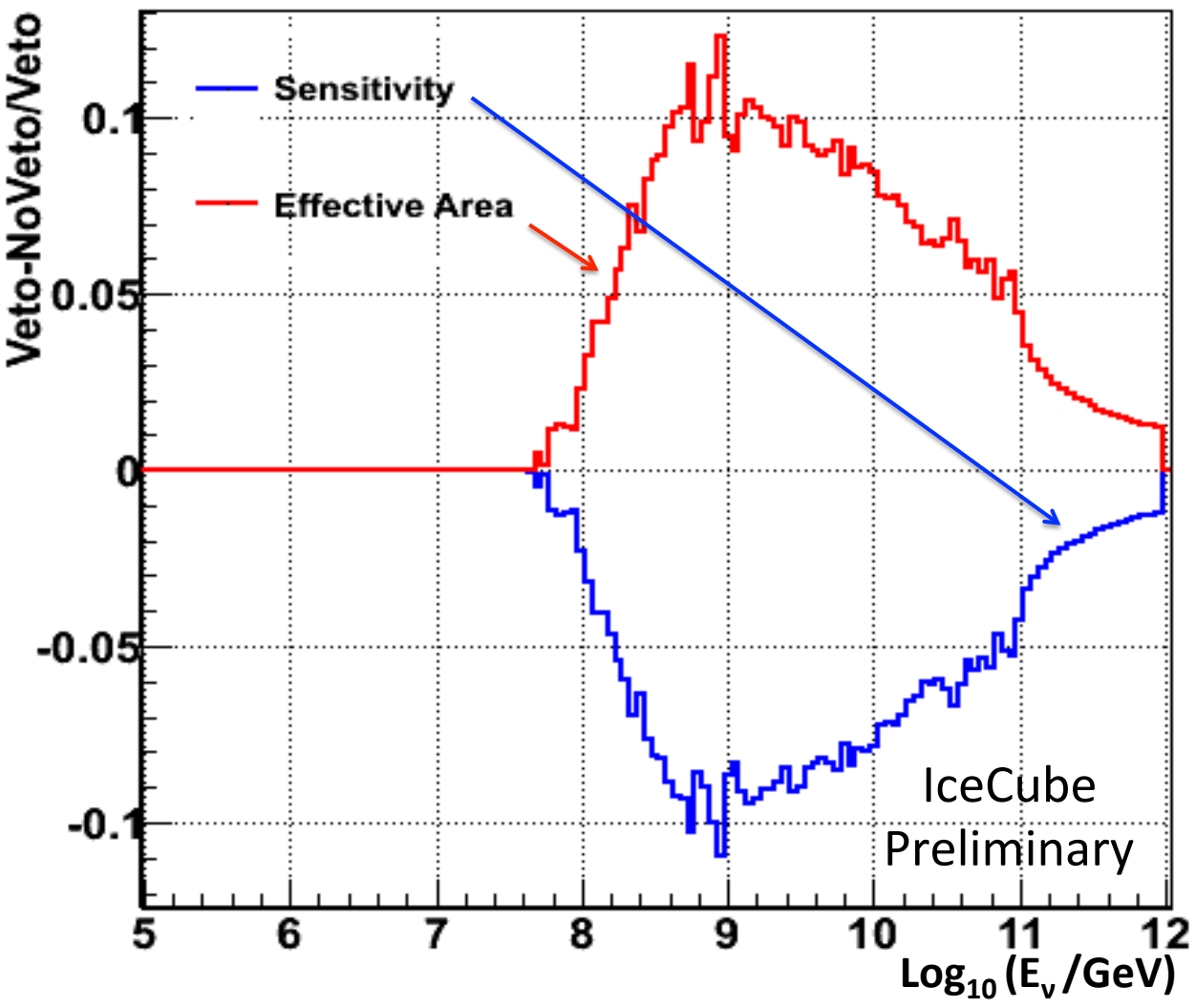

**Fig. 8**: Relative change in all-flavor sensitivity (red line) and effective area (blue line) for the cosmogenic neutrino search in IceCube with and without IceTop veto included. The sensitivity improves mostly in the neutrino energy region from $10^8$ GeV to $10^{11}$ GeV.

## References

[1] IceCube collaboration, M. G. Aartsen et al., arXiv:1304.5356 [astro-ph:HE] (2013).
[2] A. Achterberg et al., Astropart. Phys., 26 (2006) 155-173, doi: 10.1016/j.astropartphys.2006.06.007.
[3] R. Abbasi et al., Astropart. Phys., 42 (2013) 15-32, doi: 10.1016/j.astropartphys.2012.11.003.
[4] R. Abbasi et al., Nucl. Inst. Meth. A, 700 (2013) 188-220, doi: 10.1016/j.nima.2012.10.067.
[5] IceCube Coll., paper 0650 this proceedings.
[6] S. Yoshida and M. Teshima, Prog. Theor. Phys. 89 (1993) 833, doi: 10.1143/PTP.89.833.
[7] K. Mase et al., Proceedings for Recontres de Moriond (2013).
[8] D. Heck, et al., FZKA Report 6019, (1998).
[9] R. Abbasi et al. *Search for time-independent neutrino emission from astrophysical sources with 3 years of IceCube.* Paper in preparation (2013).

# IceVeto: An Extension of IceTop to Veto Horizontal Air Showers

THE ICECUBE COLLABORATION[1]

[1] *See special section in these proceedings*

*jauffenb@icecube.wisc.edu*

**Abstract:** IceCube is the world's largest high-energy neutrino observatory, built at the South Pole. It consists of photomultipliers deployed 1.5-2.5 km deep into the Antarctic ice cap and detects the trajectory of charged leptons produced during high-energy neutrino interactions in the surrounding ice. A sufficient number of background free events and high reconstruction quality are essential, for neutrino astronomy and to measure the spectrum of astrophysical neutrinos. This motivates extended studies to determine the cost and physics potential of an IceTop Extension, IceVeto to detect neutrinos in the southern hemisphere. Building on the experience of IceTop/IceCube, and Auger, the possibly most cost effective, and detection efficient way to build IceVeto is an extension of the IceTop detector with simple photomultiplier based detector modules for CR detection. First simple estimates indicate that such a veto detector will more than double the discovery potential of current point source analysis. Here we present the motivation and capabilities based on first simulations.

**Corresponding authors:**
Jan Auffenberg [1]
[1] *University of Wisconsin Madison*

## 1 IceCube and IceTop as motivation for IceVeto

The IceCube observatory [1] was built to measure high energy astrophysical neutrinos that point back to the cosmic ray accelerators. Located at the South Pole, it consists of a surface air shower array, called IceTop [2], and a deep-ice Cherenkov detector (IceCube).

IceTop is used as a veto for CR in astrophysical neutrino searches of IceCube but only for extremely vertical showers or showers of extremely high energy as the geometrical overlap of IceTop and IceCube is small [3]. IceVeto starts with the idea to increase the overlap of a surface detector for CR's with IceCube down to higher inclinations (Figure 1). There are several points that motivate extended studies to determine the physics potential of an IceTop extension, using similar detection modules like the IceTop tanks, IceVeto.

The most recent analyses of IceCube data have found neutrinos in the region of up to PeV energies. The two highest energy neutrinos are shown not to be cosmic ray-induced with a significance of $2.8\sigma$ (which is of course cosmic ray-model dependent) [4]. Both are neutrino-induced cascades that are reconstructed to be downgoing. Another analysis, known as a "starting-event search", selects only events where the neutrino had its first interaction within the IceCube volume [5]. This analysis finds an excess of neutrinos from the Southern Hemisphere, which is in clear disagreement with predictions from the cosmic ray induced atmospheric neutrino flux. Of course, any search of the southern sky would want to be sensitive to the galactic center as a potential source.

IceVeto would be able to reduce several types of background contributing to a variety of searches. Most obviously, it could tag cosmic ray air showers whose downgoing muon bundles are a background for neutrino searches in the southern sky, near the horizon, or in the northern sky where misreconstructed downgoing muons can contaminate an upgoing signal. But more interestingly, IceVeto could also reduce atmospheric *neutrino* background, which is now the dominant background for both diffuse and point source searches at high energies. As the mean free path length of muons of tens of TeV to tens of PeV muon primary energy is on the order of kilometers, the $\nu_\mu$ induced muons will reach IceCube even at high inclinations. In addition, the probability for an astrophysical $\nu_\mu$ to interact is proportion-al to the amount of matter it traversed. So accordingly, the effective area compared to the starting-event approach only increases by a factor of 2.4 in the straight downgoing region (as the detector is 1km in size at 1.45 kilometer depth) and by more than a factor of 5.5 at the height of the galactic center (zenith angle of $\theta = 62°$) where the $\nu_\mu$ has to traverse much more ice. The increase of the effective area given an IceVeto detector working from $\theta = 0° - 75°$ is about a factor of 3.5 compared to the starting event analyses [5] for muon neutrinos. The overall increase in the *all-flavor* effective area in the southern sky (assuming that $\nu_e$ and $\nu_\tau$ effective areas remain the same) will increase by a factor of about two. These additional $\nu_\mu$ events are especially useful in point source searches, as their arrival directions are the best reconstructed. The number of astrophysical $\nu_\mu$ induced muons based on an $E^{-2}$ spectrum integrated from 30 TeV to 10 PeV neutrino primary energy are estimated with the attenuation of muons taken into account for the starting-event search, IceTop and IceVeto, as shown in Figure 2.

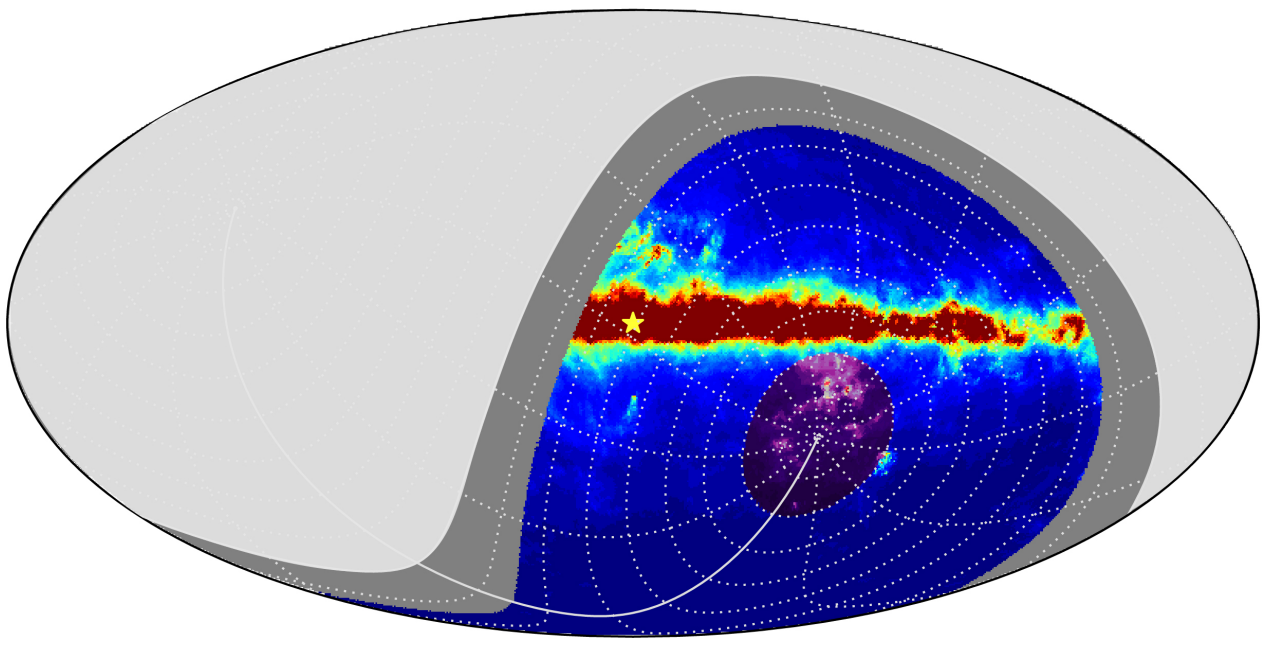

**Fig. 1**: GALPROP $\pi^0$ 1.1TeV skymap in galactic coordinates. The light grey region is the northern sky which is dominated by atmospheric neutrinos. The bright-colored region is the region of the southern sky covered by the IceVeto extension proposed in this work, which includes the galactic center (the yellow star). The purple region indicates coverage of the current IceTop detector when used as a veto. The dark grey band is above the horizon but not covered by IceVeto.

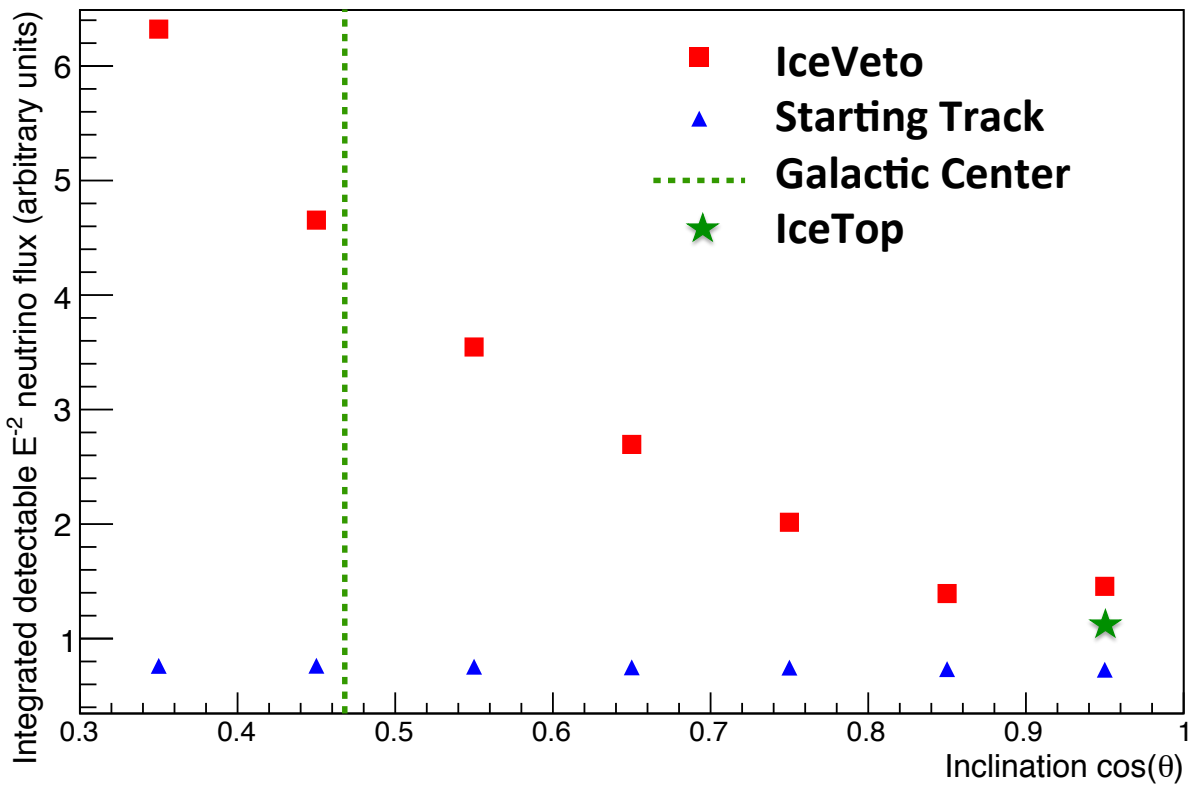

**Fig. 2**: Integrated detectable E$^{-2}$ $\nu_\mu$ induced muon flux with IceVeto compared to the starting-event selection method integrated from 30 TeV to 10 PeV. The green line indicates the position of the Galactic Center. Even in the most vertical bin ($\cos(\theta) > 0.3$) IceTop is worse than IceVeto as it already includes events that are to far away from IceTop (star).

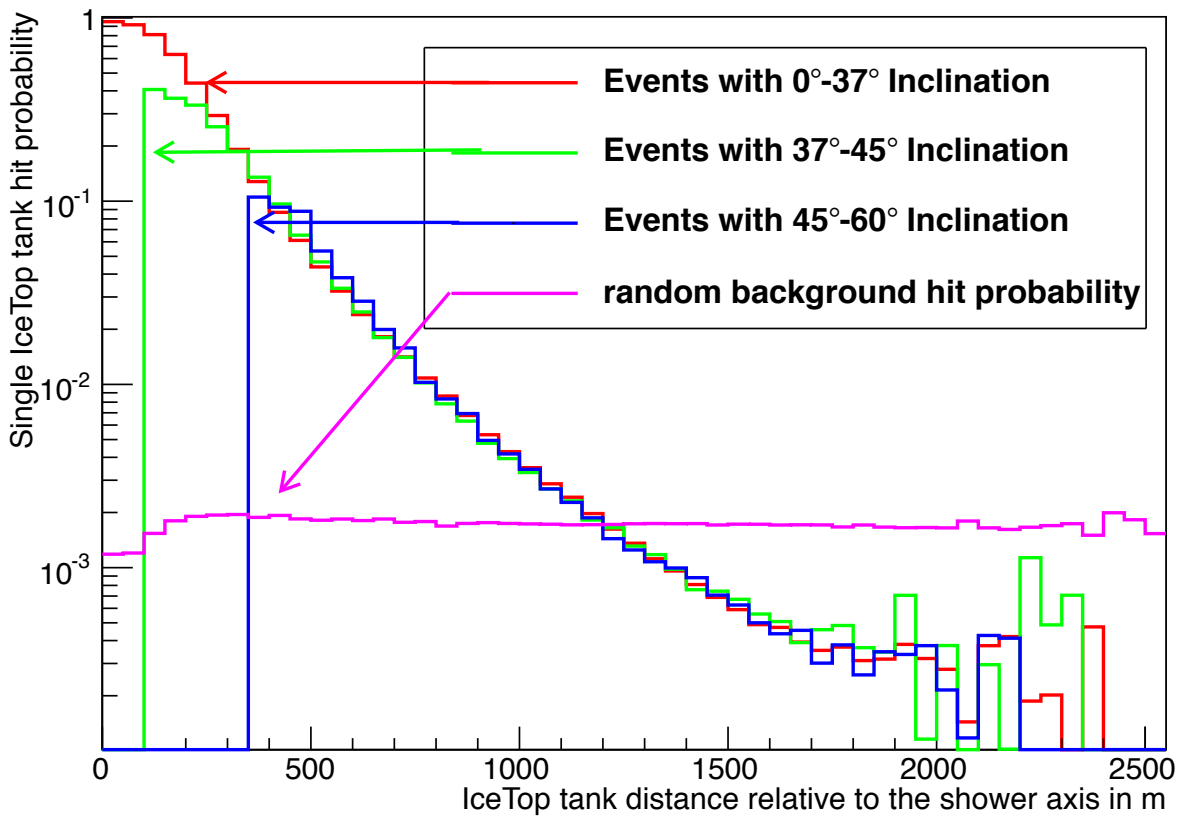

**Fig. 3**: Hit distribution of a single IceTop tank for different inclinations in the on-time veto window with the off-time window hit probability subtracted. The hit probability stays the same in the region where well-reconstructed events are dominant. The single-tank hit probability in the off-time window is constant down to the region where only misreconstructed events have no hits in IceTop tanks that are close to the axis.

Currently, atmospheric neutrinos can be removed from a cascade event sample if IceCube sees accompanying muons coming from outside the detector volume. This technique is not effective near the horizon, where muons from bundles are absorbed due to large amount of ice they have to traverse. However, such events could still be removed by a surface detector near the shower axis such as IceVeto. The electromagnetic component of the CR air shower contains orders of magnitude more detectable particles (electrons), and those particles would have traveled through about 100 times less particle density compared to those that reach IceCube.

## 2 Study of IceTop Tanks with data

To simulate a possible extension of IceTop to veto CRs for neutrino detection we need to describe the single modules the detector will be based on, in particular, to quantify the probability of a single module to get hit relative to the air shower properties (inclination, lateral distance) and relative to the neutrino energy of the particles that reach IceCube.

In order to evaluate such information we select 33.5 days of real IceCube and IceTop events from the years 2010/11 with a light deposit in IceCube of $10^3 - 10^4$ $q_{tot}$ with a bin-size of $10^3$ $q_{tot}$ and no additional cuts, where $q_{tot}$ is the light deposit roughly equal to the photomultiplier charge of one measured photon. The energy of the relativistic and charged particles that are moving through IceCube is proportional to the Cherenkov light deposit. Thus, we are selecting events with corresponding neutrino energies of about 20-400 TeV. The $q_{tot}$ distribution follows the slope of the cosmic ray spectrum and mostly populates the low-light deposit part of $10^3$ $q_{tot}$. It is running out of statistics at $10^4$ $q_{tot}$. While the neutrino energy corresponding to the light deposit in IceCube is almost independent of the inclination, the cosmic ray primary energy for a fixed light deposit in IceCube is dependent on the amount of South Pole ice the shower particles had to traverse. Thus, horizontal showers with the same light deposit in IceCube have a higher cosmic ray primary energy because they were thinned out more while traveling through more matter. This leads to a reduction of the background flux with increasing inclination. Given the fixed cut on the light deposit at $10^3 - 2\times10^3$ $q_{tot}$ in IceCube (equivalent to some 10 TeV neutrino primary energy), we are interested in the hit probability of a single IceTop tank for our simulation given a certain event topology. As for the IceTop veto, the IceVeto on-time window is chosen to be up to 1 $\mu$s after the expected air shower front. To estimate the background from accidental coincidences an off-time window of 1 $\mu$s length, 1 $\mu$s before the actual shower front is expected [3], was chosen.

The tank hit probability was found to be only dependent on the lateral distance to the shower axis but not on the inclination (see Figure 3). One can clearly see that the hit probability of close tanks in the case of a more inclined shower cannot be tested, as it is geometrically impossible for an inclined shower to hit IceCube and at the same time to be close to IceTop. There is a class of inclined events, which have an unstable reconstruction. They are IceCube corner clippers, which only pass through a shallow part of the IceCube detector far away from the surface hit position. This makes the IceTop tank hit probability drop artificially;

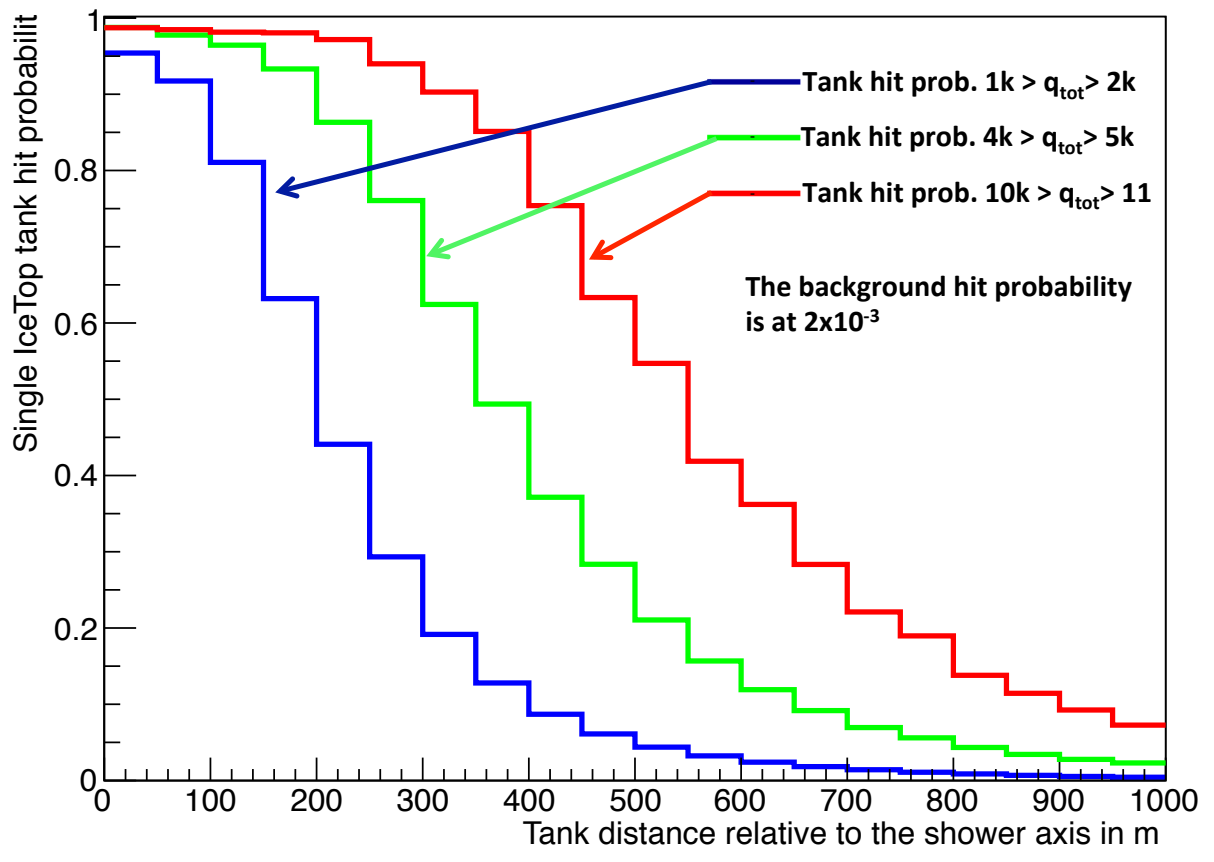

**Fig. 4**: The single-tank hit probability, derived from experimental data, is improving with increasing light deposit in IceCube. At small light deposits tanks that are close to the shower axis become more important. The background hit probability is at the level of $2 \times 10^{-3}$.

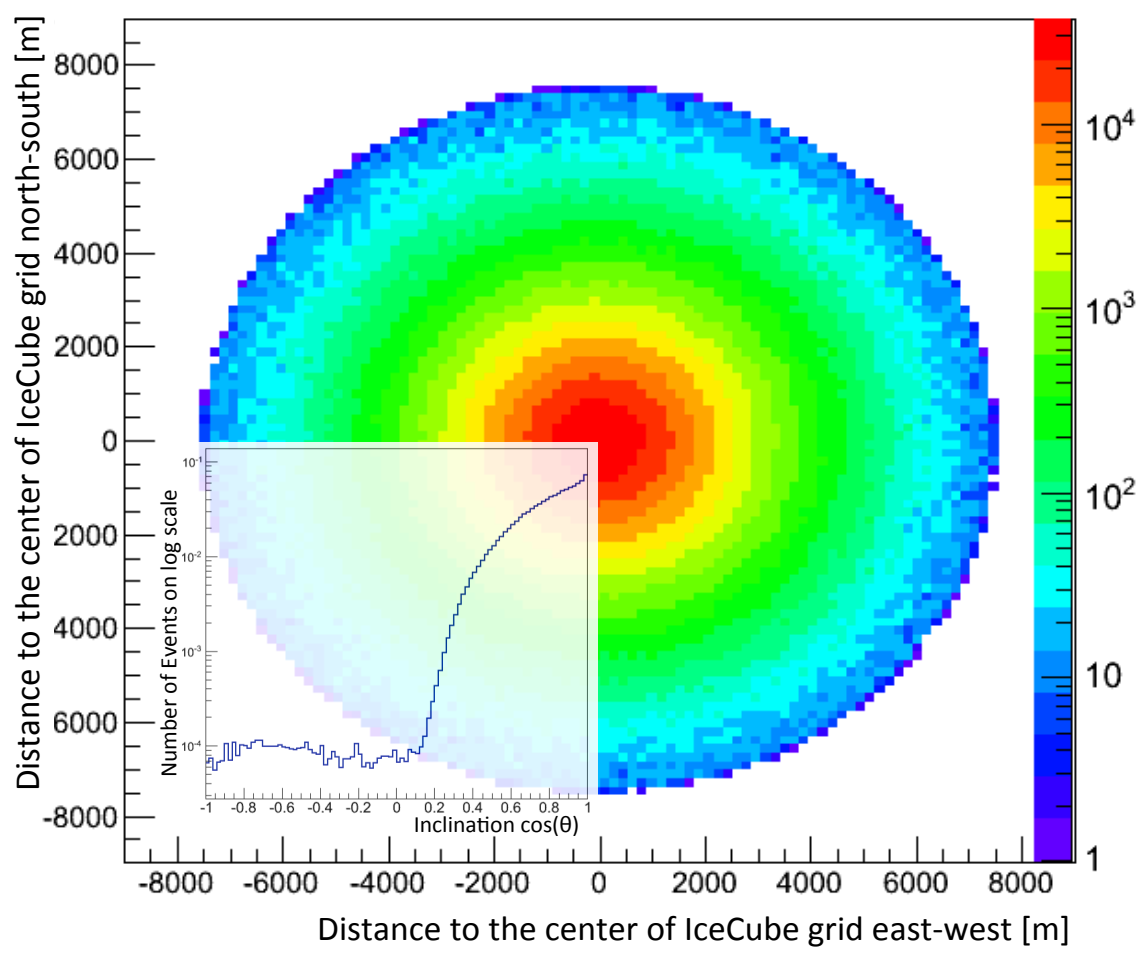

**Fig. 5**: Simulated distribution of events with $q_{tot} 10^3$ in the IceVeto array for about 8 years of lifetime and 0° to 75° inclination. The graph in the lower left corner shows the event multiplicity as a function of the inclination in $\cos(\theta)$ on a logarithmic scale. The events are rapidly thinning out as they become more horizontal.

at even closer distances of the tank to the shower axis there is no geometrical possibility left.

As a result the tank hit probability/$\mu$s time window for a cosmic ray-induced air shower for a fixed light deposit in IceCube is taken to be independent of the inclination for a given $q_{tot}$ (Fig. 4). As astrophysical neutrinos have only random cosmic ray-induced hits in IceTop, the IceTop tank-hit probability/$\mu$s is described by a time window collecting only random hits (off-time window in Figure 4) .

## 3 Simulation of a possible IceVeto configuration

Here a Monte Carlo simulation that assumes to consist of similar modules like the IceTop tanks is described. It was assumed that the cosmic ray flux is isotropic over the sky for a fixed primary energy. As the correlation of primary energy and light deposit in IceCube is changing with inclination, we correct for this from the measured distribution of the light deposit in IceCube as a function of the cosine of the inclination. The cosine conserves the assumed cosmic ray isotropy over the sky leading to an increasing radial distance of the tanks to each other with increasing distance to IceCube. This gives us the capability of simulating a realistic cosmic ray flux of IceCube events with fixed light deposit with a bin width of $10^3$ $q_{tot}$ from $10^3$ to $10^4$ $q_{tot}$ (see Figure 5). We constrain the radius of the array to be not larger than 7km. This includes the galactic center but reduces the number of tanks at the cost of losing the very inclined region for most efficient CR vetoing. The array was optimized to provide a veto efficiency high enough to suppress the CR background, comparable to the starting event selection homogeneously over the entire inclination region using an as low as possible number of modules. The geometry that uses the lowest number of modules and that still rejects 99.98% of the cosmic rays that produce an light deposit in IceCube $> 2 \times 10^3$ $q_{tot}$ while keeping 94% of the neutrino signal consists of 943 modules with a maximum distance of 6.7 km from the IceCube center (Figure 6). For a light deposit in IceCube $> 4 \times 10^3$ $q_{tot}$ the veto already rejects 99.999% of all cosmic ray-induced signals due to the increase in the hit probability. This is currently considered to be the most cost efficient way to extend IceTop to veto CRs. We expect that 1000 modules optimized to only work as a CR veto deployed on the surface are more cost effective in comparison to ideas that for example extend the detector deep in the ice and would lead to the same improvements to astrophysical neutrino detection and astronomy. First estimates indicate that IceVeto could maybe be around 20 times less expensive than IceCube or even less. Further investigation is planed in order to achieve a more precise cost estimate.

## 4 Impact on Physics Analysis

The simulation results of IceVeto lead to estimates of the impact on diffuse and point source analysis with IceCube.

### 4.1 Diffuse Flux Measurements

Diffuse neutrino flux analyses without IceVeto have only been done for the Northern Hemisphere, where the neutrinos are the only particles capable of penetrating the earth. Here, a very pure neutrino sample was found, but it is dominated by the irreducible cosmic ray-induced atmospheric neutrinos. The fact that above a certain neutrino energy, and in the geometrical overlap region of IceCube and IceVeto, we expect to be able to significantly reduce backgrounds of the cosmic ray muons and the atmospheric neutrinos, will lead to very sensitive measurements of the diffuse astrophysical neutrino flux in the Southern Hemisphere.

Comparing an $E^{-2}$ astrophysical neutrino flux simulation, based on the most stringent IceCube limit, with cosmic ray simulations based on CORSIKA, we find that cosmic ray particles dominate the spectrum in IceCube up to a light deposit in IceCube of $5 \times 10^4 q_{tot}$. After applying the IceVeto array extension, the crossover, where the signal flux exceeds the background, is at $2.8 \times 10^3 q_{tot}$, an order of magnitude lower (Figure 7). The directly observable astrophysical neutrino flux increases from about 0.5/year to 2.6/year, by a factor of five. The background levels above the energies were the signal starts to dominate are 0.06 events/year in the case without veto and 0.01 events/year in the case with IceVeto. In the latter case, 2.6 events observed in one year with IceVeto would yield a discovery of astrophysical neutrinos with a significance of over 5$\sigma$ (without systematic uncertainties). Since the veto efficiency for atmospheric neutrinos is more efficient than the cosmic ray muon veto, the atmospheric neutrino spectrum will be highly suppressed and negligible. In addition, we expect IceVeto to perform even better above $4 \times 10^3 q_{tot}$ than Figure 7 shows as it was necessary to keep the veto efficiency constant at 99.999% above $4 \times 10^3 q_{tot}$ due to insufficient statistics in data and Monte Carlo simulation.

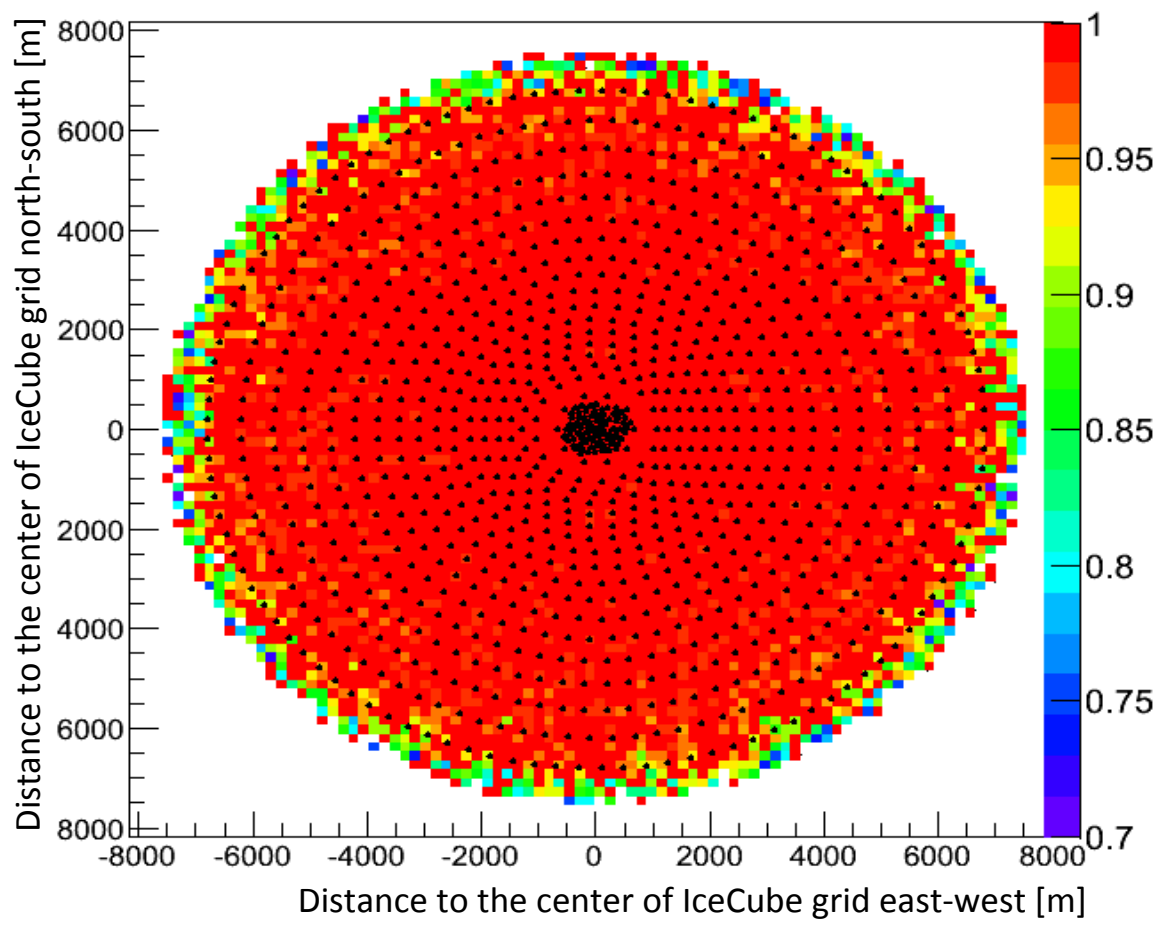

**Fig. 6**: Veto efficiency calculated with the simulation of events with 1000 $q_{tot}$ and inclinations from 0° -75°. The veto efficiency starts to thin out at the borders where no module ring is left. Each black dot indicates one station. IceTop densely populates the central part.

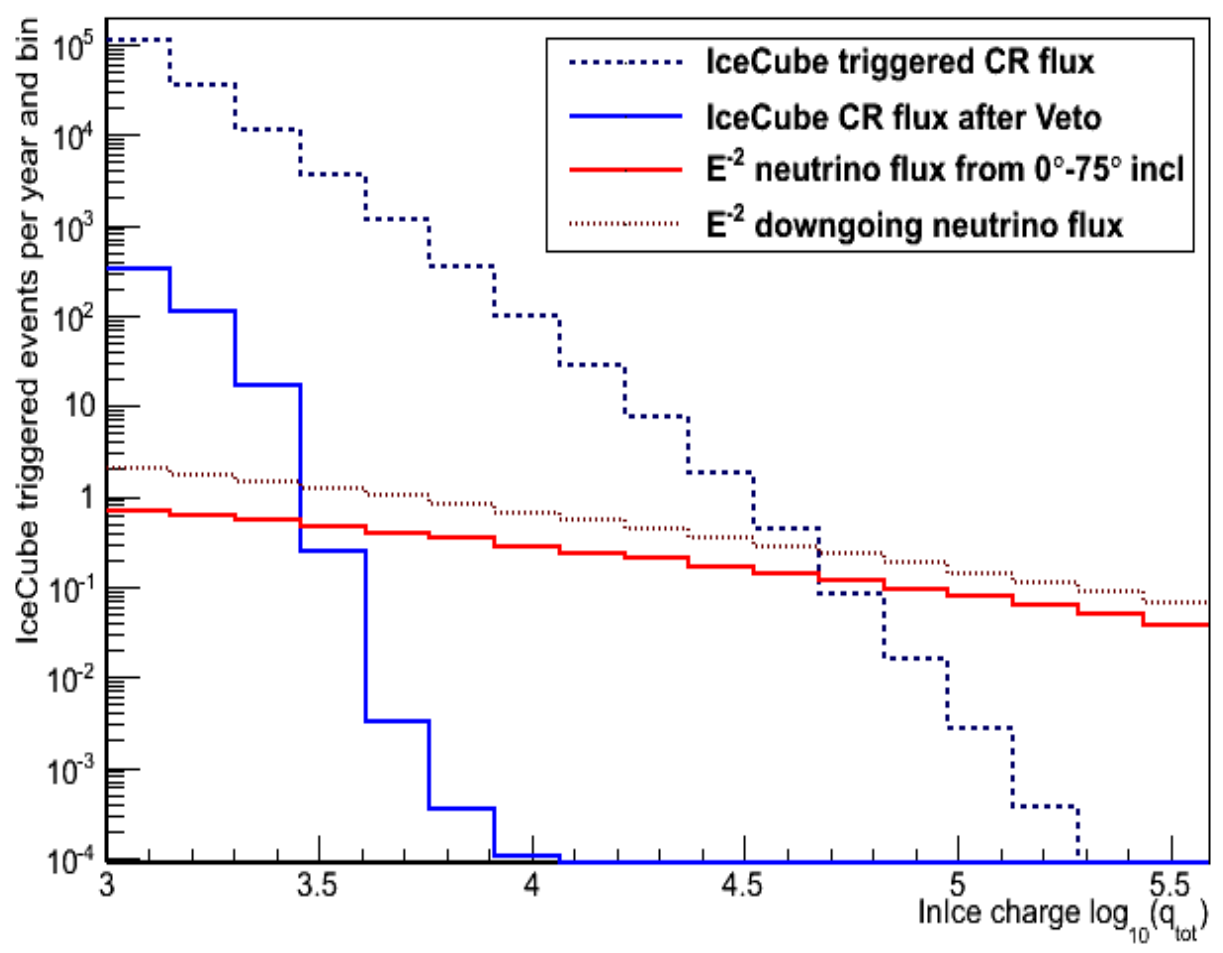

**Fig. 7**: The dashed, dropping line, shows the expected cosmic ray background in the ice, derived from simulation. The solid, faster dropping, line is the CR flux after applying the IceVeto array with $0° - 75°$ inclination. The dashed, flat line, is an astrophysical E$^{-2}$ neutrino flux ($E^2 d/\Phi/dE = 1.4 \times 10^{-8}$ GeV cm$^{-2}$s$^{-1}$sr$^{-1}$) based on the current diffuse flux limit [6] from $0° - 90°$ inclination. The solid, flat line, is the same neutrino flux from $0° - 75°$ inclination.

### 4.2 Neutrino Point Source Measurements

For point source searches with a low number of signal events it is of great importance to achieve as low background contribution as possible.

With increasing veto efficiency, the background decreases linearly when keeping the same amount of signal for IceVeto. As IceVeto provides a much larger lever arm for reconstruction, it will even help with upgoing events (northern sky) that are originally misreconstructed as downgoing. As these events are one of the main contaminations at the high-energy end of current point source analyses of the northern sky (upgoing), this will likely increase the sensitivity even here.

IceVeto focuses on excavating throughgoing astrophysical neutrino-induced muons. Since $\nu_\mu$ induced muon tracks in particular have a well-reconstructed direction, the detection of astrophysical muon-neutrinos at the 5$\sigma$ level would also imply observation or constraints on point sources in the sky. With the IceVeto extension, rejection of atmospheric neutrino background for $\nu_\mu$ induced muons can be greatly enhanced, so we could be able to look for extragalactic point sources in the southern sky with IceCube to a much higher precision than currently.

Probably most importantly we would increase the detectable neutrino flux from the galactic center. Figure 1 shows the spatial limitation of the IceTop veto (center region) and the region we gain including the veto extension.

## 5 Conclusions

IceVeto is a powerful way to extend IceCube for sub-PeV astrophysical neutrino induced muon detection by vetoing background from CRs with using less than 1000 modules similar to IceTop. Atmospheric neutrinos and muon bundles will be vetoed by the associated air shower particles. IceVeto will possibly make sub PeV $\nu_\mu$ induced muon detection at the hight of the galactic center possible with IceCube being able to detect 5.5 time more astrophysical $\nu_\mu$ induced muons (see Figure 2) and will be able to find 3.5 times more high energy astrophysical $\nu_\mu$ at the overall southern sky. They are an important key to find the sources of CR acceleration. The total gain in the flux of *all three* neutrino flavors needs further investigation but will overall be above a factor of 2 compared to the starting-event analysis for the Southern Hemisphere.

## References

[1] A. Achterberg et al., Astropart. Phys., 26 (2006) 155-173, doi: 10.1016/j.astropartphys.2006.06.007.
[2] R. Abbasi et al., Nucl. Inst. Meth. A, 700 (2013) 188-220, doi: 10.1016/j.astropartphys.2012.11.003.
[3] IceCube Coll., paper 0373 these proceedings.
[4] M. G. Aartsen et al., http://arxiv.org/abs/1304.5356 (2013).
[5] IceCube Coll., paper 0650 these proceedings contribution.
[6] A. Schukraft et al., to be published, http://arxiv.org/abs/1302.0127.

ICRC 2013

# A study of the neutrino mass hierarchy with PINGU using an oscillation parameter fit

THE ICECUBE COLLABORATION[1],

[1]*See special section in these proceedings*

*Andreas.Gross@tum.de*

**Abstract:** The determination of the neutrino mass hierarchy is among the most fundamental questions in particle physics. The recent measurement of a large mixing angle between the first and the third neutrino mass eigenstate and the first observation of atmospheric neutrino oscillations at tens of GeV with neutrino telescopes opens the intriguing new possibility to exploit matter effects in neutrino oscillations for its determination. A further extension of IceCube/DeepCore called PINGU (Precision IceCube Next Generation Upgrade) has been recently envisioned with the ultimate goal to measure this mass hierarchy. PINGU would consist of additional IceCube-like strings of optical sensors deployed in the deepest clearest ice in the center of IceCube. More densely deployed instrumentation would provide a threshold substantially below 10 GeV and enhance the sensitivity to the mass hierarchy signal in atmospheric neutrinos. Here we discuss an estimate of the PINGU sensitivity to the mass hierarchy using an approximation with an Asimov dataset and an oscillation parameter fit.

**Corresponding authors:**
Andreas Groß[1],

[1]*T.U. Munich, D-85748 Garching, Germany*

## 1 Introduction

The neutrino field has progressed through a remarkable evolution since the first evidence of oscillations in 1998 [1]. In the past 15 years nearly all mixing parameters have been measured, with the $\theta_{13}$ angle most recently. What remains are two challenging key elements: the neutrino mass hierarchy (see section 2) and the CP-phase. Given a large value for $\theta_{13}$, attention has been drawn to the potential of using the large natural flux of atmospheric neutrinos to resolve the ordering of neutrino masses, whether the hierarchy is normal or inverted [8, 9].

Neutrino telescopes are relatively new participants in the measurement of neutrino oscillation parameters [2, 3, 4, 5]. The enormous volume of these instruments make them ideal to exploit the statistics from atmospheric neutrinos. IceCube is currently the largest active neutrino telescope instrumenting a cubic-kilometer of ice at the geographic South Pole between depths of 1450 m and 2450 m with optical sensors deployed along vertical strings [7]. Neutrino identification with IceCube relies on the optical detection of Cherenkov radiation emitted by secondary particles produced in neutrino interactions in the surrounding ice or the nearby bedrock. An extension to IceCube, the DeepCore subarray, includes 8 densely instrumented strings optimized for low energies operating in conjunction with the 12 adjacent standard IceCube strings. Depending on the details of the analysis strategy, the energy threshold of DeepCore ranges from slightly below 10 GeV to 20 GeV.

With the first neutrino oscillation parameter measurements from DeepCore and ANTARES [2, 3, 4, 5], the "low-energy" particle physics potential of neutrino telescopes became evident. Estimates of the full potential of DeepCore, extrapolated from its current performance, show the detector may become competitive with current leading experiments in the field [6]. It is also evident that the DeepCore energy threshold near 10 GeV is not constrained by the detector medium. A further infill of the deep clear ice within the DeepCore volume, called the Precision IceCube Next Generation Upgrade (PINGU), would result in a further reduction of the energy threshold to a few GeV, thereby covering a region with high sensitivity for the measurement of the mass hierarchy with atmospheric neutrinos [9]. The herein proposed detector is currently being optimized to provide a first definitive measurement of this key quantity within the coming decade. The deployment of PINGU to the Antarctic ice is expected to be possible in two subsequent Antarctic summer seasons assuming 20 strings (see section 3), larger configurations would require (at least) 3 seasons.

## 2 Physics of Neutrino Mass Hierarchy

Neutrino oscillations occur because the neutrino mass eigenstates $\nu_1, \nu_2$, and $\nu_3$ differ from the neutrino flavor eigenstates $\nu_e, \nu_\mu$, and $\nu_\tau$. They can be parametrized by two mass differences $\delta m^2_{\text{solar}} = m^2_2 - m^2_1$ and $\Delta m^2_{\text{atm}} = m^2_3 - 1/2(m^2_2 + m^2_1)$, 3 mixing angles $\theta_{12}, \theta_{23}$ and $\theta_{13}$ and a CP violating phase $\delta$. While the mixing angles have been measured with good precision and the sign of $\delta m^2_{solar}$ has been measured through matter effects in the neutrino oscillations in the sun [13], the sign of $\Delta m^2_{\text{atm}}$ is unknown and $\delta$ is unbounded between 0 and $2\pi$.

The measurement of the neutrino mass hierarchy with PINGU, i.e. the determination of the sign of $\Delta m^2_{\text{atm}}$, relies on the modification of vacuum oscillations due to interactions with matter in the Earth described by the MSW effect [10, 11]. The size of these matter effects depends on the realized mass hierarchy, where the most relevant channel for atmospheric neutrinos in PINGU is muon neutrino disappearance.

## 3 PINGU simulation and reconstruction

The design of PINGU takes advantage of the elements learned from the construction and operation of IceCube and DeepCore. This includes the simulation of PINGU events which is based on the IceCube simulation software. Neutrino events are generated using the GENIE simulation package [14]. Particles generated at the vertex of charged current and neutral current interactions, and the subsequent hadronic and electromagnetic showers, are propagated by a full GEANT4 simulation [15, 16]. Each of the resultant Cherenkov photons from any charged secondary particles are then directly propagated within the detector volume. The detector geometry that will be selected for PINGU is that which achieves the optimal sensitivity to the mass hierarchy measurement.

Geometries under consideration, and subsequently simulated as outlined above, include an additional 20 – 40 detector strings, each holding 60 to 120 light detectors called Digital Optical Modules (DOMs), deployed within the DeepCore volume (see Fig. 1 for examples). The effective neutrino volume obtained from the simulation of PINGU events for the geometry shown in Fig. 1 (top) is shown in Fig. 2, demonstrating that a detector of several megatons is potentially achievable for low energy neutrinos of a few GeV. We also include detailed numerical simulations to calculate the oscillation probability of each atmospheric neutrino traversing the Earth, taking into account 3 flavor effects, where the matter density profile is parametrized according to the Preliminary Reference Earth Model (PREM) [12].

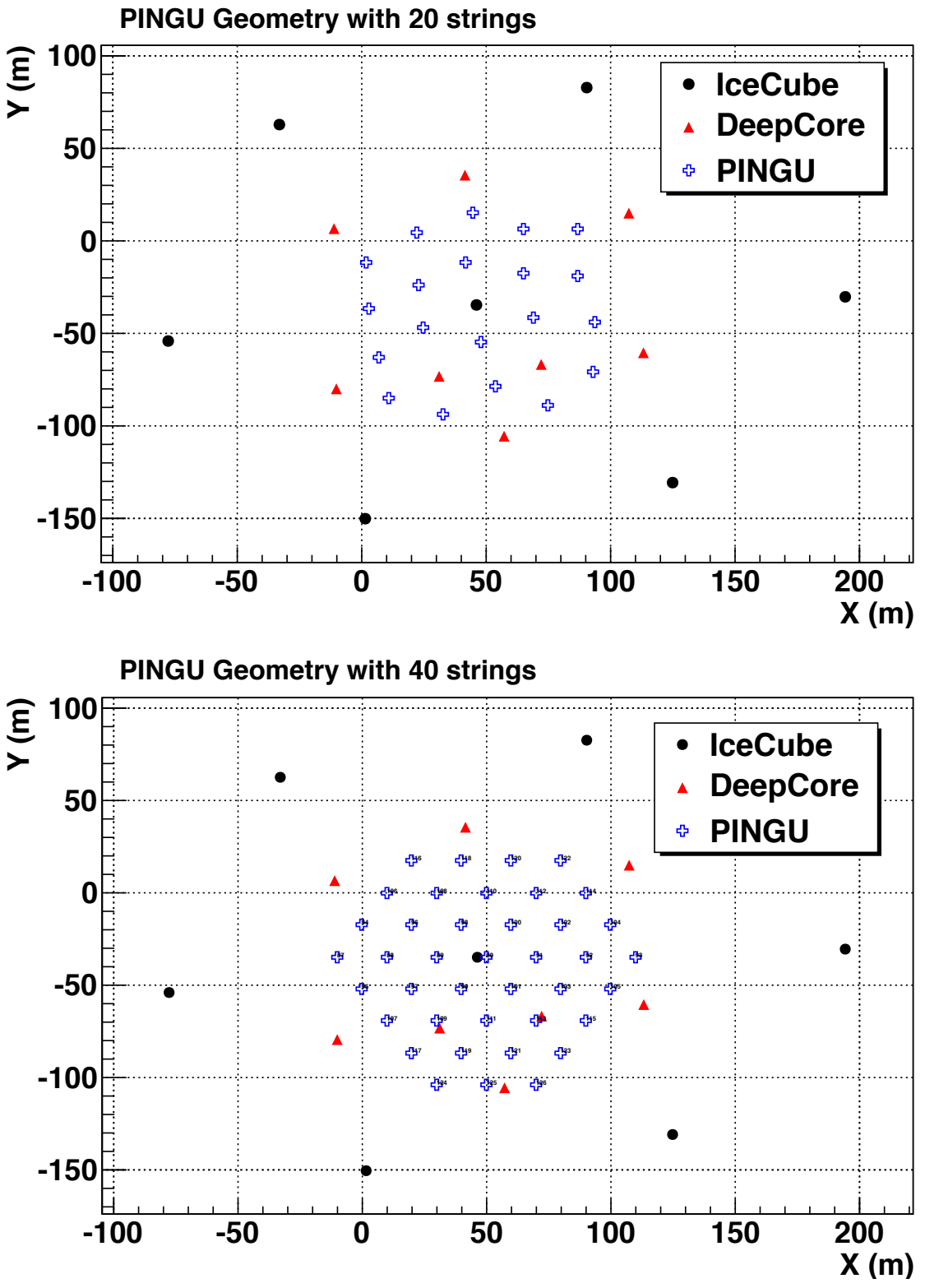

**Figure 1**: Overhead view of two geometry layouts currently under study for PINGU with 20 and 40 additional detector strings, respectively.

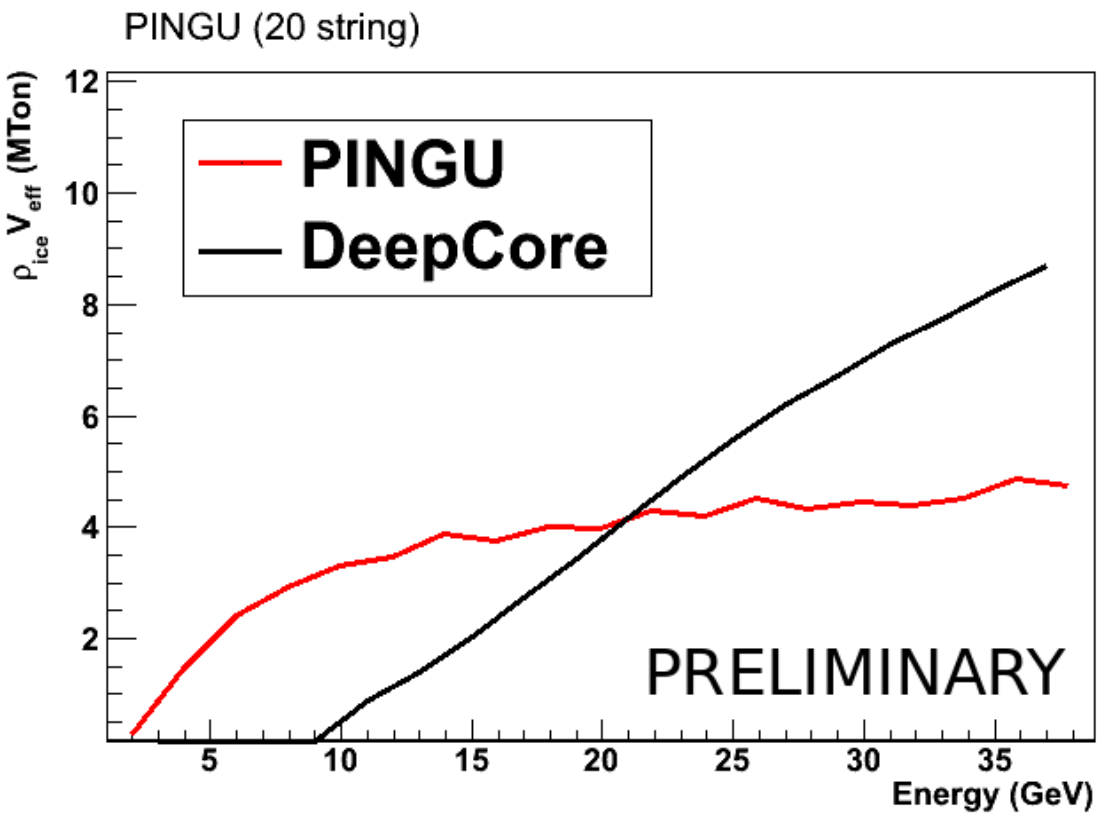

**Figure 2**: Effective volume for PINGU in the baseline configuration (20 strings) in comparison with DeepCore. For rejection of the atmospheric muon background, a fiducial volume defined by a cylinder with radius of 75 m and a height of 332 m is assumed. A threshold a threshold of 20 DOMs hit per event was assumed to take into account reconstruction inefficiencies.

The PINGU simulated events are processed through a series of IceCube-DeepCore tools that include quality measures for event selection and reconstruction. IceCube has produced algorithms specifically designed to reconstruct lower energy events (< 100 GeV) in the ice. Some are relatively fast independent reconstructions of energy (Monopod) and muon direction (SANTA), and one is a more sophisticated reconstruction that simultaneously reconstructs an 8-dimensional likelihood (HybridReco) which describes a cascade at the interaction vertex and the outgoing muon (if present). SANTA [5] is an algorithm originally developed by the ANTARES collaboration [17] that identifies the hits from the muon which have not been delayed due to scattering. Monopod reconstructs the maximum likelihood value of the energy from the light emitted by the hadronic and electromagnetic cascade produced at the neutrino interaction vertex. HybridReco performs a simultaneous fit for the 8 parameters of interest for an event ($x, y, z, t, E$ of the cascade, muon track length, zenith and azimuth) and therefore involves a more sophisticated minimization. The preliminary performance of the Monopod and SANTA algorithms are shown in Figs. 3 and 4 for the 40 string PINGU geometry simulated events.

## 4 PINGU and the neutrino mass hierarchy

In order to ensure that statistical statements are robust, several approaches for determining the mass hierarchy with PINGU are pursued. Here, we elaborate on a study that uses a fit of neutrino oscillation parameters treating $\Delta m^2_{\rm atm}$ as a signed parameter. The aim of the analysis is to completely constrain the allowed parameter space to $\Delta m^2_{\rm atm} > 0$ (normal hierarchy) or $\Delta m^2 < 0$ (inverted hierarchy) and in this way to rule out the other hierarchy. Due to the parameter scan, the physics result is independent of the conventions used for the definition of the oscillation parameters.

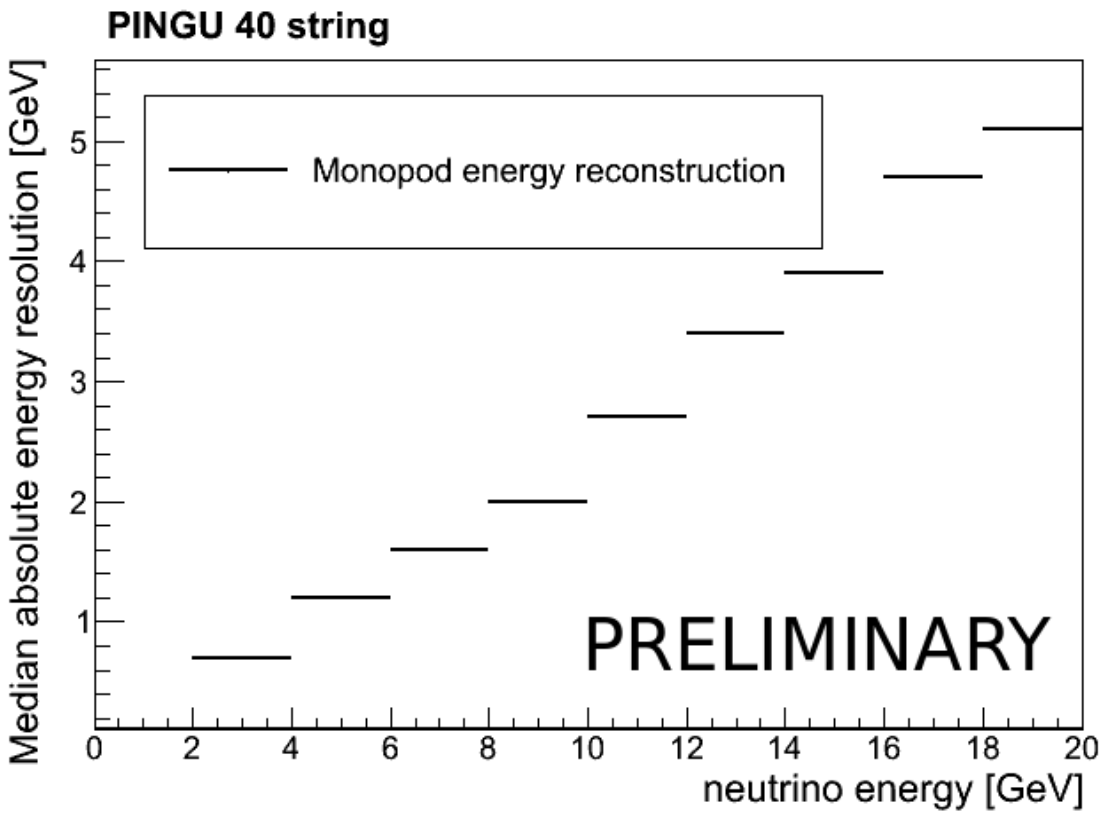

**Figure 3**: Median neutrino energy resolution as a function of true energy for the 40 string configuration.

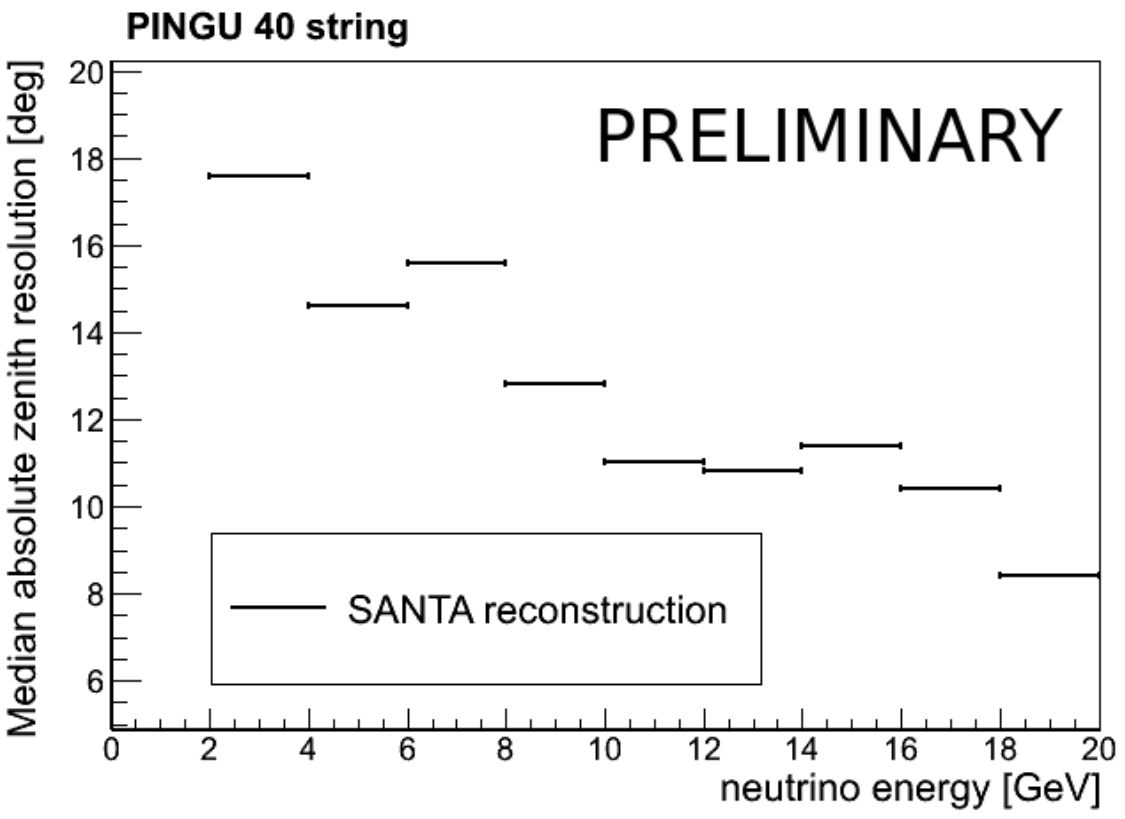

**Figure 4**: Median neutrino zenith resolution as a function of true energy.

The approach presented here uses a binned analysis in $\cos(zenith)$ and $\log_{10}(energy)$ as variables with a $\chi^2$ statistic using pulls for systematic uncertainties, a method used and described in [2]. The $\chi^2$ between prediction and (pseudo-) data is calculated as a function of the considered oscillation parameters. The test statistics for the purpose of hierarchy measurement is defined by

$$\Delta\chi^2 = \min\{\chi^2|\Delta m^2 > 0\} - \min\{\chi^2|\Delta m^2 < 0\}.$$

If $\Delta\chi^2 > 0$, the inverted hierarchy is favored, while for $\Delta\chi^2 < 0$ the data favors the normal hierarchy. We evaluated the 40 string PINGU configuration with 60 DOMs on each string (see Fig.1 (bottom) ) using the SANTA and Monopod reconstruction algorithms applied to all events. Hence, the detector resolution, including the tails of the distributions, is fully taken into account. Based on the reconstruction performance studies, the application of HybridReco is expected to improve the sensitivity towards neutrino mass hierarchy.

An approximation for the median sensitivity of the detector is provided by the analysis of a representative dataset, also known as the Asimov data set [18]. This pseudo-data set is defined by the average expected number of events obtained from Monte Carlo simulations to which the analysis is applied. The analysis of an Asimov data set is shown, for example, in Fig. 5, where the true oscillation parameters were set to $\Delta m^2_{\rm atm} = -2.4\cdot 10^{-3}$ eV$^2$ (inverted hierarchy), $\sin^2(\theta_{13}) = 0.024$ and $\sin^2(\theta_{23}) = 0.35$. The plot shows the distribution of the $\chi^2$ of this Asimov data set as a function of the oscillation parameters. On the left side of Fig. 5, negative values for $\Delta m^2_{\rm atm}$ were considered (inverted hierarchy), while on the right positive values were assumed (normal hierarchy). The value of 12.1 for the test statistics $\Delta\chi^2$ for this Asimov data set results from the difference of the minimum $\chi^2$ of these sub-figures. The denser geometry with 40 detector strings (see Fig. 1) was used, and the assumed livetime is one year. Backgrounds from downwards going air shower muons and from $\nu_e$ and $\nu_\tau$ events are not taken into account. The appearance of $\nu_\mu$ due to $\nu_e \to \nu_\mu$ is included.

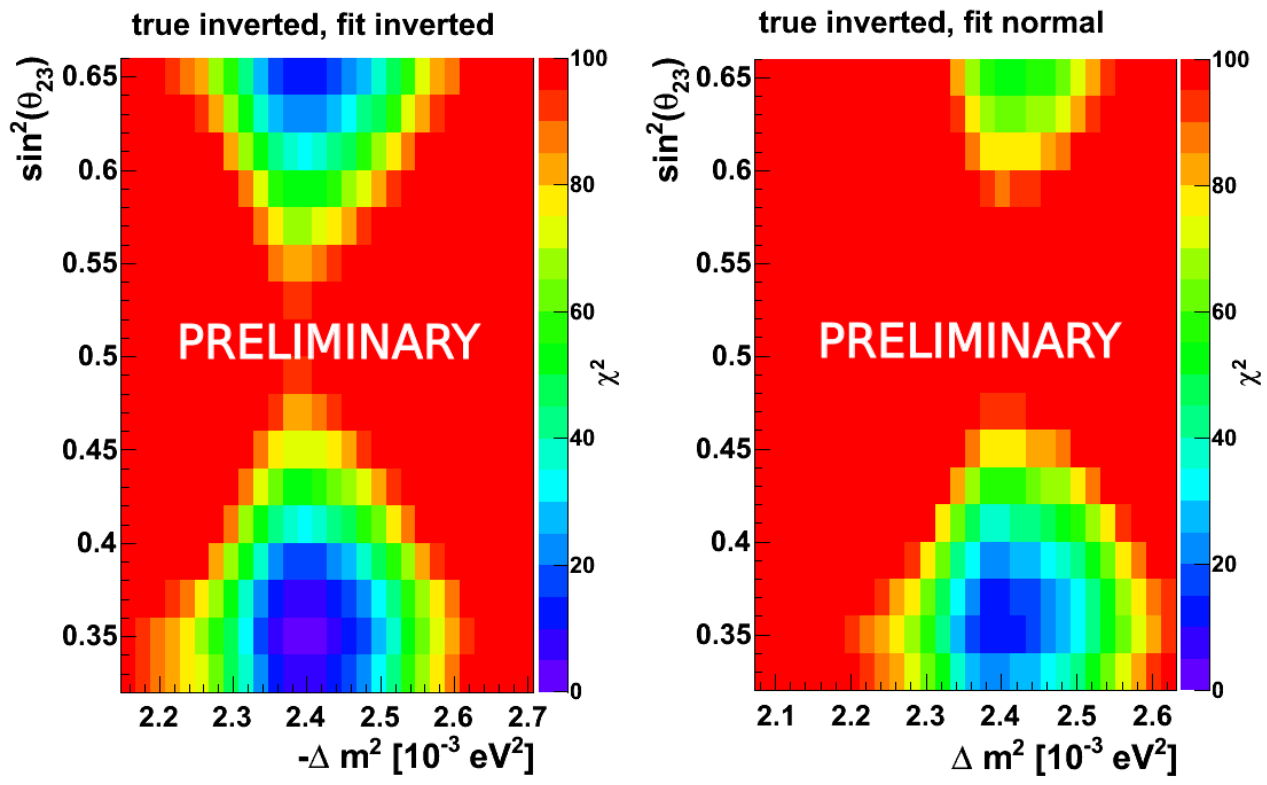

**Figure 5**: Example for the analysis of an Asimov data set: $\chi^2$ as a function of $\Delta m^2_{\rm atm}$ and $\sin^2(\theta_{23})$. The true oscillation parameters were set to $\Delta m^2_{\rm atm} = -2.4\cdot 10^{-3}$ eV$^2$ (inverted hierarchy), $\sin^2(\theta_{13}) = 0.024$ and $\sin^2(\theta_{23}) = 0.35$.

We have tested the interpretation of the $\Delta\chi^2$ obtained in the Asimov approach as the median significance via a $\chi^2$ distribution with 1 degree of freedom with a full ensemble simulation with a large number of pseudo-experiments. Here the event numbers in each bin were modified according to Poisson fluctuations around their expectation values. In cases where $\Delta\chi^2 > 0$, i.e. the inverted hierarchy is favored, the p-value p($\Delta\chi^2$) for rejecting a single parameter point (mass splitting, mixing angle) with the normal hierarchy is defined as the fraction of pseudo-experiments which favor normal hierarchy by more than $\Delta\chi^2$. The p-value for the rejection of the normal hierarchy hypothesis is then defined as the maximum p-value obtained for any true parameter point with the normal hierarchy. In the case where the normal hierarchy is favored, the p-value for the inverted hierarchy rejection is defined in an analogous way. The p-value as a function of $\Delta\chi^2$ obtained in this way is shown in Fig. 6 in comparison with the expectation from the Asimov approach given by a $\chi^2$ distribution. Pseudo-experiments were generated for 7 different assumptions of the true oscillation parameters. Fig. 6 shows the maximum p-value of these, corresponding to the above defined p-value for the rejection of the hierarchy hypothesis. We find the $\chi^2$ distribution (assumed for the Asimov approach) to deliver a higher p-value over most of the $\Delta\chi^2$ range, indicating that the Asimov approach provides a conservative estimate of the median sensitivity in most cases.

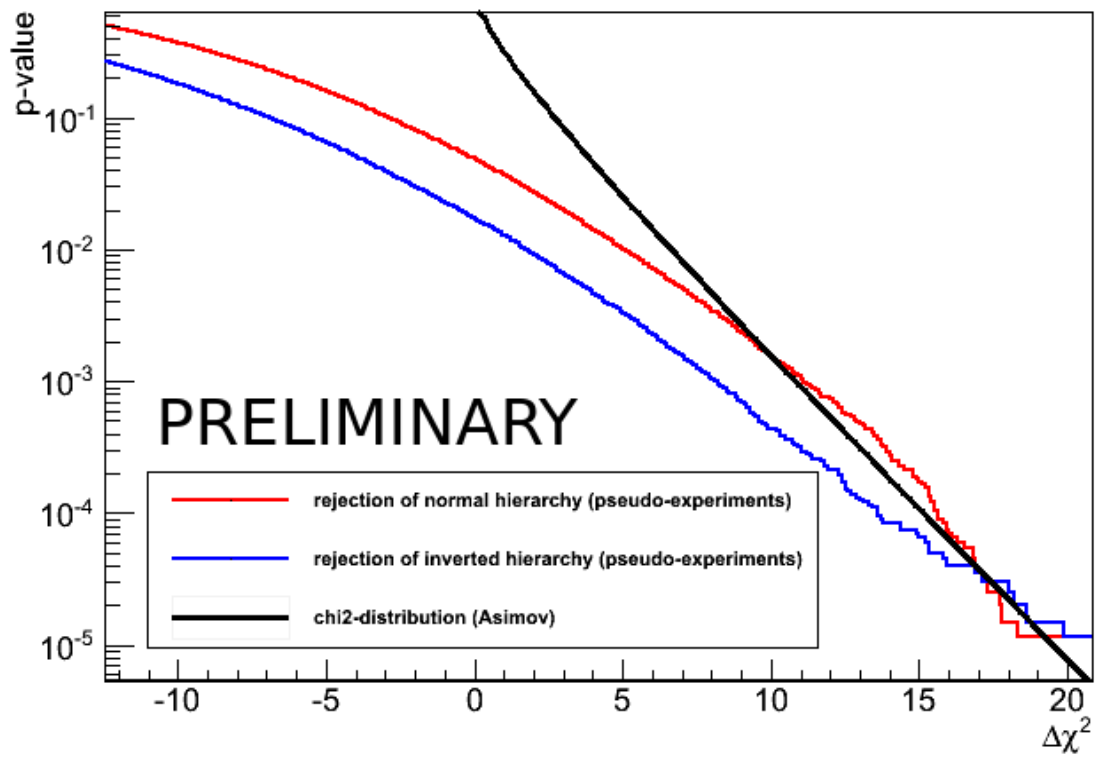

**Figure 6**: The p-value as a function of $\Delta\chi^2$ from pseudo-experiments for normal hierarchy (red) and inverted hierarchy (blue) in comparison to the cumulative $\chi^2$ distribution with 1 degree of freedom (black). Pseudo-experiments were performed for 7 different values of the true oscillation parameters, the curves show the highest p-value obtained for any of these.

The results of the different approaches to calculate PINGU's sensitivity are summarized in Fig. 7, see also [20]. The upper curve represents the results of the approach discussed here, the lower curve represents the result from a second approach based on the use of likelihood ratios. Similar to [19], patterns are defined as the expectation of the number of events in each bin of cos(zenith) and $\log_{10}$(energy) for normal and inverted hierarchy and the likelihood of the (pseudo-) data is calculated for these patterns. Various patterns are considered corresponding to different assumptions of the true $\Delta m^2_{\rm atm}$. The likelihood ratio between the normal and inverted hierarchy patterns is used as the test statistics. The distribution of the likelihood ratio is obtained from pseudo-experiments. The expected significance according to this study is smaller, since a 20-string configuration and a smaller signal efficiency were asumed here.

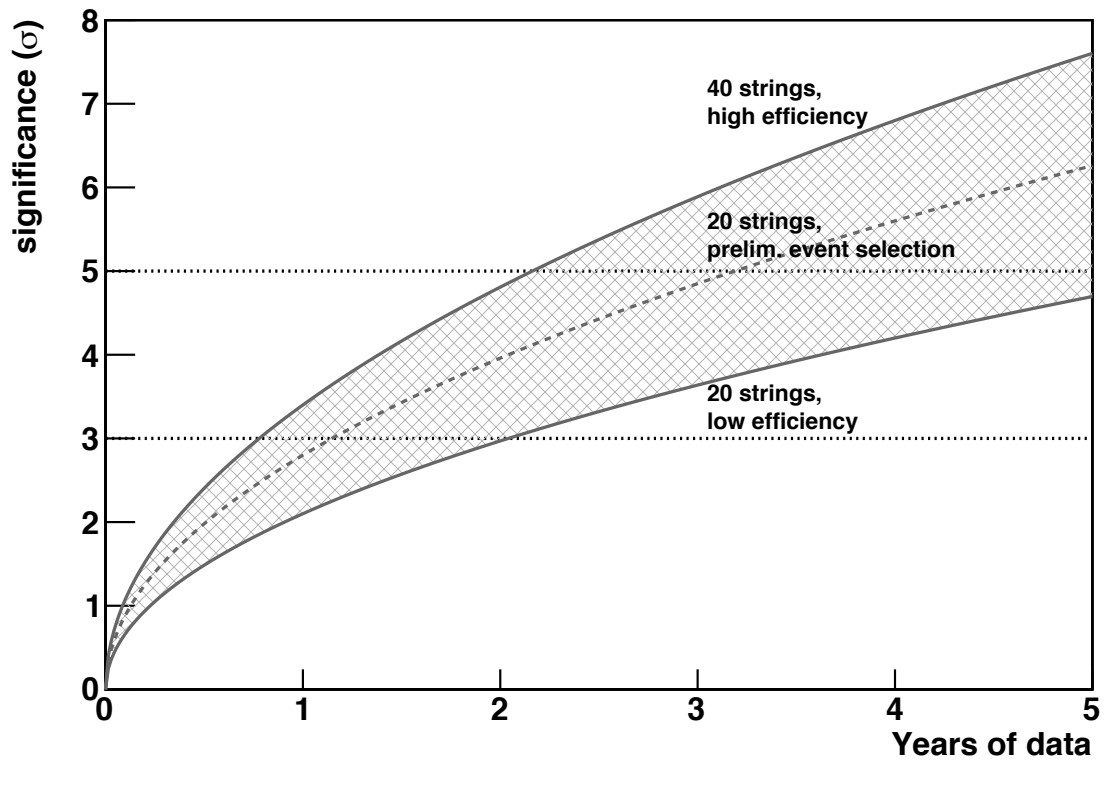

**Figure 7**: Expected significance for the neutrino mass hierarchy by PINGU as a function of time [20].

## 5 Further physics topics for PINGU

Beyond the measurement of the neutrino mass hierarchy, PINGU has a rich physics potential. In the field of neutrino flavor physics, this includes precision measurements of mixing parameters which could aid in solving the question of whether $\theta_{23}$ is non-maximal, i.e. $\theta_{23} \neq 45°$, and if so what is the octant. PINGU's potential, however, is not restricted to neutrino flavor physics. The Earth core composition influences neutrino oscillation patterns for neutrinos traversing the Earth. Utilizing the arguments made in the oscillation analysis, PINGU may provide constraints on the composition of the Earth core which would be relevant for the geophysics community. Further, this new detector array would extend IceCube's and DeepCore's dark matter searches to WIMP masses below 20 GeV with significantly improved sensitivity; a region where hints of a possible dark matter signal have been reported by DAMA, CoGeNT, CRESST and most recently by the CDMS-Si data. Finally, PINGU also has improved sensitivity in the detection of low-energy supernova neutrinos. While not sufficient to detect supernovae from distant galaxies, it will improve the measurement of any supernova in our galaxy, provide some informaton on the neutrino energy spectrum and track the time-dependence of the average neutrino energy [21].

## References

[1] J. Beringer et al., Phys. Rev. D86 (2012) 010001
[2] M. Aartsen et al., accepted by Phys. Rev. Lett., arxiv.org/1305.3909.
[3] S. Adrian-Martinez et al., Phys. Lett. B714 (2012) 224-230, doi:10.1016/j.physletb.2012.07.002.
[4] IceCube collaboration, paper 0848 these proceedings.
[5] IceCube collaboration, paper 0450 these proceedings.
[6] IceCube collaboration, paper 0460 these proceedings.
[7] A. Achterberg et al., Astropart. Phys. 26, 155 (2006).
[8] O. Mena, I. Mocioiu and S. Razzaque, Phys. Rev. D78 (2008) 093003.
[9] E. Kh. Akhmedov, S. Razzaque and A. Yu. Smirnov, JHEP 02 (2013) 082, doi:10.1007/JHEP02(2013)
[10] L. Wolfenstein, Phys. Rev. D7 (1978) 2369, doi:10.1103/PhysRevD.17.2369
[11] S. P. Mikheev and A. Yu. Smirnov, Sov. J. Nucl. Phys. 4 (1985) 913
[12] A. M. Dziewonski and D. L. Anderson, Physics of the Earth and Planetary Interiors 25 (1981) 297, doi:10.1016/0031-9201(81)90046.
[13] Q. R. Ahmad et al, Phys. Rev. Lett. 89 (2002) 011302
[14] C. Andreopoulos et al., Nucl. Instrum. Meth. A614 (2010) 87
[15] S. Agostinelli et al., Nucl. Instrum. Meth. A506 (2003) 250
[16] J. Allison et al., IEEE Transactions on Nuclear Science 53 (2006) 270, doi:10.1109/TNS.2006.869826
[17] J. A. Aguilar et al., Astropart. Phys. 34 (2011) 652, doi:10.1016/j.astropartphys.2011.01.003.
[18] G. Cowan et al., Eur. Phys. J. C 71 (2011) 1554.
[19] D. Franco et al., JHEP 1304 (2013) 008
[20] M. Aartsen et al., arXiv:1306.5846 [astro-ph.IM]
[21] M. Salathe, M. Ribordy and L. Demirrs, Astropart. Phys. 35 (2012) 485

# Evidence of optical anisotropy of the South Pole ice

THE ICECUBE COLLABORATION[1]

[1] *See special section in these proceedings*

*dima@icecube.wisc.edu*

**Abstract:** In our continued investigations of the optical properties of the South Pole ice, the IceCube collaboration has discovered evidence of a slight azimuthal dependence of the light propagation properties, which can be attributed to an apparently smaller amount of scattering in one direction. We developed a phenomenological model of such anisotropic scattering and fitted it to in-situ light source data. The model that includes the anisotropic scattering significantly improves the description of the calibration data when compared to a model without anisotropy. We have also observed evidence of the anisotropy in the normal muon data.

**Corresponding authors:** Dmitry Chirkin[1]
[1] *Dept. of Physics and WIPAC, University of Wisconsin, Madison, WI 53706, USA*

## 1 Introduction

IceCube is a cubic-kilometer neutrino detector installed in the ice at the geographic South Pole [1] instrumenting depths between 1450 m and 2450 m. Detector construction started in 2005 and finished in 2010. Neutrino reconstruction relies on the optical detection of Cherenkov radiation emitted by secondary particles produced in neutrino interactions in the surrounding ice or the nearby bedrock.

The optical properties of ice surrounding the detector are described with a table of absorption and effective scattering coefficients describing average ice properties in 10 m-thick ice layers. These properties were determined with a dedicated calibration measurement as described in [2]. The data for this measurement were collected in 2008 with the 40 string detector configuration shown in figure 1. Every optical sensor (digital optical module, or DOM) on string 63 was operated in "flasher" mode to emit light from on-board LEDs in an approximately azimuthally-symmetric pattern, which was observed by the DOMs on the surrounding strings.

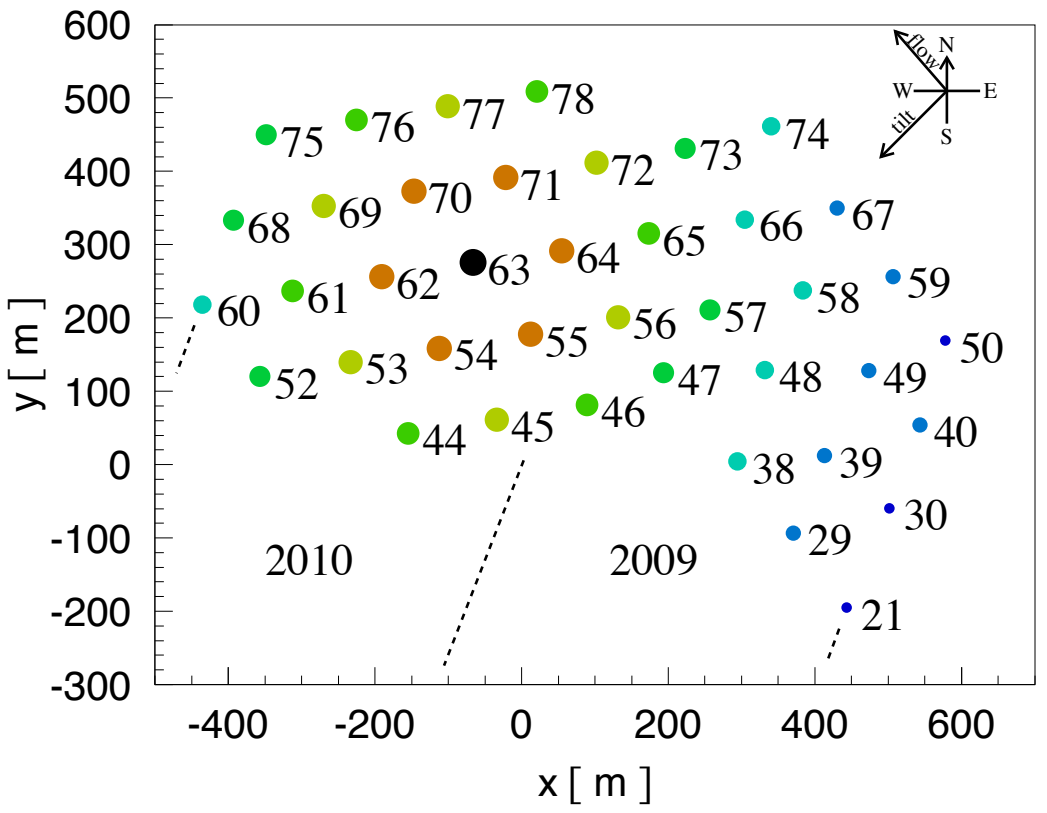

**Figure 1**: Top view layout of IceCube in the 40-string configuration in 2008. String 63, for which the DOMs emitted flashing light in the study presented here, is shown in black. The nearest 6 strings are shown in brown. The dashed lines and numbers 2009 and 2010 in the left figure indicate the approximate location of the detector parts deployed during those years.

## 2 Anisotropy of South Pole Ice

Shortly after the study of [2] was complete, we noticed a consistent azimuthal asymmetry in charge collected on the strings surrounding the flashing string that depends on the direction and distance to receiving string. Figures 2, 3, and 4 demonstrate the observed effect: more light is observed in the direction of strings 70 and 55 than on average over all directions, by on average about 16% per 100 m of distance from the emitting string 63.

It appears that the in-situ light source data collected by IceCube contains evidence of ice anisotropy, i.e., different photon propagation properties in different directions of the *xy* plane. It additionally appears that these properties are, to a large extent, the same in the directions $\vec{n}$ and $-\vec{n}$ for any $\vec{n}$ in the *xy* plane. This observation is important as it precludes a possibility that the location of the hole ice (ice re-frozen after the string deployment or otherwise impacted by the deployment) or the supporting cable with respect to the DOMs can create the effect present in data. It is highly unlikely that any effect from the hole ice or the cable would have a consistent directional behavior for all the DOMs on the emitting string and for all receiving strings. Therefore, we must ascribe the observed effect, to at least some extent, to the inherent properties of the surrounding ice.

Although one can calculate the scattering and absorption properties of individual dust particles, whatever their shape, the positions and orientations of all dust particles at any depth in the volume of the detector are unknown. Perhaps the observed effect is caused by the preferential alignment of the ice crystals, possibly resulting in the preferential alignment of the embedded dust particles. The microscopic cause of the observed effect being unknown we nevertheless note that it should be possible to specify the anisotropic properties of ice in some useful macroscopic

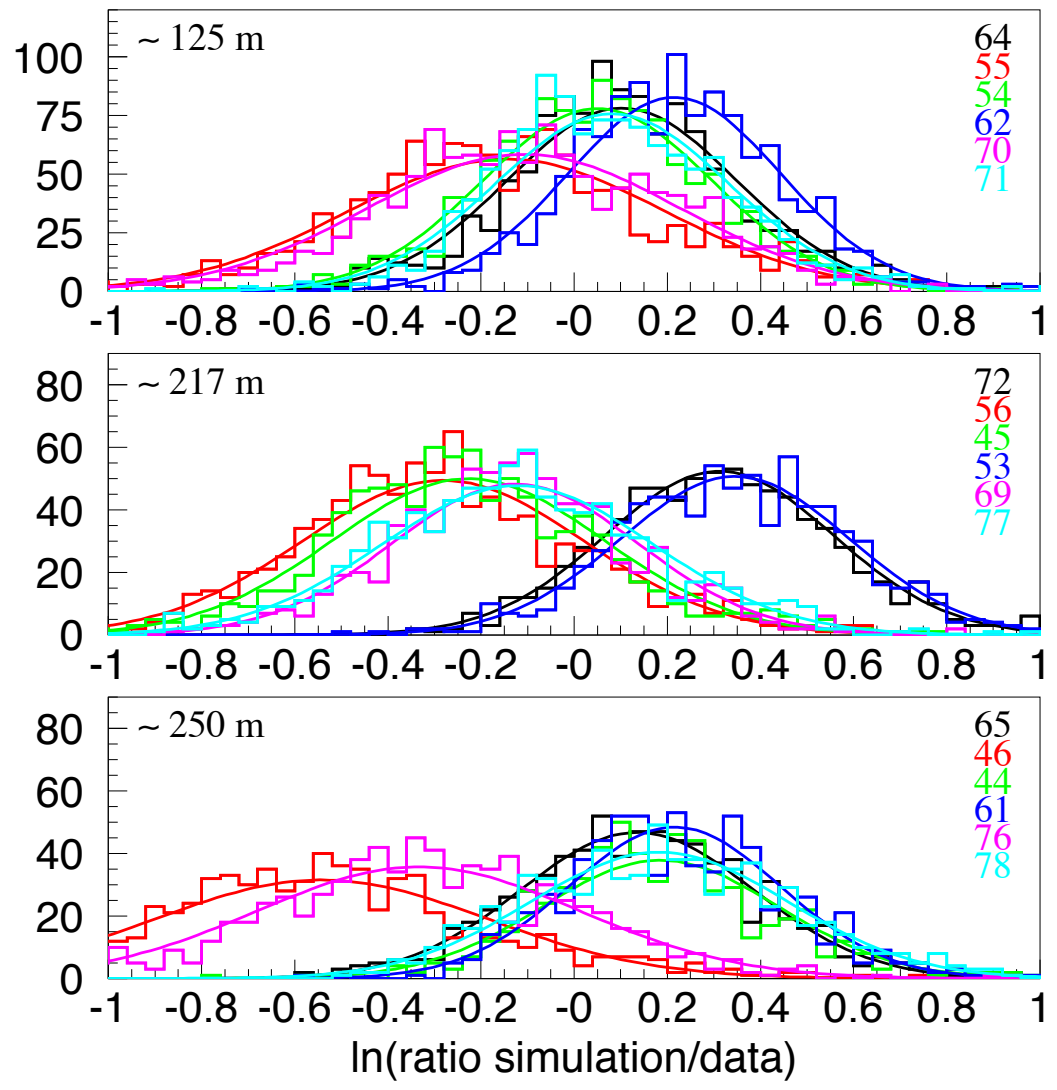

**Figure 2**: Ratio of simulation to data of total charge collected in DOMs on strings surrounding string 63: nearby strings 64, 55, 54, 62, 70, and 71 ($\sim$ 125 m away), next-to-near strings 72, 56, 45, 53, 69, and 77 ($\sim$ 217 m away), and even further ring of strings 65, 46, 44, 61, 76, and 78 ($\sim$ 250 m away). Each histogram contains entries for all emitters on string 63 and receivers on the denoted string such that the total received charge is greater than 10 photoelectrons. The ice model used in simulation is that of [2] and lacks anisotropy.

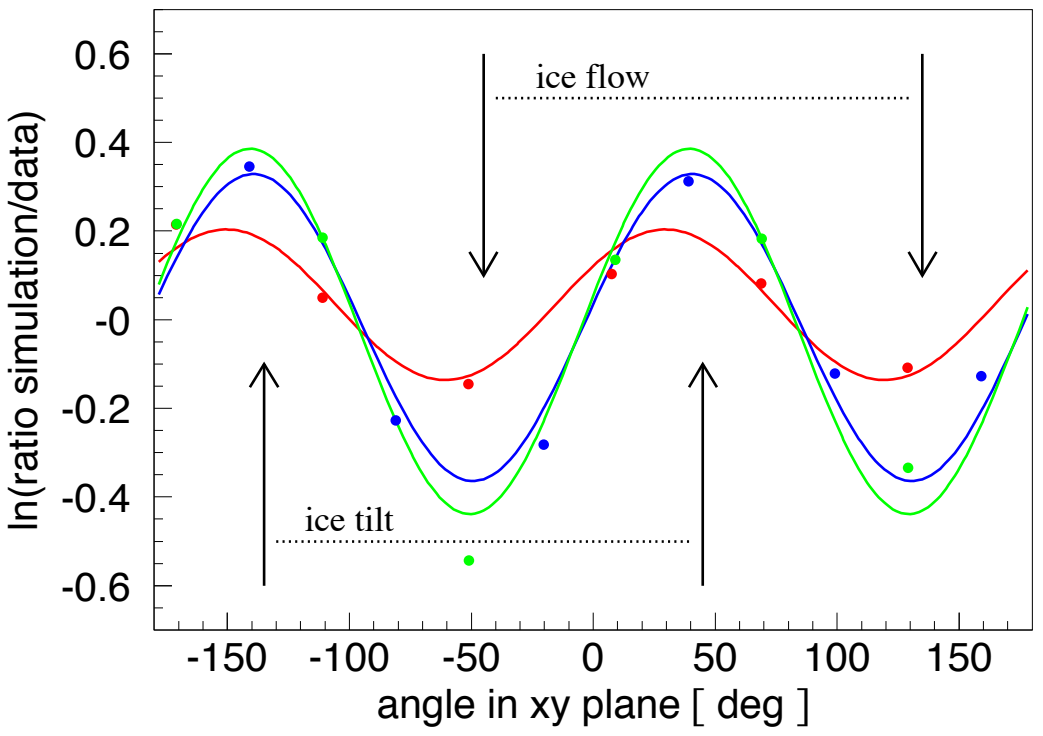

**Figure 3**: The maxima of histograms in figure 2 is plotted vs. azimuth of the direction from the emitting string 63 to the corresponding receiving strings. The points approximately fall on the fitted sines of twice the azimuth angle with phase that is approximately the same for all 3 curves and amplitude that increases with distance to the corresponding strings ($\sim$ 125 m, 217 m, or 250 m). Also shown are the gradient direction of the ice tilt (see [2]) and direction in which the ice moves at the South Pole at a rate of about 10 m/year (ice flow).

way.

One simple approach is to specify that the scattering coefficient depends on the photon direction in the $xy$ plane as $b_e(\vec{n}) = b_e \cdot \varsigma(\phi)$, where $\phi$ is the azimuth angle of the photon direction $\vec{n}$. We, however, note, that this alone will not lead to a consistent description of the observed ice properties as the following relationship on the scattering cross section is not satisfied: $\sigma(\vec{n}_{in}, \vec{n}_{out}) = \sigma(\vec{n}_{out}, \vec{n}_{in})$. This relationship follows from combining the generic time-reversal symmetry condition $\sigma(\vec{n}_{in}, \vec{n}_{out}) = \sigma(-\vec{n}_{out}, -\vec{n}_{in})$ and the $\vec{n} \leftrightarrow -\vec{n}$ symmetry that we noted earlier: $\sigma(\vec{n}_{in}, \vec{n}_{out}) = \sigma(-\vec{n}_{in}, -\vec{n}_{out})$ (generalized here to all directions for all microscopic scattering events).

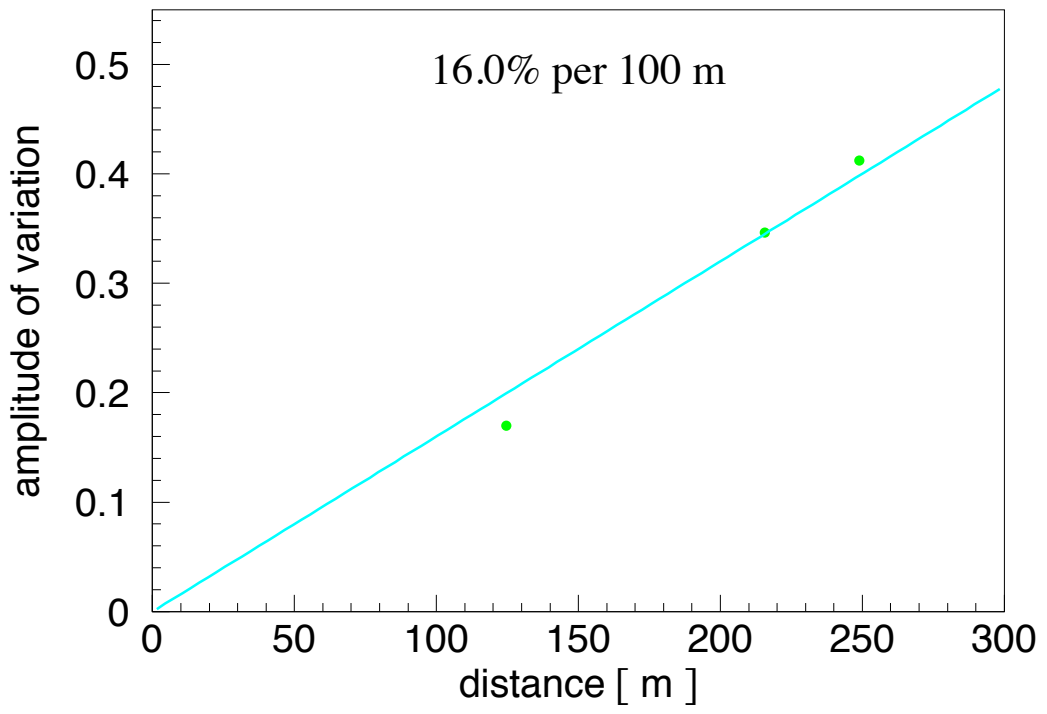

**Figure 4**: The amplitudes of sines fitted in figure 3 vs. distance. The 16% per 100 m fitted here describes the average behavior at all depths in the detector. This value, when computed for various depths, ranges from 10% in the clearest ice to 23% at the top of the detector.

The following description was eventually used, as it is consistent with the above condition on the cross section. Instead of modifying the scattering coefficient $b_e$, we modify the scattering function $f(\cos\theta)$, which describes the probability that the photon changes direction by an angle $\theta$ when scattered:

$$f(\vec{n_{in}} \cdot \vec{n_{out}}) \to f(\vec{k_{in}} \cdot \vec{k_{out}}), \quad \vec{k}_{in,out} = \frac{A\vec{n}_{in,out}}{|A\vec{n}_{in,out}|}.$$

The matrix $A$ can be diagonalized to

$$A = \begin{pmatrix} \alpha & 0 & 0 \\ 0 & \beta & 0 \\ 0 & 0 & \gamma \end{pmatrix} = \exp\begin{pmatrix} \kappa_1 & 0 & 0 \\ 0 & \kappa_2 & 0 \\ 0 & 0 & \kappa_3 \end{pmatrix}$$

in a basis of the direction of the largest scattering in the $xy$ plane, the direction of the smallest scattering in the $xy$ plane, and $z$. If the ice is isotropic, $\alpha = \beta = \gamma = 1$, and we get back the scattering function that only depends on a product $\vec{n}_{in} \cdot \vec{n}_{out}$. If there is anisotropy we can always assume that $\alpha\beta\gamma = 1$ (or $\kappa_1 + \kappa_2 + \kappa_3 = 0$), since descriptions with a matrix $A$ and $cA$ ($c$ being any number $\neq 0$) are equivalent to each other. This can be seen from the expression for $\vec{k}$, from which $c$ cancels out.

With this description of scattering the geometric scattering coefficient $b$ is constant for all directions, while the effective scattering coefficient $b_e = b \cdot (1 - \langle\cos\theta\rangle)$ receives some dependence on the direction of the incident photon via the direction-dependent term $1 - \langle\cos\theta\rangle$. In the following we derive the small-angle scattering approximation for this term, which clarifies this dependence, and can be useful if we choose to modify the absorption coefficient using the empirical relation $a \propto b_e$, thereby adding anisotropy also to the absorption.

First, we note that since the scattering function $f(\vec{k}_{in} \cdot \vec{k}_{out})$ depends only on the product $\vec{k}_{in} \cdot \vec{k}_{out}$, for the difference $\delta\vec{k} = \vec{k}_{out} - \vec{k}_{in}$, the following holds:

$$\langle\delta k^i\delta k^j\rangle = \frac{1-h}{2}\cdot\delta_{ij} - (2g - \frac{3h+1}{2})\cdot k^ik^j,$$

$$g = \langle\vec{k}_{out}\vec{k}_{in}\rangle,\quad h = \langle(\vec{k}_{out}\vec{k}_{in})^2\rangle,$$

where $g$ is a parameter of the scattering function. In this and the following expressions we omit the index "*in*": $\vec{k}_{in} = \vec{k}$. The brackets $\langle\rangle$ denote averaging over all possible final directions $\vec{k}_{out}$ after a single scatter with probabilities prescribed with the scattering function. The above relationship can be proven by evaluating it in a basis of $\vec{k}$, and any two vectors, perpendicular to $\vec{k}$ and to each other. In this basis,

$$\vec{k}_{in} = (0,0,1),\quad \vec{k}_{out} = (\sin\theta\cos\phi, \sin\theta\sin\phi, \cos\theta)\quad\rightarrow$$

$$\delta\vec{k} = (\sin\theta\cos\phi, \sin\theta\sin\phi, \cos\theta - 1).$$

The averaging over the final directions is performed with integration

$$\int_0^{2\pi} d\phi\int_{-1}^{1} f(\cos\theta)d(-\cos\theta)$$

The off-diagonal terms of $\langle\delta k^i\delta k^j\rangle$ are zero due to integration over $\phi$. The diagonal terms evaluate to the expression given above. The trace of $\langle\delta k^i\delta k^j\rangle$ evaluates to $\langle\delta\vec{k}^2\rangle = \langle(\vec{k}_{out} - \vec{k}_{in})^2\rangle = 2\cdot(1-g)$.

We note that

$$\vec{n} = |A\vec{n}|\cdot A^{-1}\vec{k},\quad \vec{n}^2 = |A\vec{n}|^2\cdot|A^{-1}\vec{k}|^2 = 1$$

$$\rightarrow\quad \vec{n} = \frac{A^{-1}\vec{k}}{|A^{-1}\vec{k}|} = \frac{B\vec{k}}{|B\vec{k}|}.$$

For brevity we use $B = A^{-1}$. We can now evaluate the derivative

$$\frac{\partial n^i}{\partial k^n} = \frac{\partial}{\partial k^n}\frac{B_{ij}k^j}{\sqrt{B_{kl}k^lB_{km}k^m}} = \frac{B_{in}}{|B\vec{k}|} - \frac{B_{ij}k^jB_{kl}k^lB_{kn}}{|B\vec{k}|^3} =$$

$$\frac{B_{in} - n^in^kB_{kn}}{|B\vec{k}|}.$$

Now we can evaluate the $\langle\vec{n}_{in}\vec{n}_{out}\rangle$ from

$$2\cdot(1 - \langle\vec{n}_{in}\vec{n}_{out}\rangle) = \langle(\vec{n}_{out} - \vec{n}_{in})^2\rangle = \langle\delta\vec{n}^2\rangle \approx$$

$$\frac{B_{in} - n^in^kB_{kn}}{|B\vec{k}|}\cdot\frac{B_{im} - n^in^lB_{lm}}{|B\vec{k}|}\cdot\langle\delta k^n\delta k^m\rangle.$$

The second term proportional to $k^nk^m$ in the expression for $\langle\delta k^n\delta k^m\rangle$ leads to zero contribution in the above expression. Only the first term proportional to $\delta_{nm}$ contributes, resulting in

$$2\cdot(1 - \langle\vec{n}_{in}\vec{n}_{out}\rangle) \approx \frac{B_{in}B_{in} - n^iB_{in}n^jB_{jn}}{|B\vec{k}|^2}\cdot\frac{1-h}{2} =$$

$$(B_{in}B_{in} - n^iB_{in}n^jB_{jn})\cdot|A\vec{n}|^2\cdot\frac{1-h}{2}.$$

In the simple case when $B_{ij} = \delta_{ij}$, $\vec{n} = \vec{k}$, and we should get back

$$2\cdot(1-g) \approx 1-h.$$

Whether this condition is satisfied depends on the properties of the scattering function near its maximum. For $1-g = 0.1$, (1-h)/2=0.090 for simplified Liu (SL) and (1-h)/2=0.063 for Henyey-Greenstein (HG) scattering functions (see [2] for definitions). As a further approximation, we take this condition for granted, and derive the expression for the $1 - \langle\cos\theta\rangle$, which gives us the directional dependence of the effective scattering:

$$1 - \langle\cos\theta\rangle = (1-g)\cdot\frac{1}{2}\cdot(B_{in}B_{in} - n^iB_{in}n^jB_{jn})\cdot|A\vec{n}|^2.$$

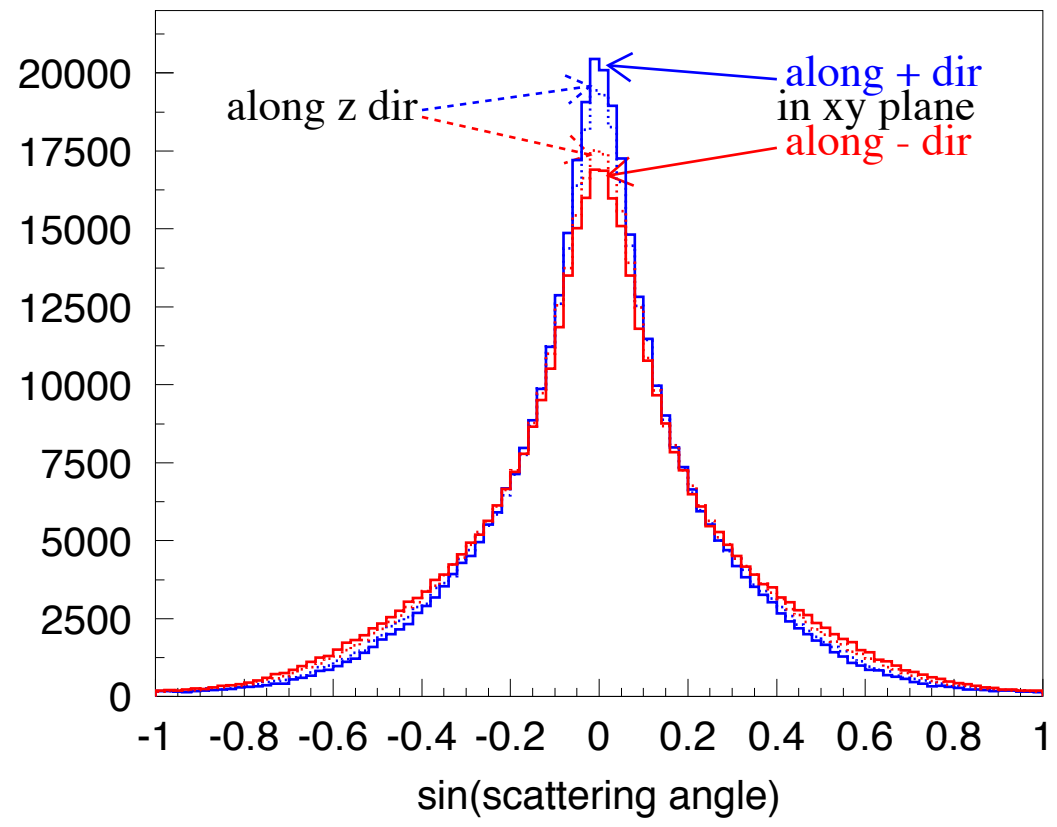

**Figure 5**: Distributions of the photon direction component after scatter along a direction perpendicular to the initial direction: either in xy plane (solid), or along the z axis (dashed). The -8% anisotropy along the main anisotropy axis ($\kappa_1 = -0.08$) is assumed. The photons with initial direction along the main anisotropy axis (the "-" direction) scatter less than the photons with initial directions along the minor axis (the "+" direction).

## 3 Results

We performed fits for the coefficients $\alpha = \exp(\kappa_1)$ and $\beta = \exp(\kappa_2)$ using the two methods of [2]: first, using only the integrated charge on the receiving DOMs and, second, using the time-binned data. The likelihood description used by the fit was updated according to [3]. The main axes of the diagonalized matrix $A$ describing anisotropy were chosen: one along the $z$-axis, and the other two in the $xy$ plane. The azimuth angle of the axes in the $xy$ plane was also fitted. The two methods yield anisotropy coefficient values that are within 20% of their average: $\kappa_1 = -0.082$ and $\kappa_2 = 0.040$, as shown in figure 6. Taking these as the result the third coefficient is $\kappa_3 = -\kappa_1 - \kappa_2 = 0.042$.

The large discrepancy between the two methods is possibly due to effects unaccounted yet in this fit, such as depth dependence of the anisotropy matrix. Given this we can assume that $\kappa_2 = \kappa_3$, and, thus, that there is a symmetry between the to directions described by $\kappa_2$ and $\kappa_3$, and that main axis of anisotropy is described by $\kappa_1$. The direction of this axis was fitted to 126 degrees (within 5 degrees of the direction of the ice flow). Figure 5 demonstrates the effect of anisotropy on photon scattering.

We repeated the entire ice model fit procedure described in [2], additionally fitting for the anisotropy (the two coefficients $\kappa_1$, $\kappa_2$, and the direction of the axis corresponding to $\kappa_1$). The resulting absorption and effective scattering are in good agreement with the result reported in [2], as shown in figure 8. The oscillating behavior in the ratio of data to simulation vs. direction to receiving string is substantially reduced, as shown in figure 7.

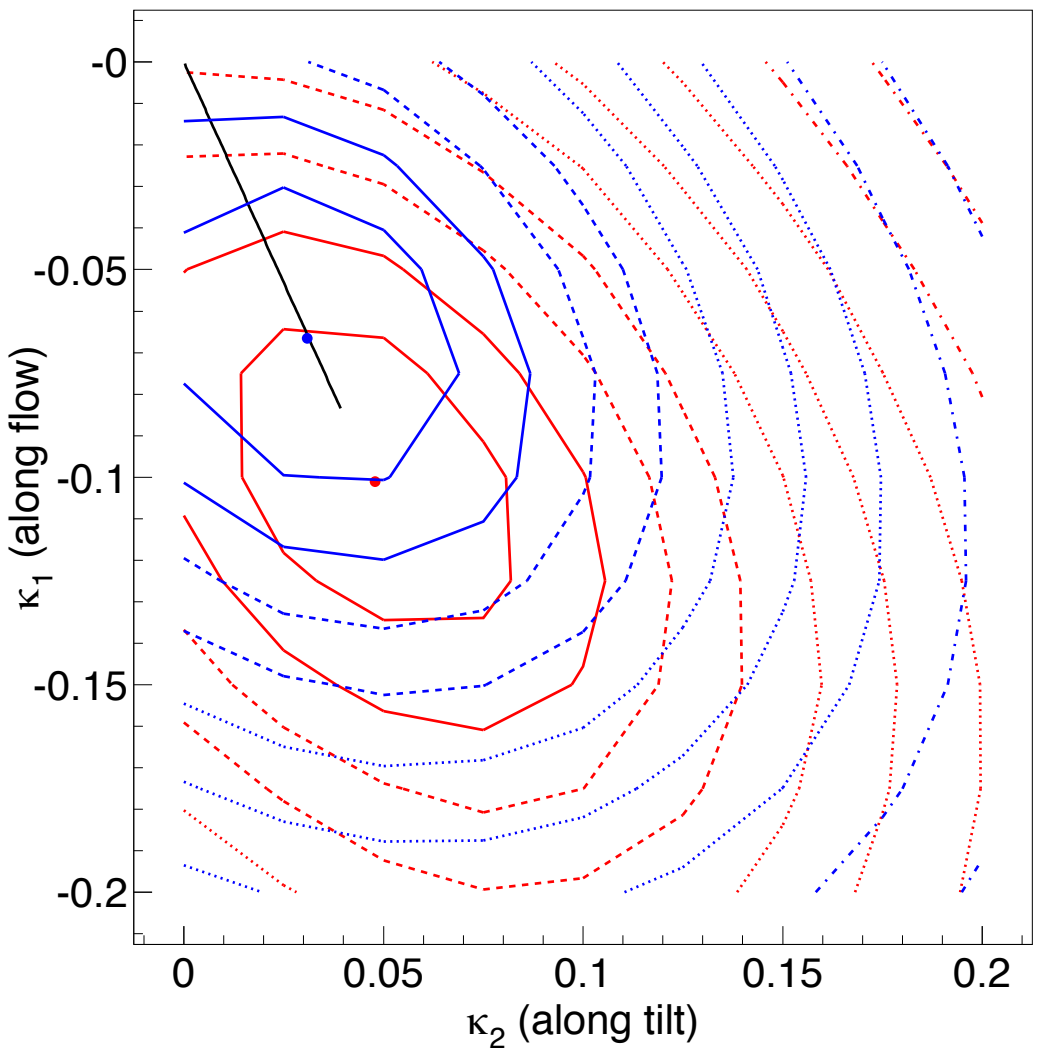

**Figure 6**: Likelihood function in the vicinity of the minimum using only charge information (red) and using time-binned data (blue). The values are shown with contours on a log scale. The two dots in each plot show positions of the minima in both cases. The line is drawn from (0,0) to the average of the two dots and shows that the ratio of $\kappa_1/\kappa_2$ is approximately the same in both cases.

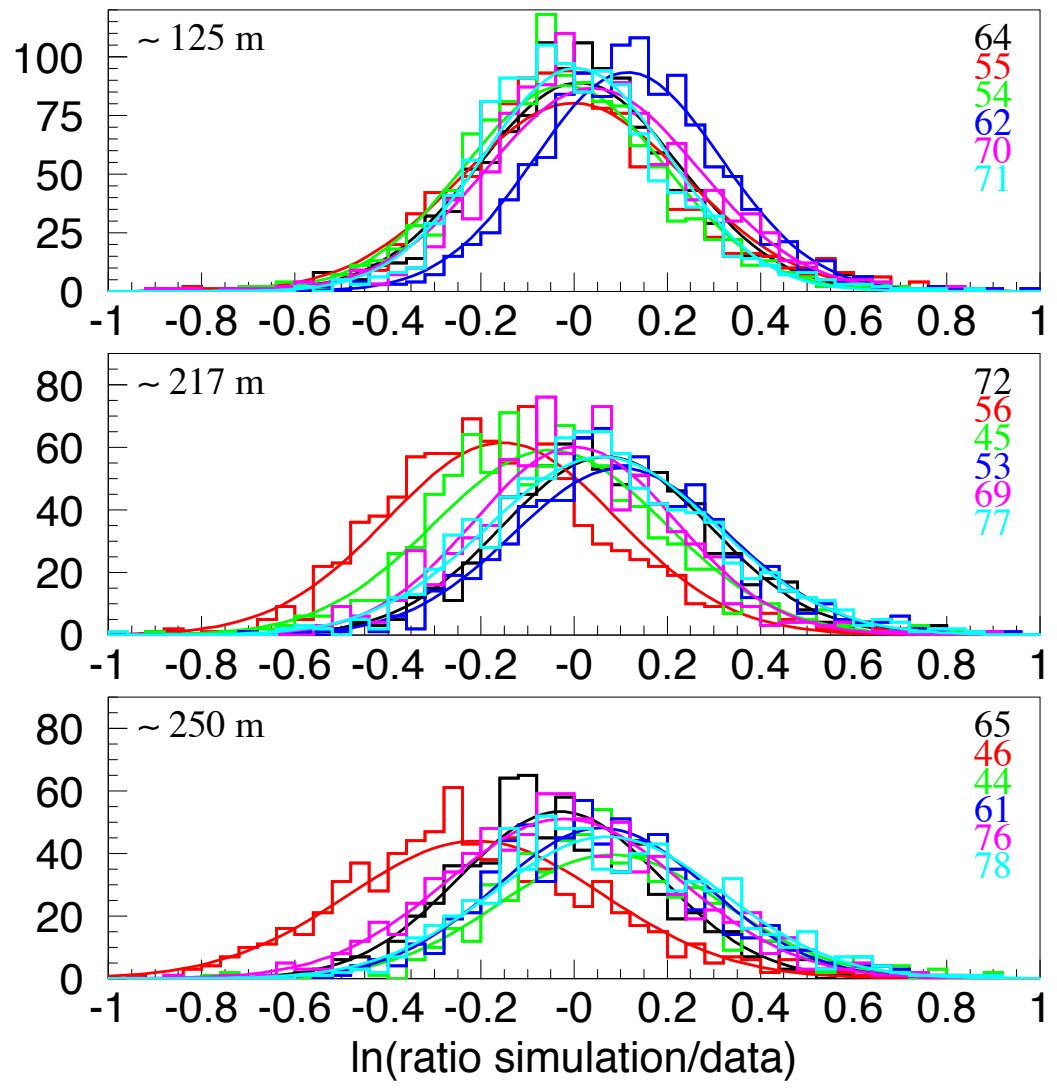

**Figure 7**: Ratio of (updated) simulation to data of total charge collected in DOMs on strings surrounding string 63, same notations as in figure 2. The ice model used in this simulation includes the anisotropy fit result of this paper.

The ice anisotropy reported here has also been confirmed with a study that employed well-reconstructed downgoing muons, where the charge collected in a direction of the main anisotropy axis showed an average $\sim 14\%$ excess at 100 m away from the muon with respect to the average over all directions (see figure 9).

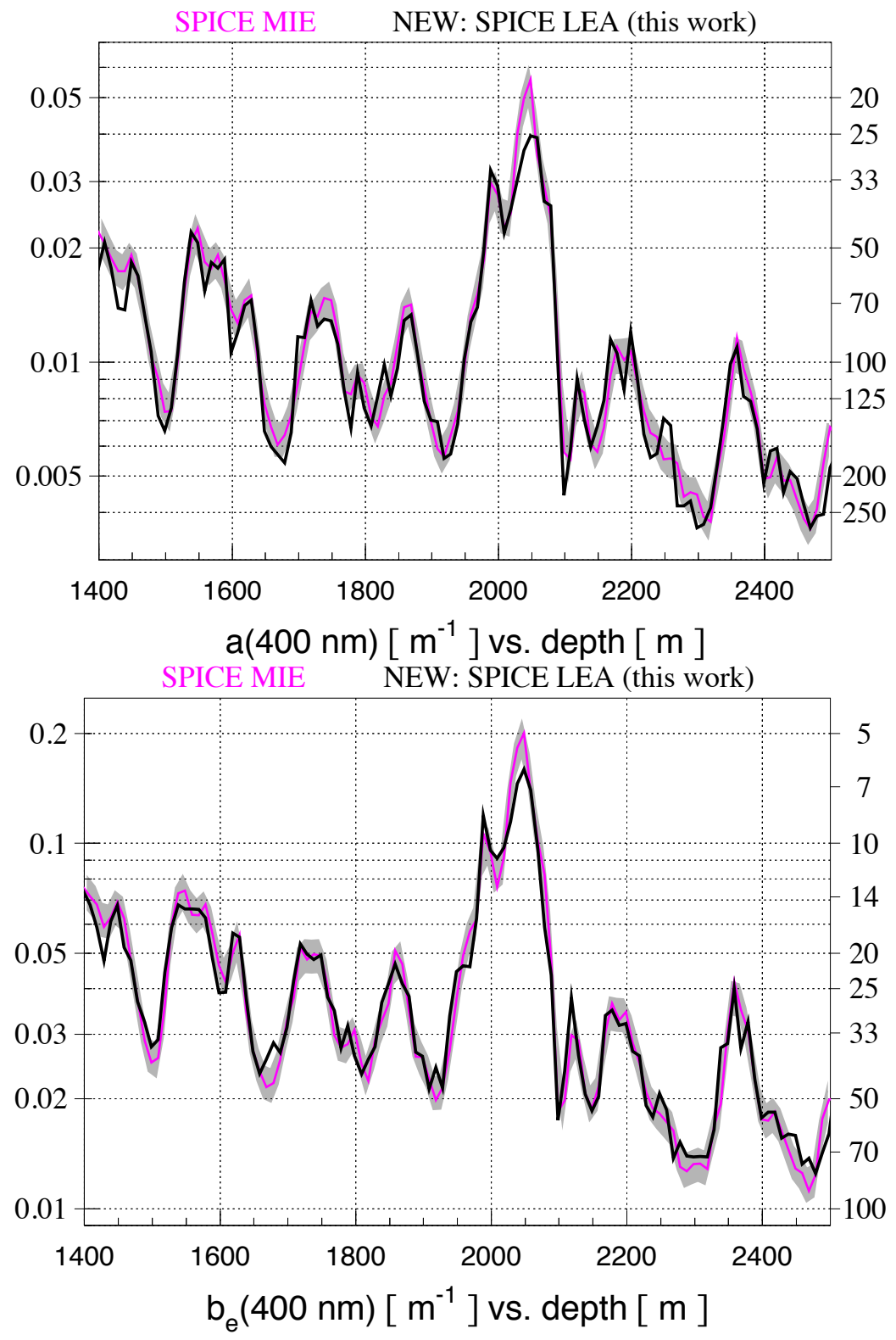

**Figure 8**: New absorption and effective scattering parameters (LEA) compared with the result reported in [2] (MIE). The grey band around the MIE result shows the uncertainties reported in [2].

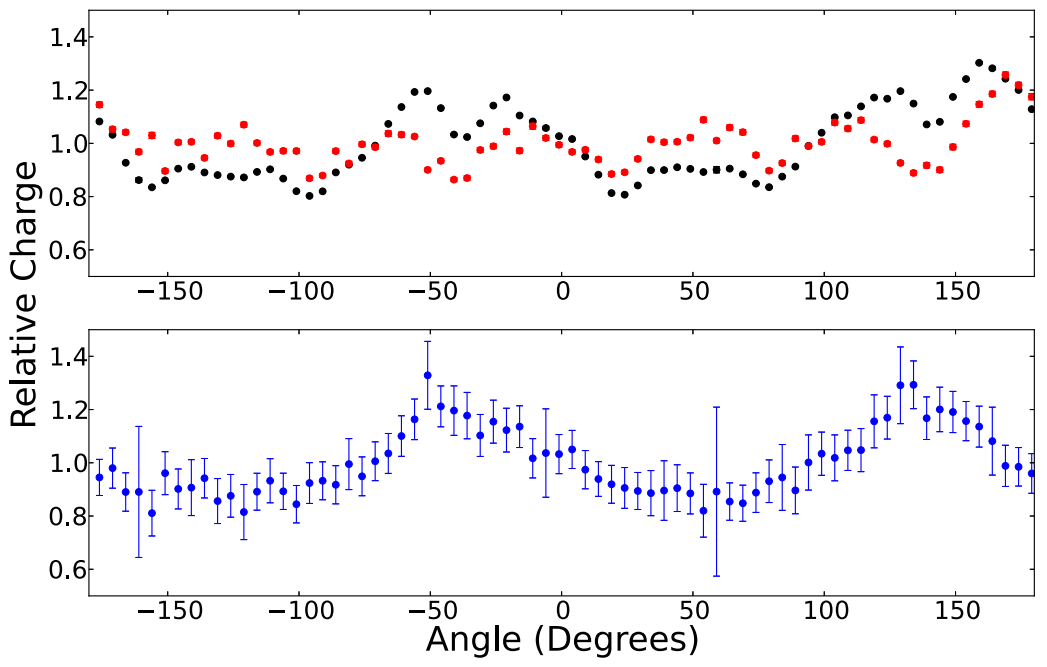

**Figure 9**: Top: Variations in charge collected 100-150 m away from the reconstructed muon tracks in data (black) and simulation (red) based on ice model of [2] lacking anisotropy (for which some variation is expected due to hexagonal detector geometry). Bottom: ratio of data to simulation curves of the plot above. The angle shown on the x-axis is the same as in figure 3. The main axis of anisotropy is at 126 (and -54) degrees, same as in figure 3.

## References

[1] A. Achterberg et al., Astropart.Phys., 26 (2006) 155.
[2] M. Aartsen et al., NIM-A, 711 (2013) 73, arXiv:1301.5361.
[3] D. Chirkin, arXiv:1304.0735.

# Event reconstruction in IceCube based on direct event re-simulation

THE ICECUBE COLLABORATION[1]

[1]*See special section in these proceedings*

*dima@icecube.wisc.edu*

**Abstract:** We present a new method of event reconstruction in IceCube that searches through an extensive set of re-simulated events to find the best match to data. Several search methods, including the stochastic approximation and localized random search, were tried. We demonstrate the capability of the new method to reconstruct cascade-like as well as track-like events. While significantly slower than the parametrized probability density function (PDF) -based reconstructions, the new method is able to take into account all of the nuances present in the simulation that are sometimes omitted from the PDF-based description of data.

**Corresponding author:** Dmitry Chirkin[1]
[1] *Dept. of Physics and WIPAC, University of Wisconsin, Madison, WI 53706, USA*

## 1 Introduction

IceCube is a cubic-kilometer neutrino detector installed in the ice at the geographic South Pole [1] between depths of 1450 m and 2450 m. Detector construction started in 2005 and finished in 2010. Event reconstruction relies on the optical detection of Cherenkov radiation emitted by secondary particles produced in neutrino interactions in the surrounding ice or the nearby bedrock.

The development of the method described in this paper was prompted by our recognition that ice surrounding the detector does not have the azimuthal symmetry previously ascribed to it (see [2], [3]) and still used in our probability density function (PDF) -based event reconstructions. While it is relatively easy to add a description of almost any conceivable detail to the detector simulation, adding even one new dimension to the existing photon propagation probability tables is prohibitive. This is due to the large computing time necessary to create the PDF tables, and also due to the excessive storage requirements of and the speed of access to the stored tables (the latter becomes slow if the tables cannot be fitted into the active computer memory).

In section 2 we discuss reconstruction of cascades, i.e., relatively short particle showers, which happen when neutrinos interact via the neutral current process, or when $\nu_e$, $\bar{\nu}_e$ and $\nu_\tau$, $\bar{\nu}_\tau$ interact via the charged current process. In these cases all visible particles produced in the interaction are short-lived, either decaying or losing energy very quickly, producing an initial light pattern that can often be approximated as emanating from a point, when compared to the large scale of the IceCube detector.

In contrast, the charged current interactions of $\nu_\mu$, $\bar{\nu}_\mu$ create a muon (in addition to the initial cascade), which can propagate a great distance away from the interaction point, leaving behind a characteristic "track"-like pattern of light. We call these events "tracks" and discuss the application of our method to the track-like events in section 3.

## 2 Cascade reconstruction

We start by describing the cascade reconstruction, the purpose of which is to ascertain the parameters of the cascade-like event observed in data: energy ($E$), time and position ($t_0, x, y, z$), and direction ($\theta$, $\phi$). The algorithm works by simulating many cascades with various sets of parameters and choosing the one (or a few) that look most like data.

We compare the time-binned photo-electron charges registered by optical sensors between data and simulation with a likelihood function described in [4], which takes into account statistical fluctuations in both data and simulation. A systematic disagreement between data and simulation is modeled with a log-normal distribution (describing the "model error" of the ice model, as explained in [2] and reduced to 20% in [3]).

We obtained the best performance with the "localized random search"-based optimization method. The reconstruction is started from a first guess with the coordinates of the center of gravity of hits in the detector, a random direction, the energy of $E = 10^5$ GeV, and the start time $t_0$ coincident with the time of the first hit in the event. We also specify the widths of the proposal distributions for space coordinates ($\Delta r$=10 m) and angular coordinates ($\Delta\Psi$=30 degrees). The proposal distributions are the 3D Gaussian and 2D Gaussian on a sphere (von-Mieses-Fisher distribution).

At each iteration 25 possible sets of space and angular coordinates ($x$, $y$, $z$, $\theta$, $\phi$) are sampled from the proposal distribution around the best cascade from the previous iteration (or the initial guess, if this is the first iteration). For each such set the photon simulation of [6] is run with the best values of $E$ and $t_0$ from the previous iteration, producing Monte-Carlo hits. Then the best values of energy $E$ and start time $t_0$ that maximize the likelihood function are found. This is done by scaling the simulation (varying $n_s$ of [4]) and shifting it in time. Finally, each set is re-simulated with the optimized values of $E$ and $t_0$. The final likelihood values for all 25 sets are computed and the event with the best value is selected. This best value and corresponding cascade parameters are saved and used as a starting point for the next iteration.

This process is repeated 400 times. Every 40 iterations the proposal distribution width parameters $\Delta r$ and $\Delta\Psi$ are updated to the RMS values of space and angular coordinates calculated from the best 20 of the previous 40 iterations. Finally, the solution is calculated by averaging the best 160 iteration results. Their spread can be used as a rough measure of statistical uncertainties, and is immediately available as it is calculated during reconstruction (unlike the more rigorous procedure described in section 4). The specific numbers listed above (25 sets per iteration, 400 iterations, initial parameters of the proposal distribution) resulted in a robust, yet relatively fast reconstruction.

In addition to the "localized random search" method described above we have tried two more approaches, which yielded somewhat less robust performance. The first of these is the simultaneous perturbation stochastic approximation [5] and its variations, which is a derivative-based maximization search method, where the derivative is calculated "stochastically" from consecutive iterations, and the step size is reduced in a controlled manner. The second method is the Markov chain Monte-Carlo with each proposed step either taken or rejected based on whether the calculated value of the likelihood at the new coordinates is better or worse than that at the current point. Since the likelihood value at given coordinates is calculated from a simulation performed anew for each new evaluation, it is a "stochastic" quantity, forming a distribution, rather than assuming a single value. So, it is possible to jump from a point with parameters providing a (on average) better description of data to a point with parameters providing a (on average) worse description of the data. Although the iterations of the method do, in fact, form a Markov chain, this Markov chain is not expected to be reversible, with no obvious probabilistic interpretation of the resulting sequence of iterations.

## 3 Track reconstruction

The algorithm described in the previous section also works for track reconstruction with the following modification, which consists of one or more "passes". We start by describing the first pass. For each proposed set of space and angular coordinates defining a track ($x$, $y$, $z$, $\theta$, $\phi$), small cascades of equal energy are simulated every 7.5 meters along the part of the track that is inside the detector with a fixed combined energy of $10^6$ GeV. Each such cascade (numbered with index $j$) produces a pattern of hits in the detector, creating a column in a matrix $A_{ij}$, where $i$ represents a time bin in a DOM ($A_{ij}$ being the photo-electron charge in that bin left by the cascade $j$).

We now have an unfolding problem similar to that discussed in [7], where the energy of each cascade $j$ needs to be scaled by "weights" $w_j$ so that the sum $\sum_j A_{ij} w_j$ best matches the hit counts in data $d_i$. This problem is solved taking into account the Poisson nature of both data and simulation sets of each cascade (as described in [4]), resulting in the set of unfolded simulation weights $w_i$. The first guess is obtained by solving the $\sum_j A_{ij} w_j = d_i$ with a non-negative least squares (NNLS) algorithm, and the solution is then refined for the full likelihood description with a derivative method (BFGS2 of GSL, [8]).

The unfolding procedure is repeated for various values of $t_0$, and the best value of $t_0$ is chosen to maximize the likelihood comparing the data with a superposition of cascades weighted with the unfolded weights $w_j$ determined for that $t_0$. The unfolded pattern of weights corresponding to the best value of $t_0$ is applied to the energies of cascades along the track, and the event is re-simulated.

The procedure described above is optionally repeated one more time (second pass) starting with small cascades of energies proportional to $w_j$ (but still with a total sum of $10^6$ GeV), adding simulation to the pool created during the first pass (i.e., simulation sets from both passes are used for the energy loss unfolding part of the calculation, and for the fit to $t_0$). The final pattern of cascade energies is re-simulated one more time and the final likelihood value for this iteration is computed.

The entire calculation described in this section is performed for each of the 25 trial sets of each of the 400 iterations of section 2. Fig. 4 demonstrates the convergence behavior of the method for a track event.

## 4 Estimating uncertainties

Statistical uncertainties can be estimated with an Approximate Bayesian Computation (ABC) method as in [9, 10].

The idea is to substitute the probability of data $D$ given event parameters $\theta$, $p(D|\theta)$, in the Bayes expression for the conditional probability density

$$p(\theta|D) = \frac{p(D|\theta)\cdot p(\theta)}{p(D)},$$

with a probability to obtain data events that are sufficiently similar to the event under investigation. The similarity is defined with a "distance" between the other possible data events $\hat{D}$ and our data event $D$

$$\rho(\hat{D}, D) \leq \varepsilon,$$

which we substitute with the minus log likelihood expression of [4], comparing the time-binned photo-electron charges between data and simulation.

Since the prior of the track parameters $x$, $y$, $z$, $\theta$, $\phi$ can be often assumed to be flat in 3D space of coordinates and on a 2D sphere of directions (as neutrinos interact uniformly anywhere inside the detector), the sampling method of choice is a Markov Chain (rather than rejection), where steps are sampled from a proposal distribution and accepted when the "distance" condition is satisfied. For more details of this procedure (including a proof), see [9]. The parameters of the proposal distribution and $\varepsilon$ in the "distance" expression are estimated from the best steps of the minimization procedure of sections 2 and 3.

Estimating statistical uncertainties using this method takes approximately the same amount of time as the reconstruction (two days on a specially-configured computer). It is unclear if adding nuisance parameters to this procedure to additionally account for the systematic uncertainties is realistic as the step acceptance rate is lowered with each additional parameter (leading to a "curse of dimensionality").

## 5 Reconstruction Examples

The method of this report was applied to the 28 high energy events of [11]. Figure 1 compares the reconstructed energy and zenith with the main reconstruction method used in [11]. Most of the points are within the average estimated uncertainties of 10% in energy and 10 degrees in the zenith angle. The energy of the outlier is overestimated by the reconstruction method of [11], since it did not include the effect of the ice layer tilt (change in depth of a given ice layer at different $x$, $y$ across the detector, see [2]) in the PDF used by the reconstruction, whereas the method of this report used our best description of the ice, including the tilt. This particular event was affected more than others because it lay at coordinates just above the dust peak (layer of ice inside the detector volume containing the most dust, at around 2000 m, see [2]), and where the correction due to tilt was significant, affecting whether the event appears to be inside or outside the dust peak ice layer. The uncertainties for this event were estimated in [11] by reconstructing events similar to this one, simulated with the best knowledge of ice. The median of the energy uncertainty range was found to be a good agreement with the energy reconstructed with the method of this report. The outlier in zenith is for a cascade event with a particularly poor reconstructed angular resolution.

Out of the 28 events only one reconstructed as a track, with many reconstructed energy losses along its path (as shown in figure 2, also see figure 3). In the other 6 events that have a muon in addition to the larger cascade the muon part of the event is largely ignored by this reconstruction. We believe this happens when the contribution to the likelihood from the few hits left by the muon is so small that it is largely "washed out" by the fluctuations of the much larger contribution from the cascade part of the event. Finally, figure 4 illustrates how the method converges to the solution, and figure 5 shows a plot of reconstructed uncertainties in direction and energy vs. reconstructed energy.

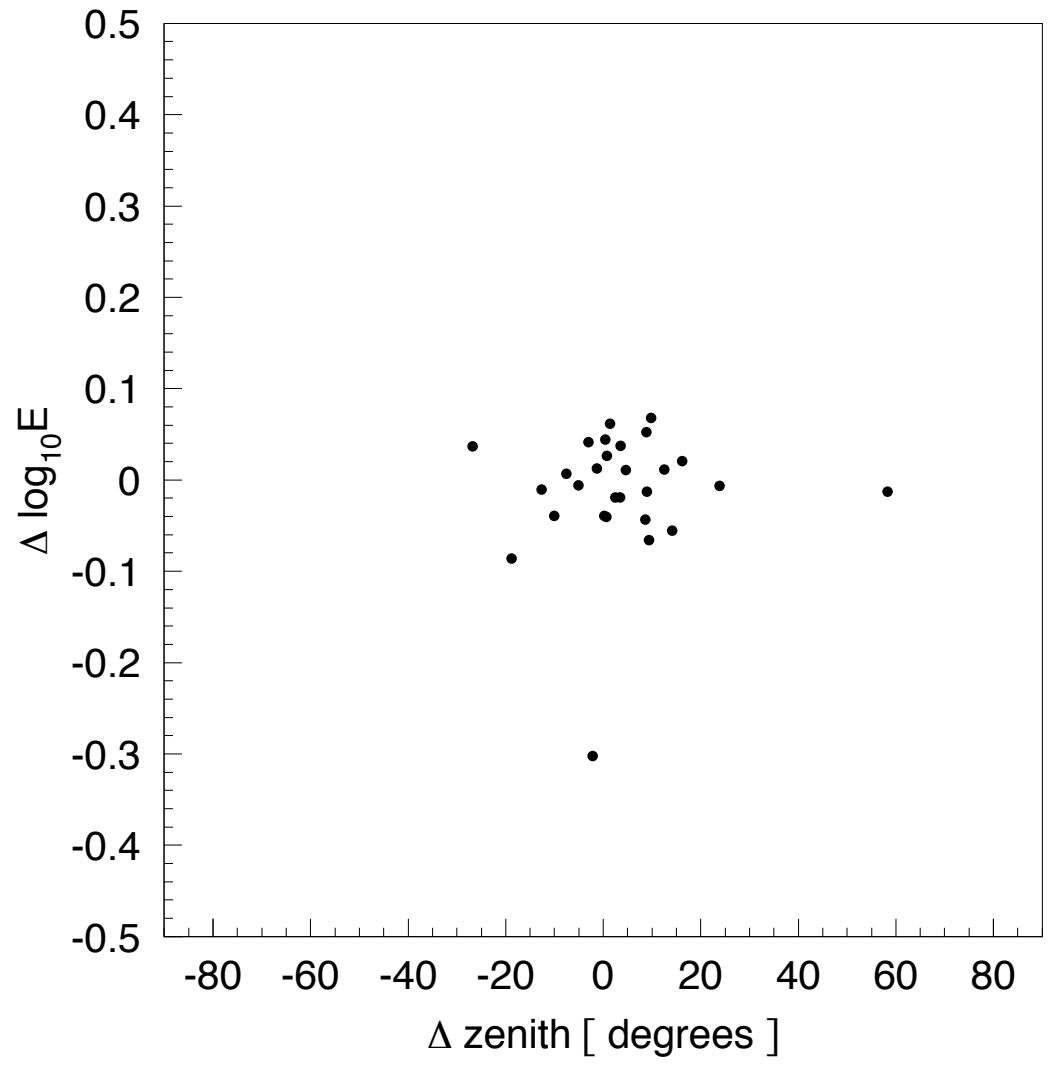

**Figure 1**: Comparison between the results of the reconstruction method described in this report with the main reconstruction used in [11] for the 28 high energy events of [11].

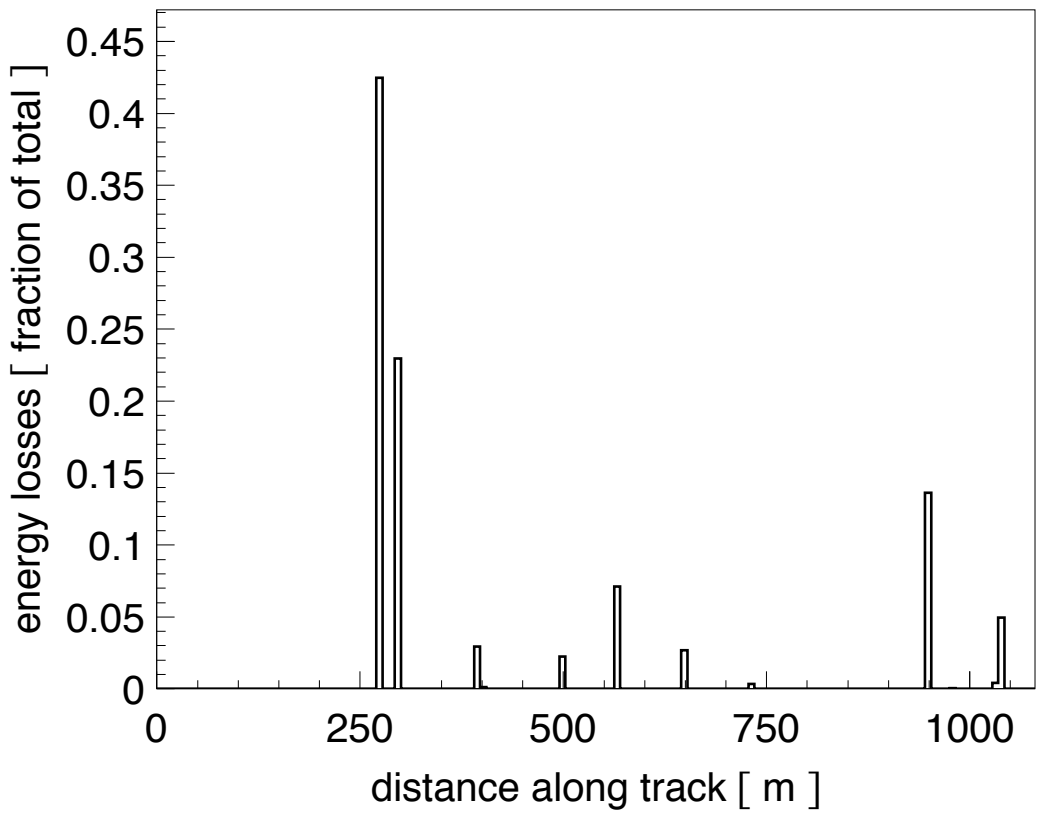

**Figure 2**: Reconstructed energy losses along the best-fit track of a track-like event shown in figure 3 (one of the 28 high energy events of [11]). The losses are normalized to the total energy lost in the volume of the detector.

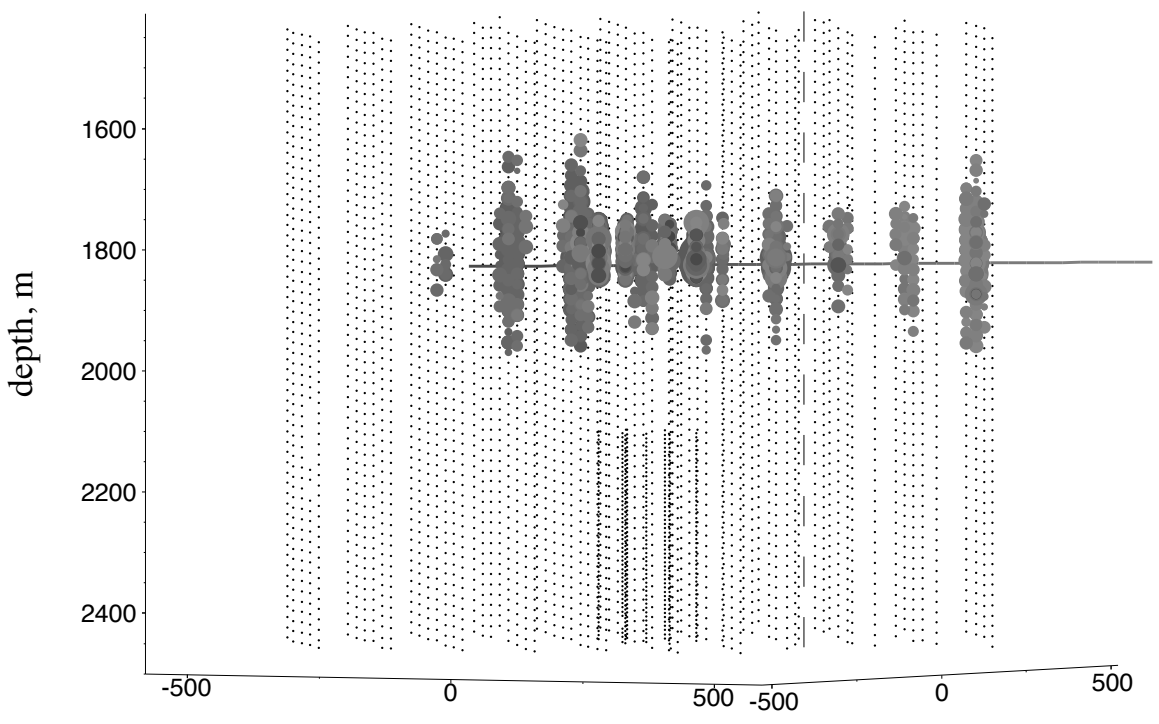

**Figure 3**: Illustration of a track-like event, one of the 28 high energy events of [11].

## References

[1] A. Achterberg et al., Astropart. Phys., 26 (2006) 155.
[2] M. Aartsen et al., NIM-A, 711 (2013) 73, arXiv:1301.5361.
[3] IceCube Coll., paper 0580 these proceedings.
[4] D. Chirkin, arXiv:1304.0735.
[5] http://en.wikipedia.org/wiki/Simultaneous_perturbation_stochastic_approximation
[6] D. Chirkin for the IceCube Collaboration, accepted by NIM-A, DOI: 10.1016/j.nima.2012.11.170.
[7] N. Whitehorn, Ph.D thesis, UW-Madison (2012).
[8] GNU Scientific Library, http://www.gnu.org.
[9] P. Marjoram et al., PNAS, 100-26 (2003) 15324.
[10] E. Cameron, A. Pettitt, MNRAS, 425-1 (2012) 44.
[11] IceCube Coll., paper 0650, these proceedings.

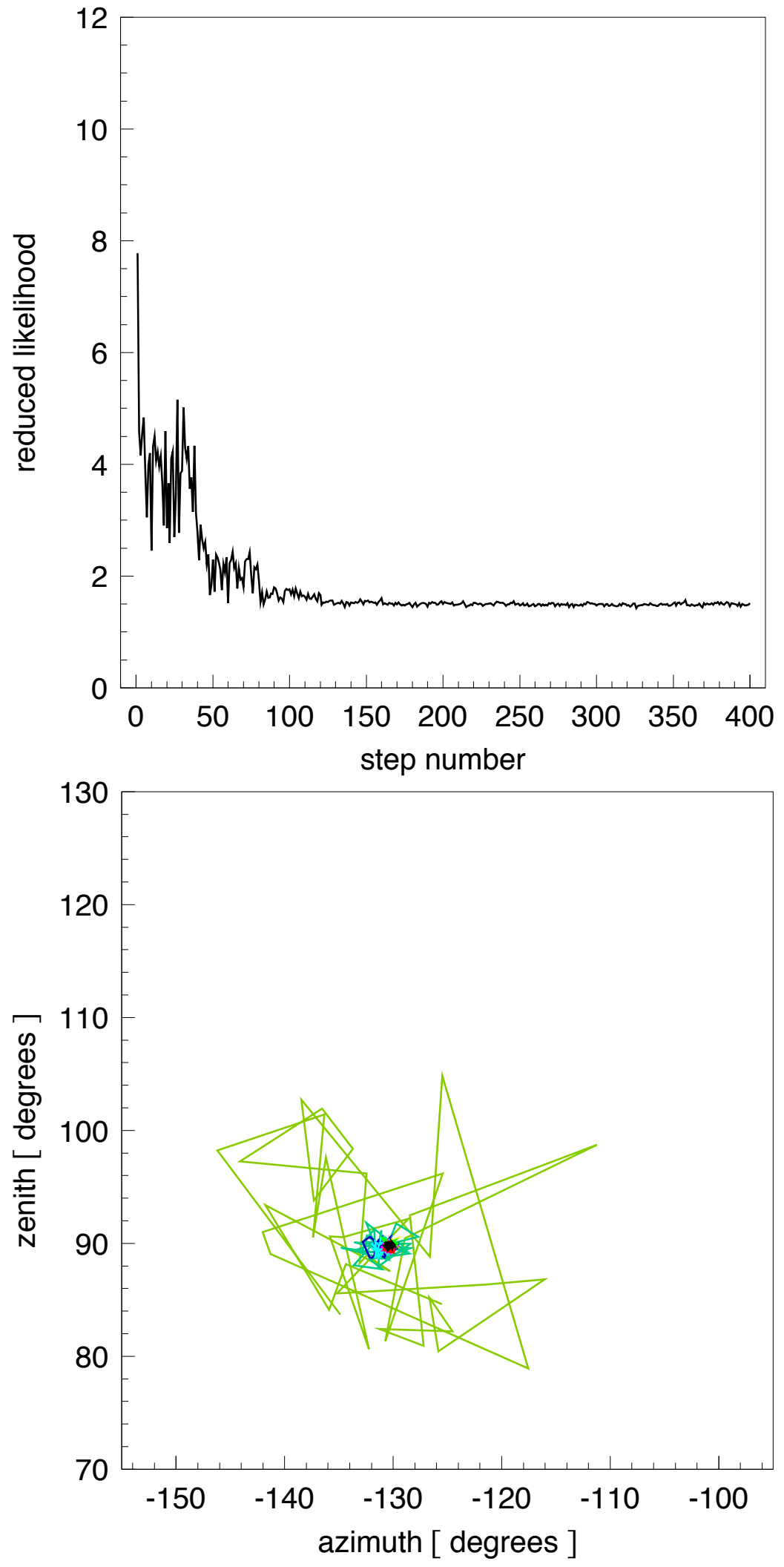

**Figure 4**: Top: Reduced log likelihood (log likelihood divided by the number of degrees of freedom) after each of the 400 iterations. Some stepping structure evident in the plot is related to the fact that the parameters of the proposal distribution are updated every 40 steps. Bottom: Convergence of the best fit direction: results of successive iterations are connected by line segments for visual clarity. This is a reconstruction of a well-defined track, and results in the method wandering around in a very narrow range of zenith and azimuth. This reconstruction took about 2 days on a computer with 5 GPUs. Most of the calculation time was spent on the GPUs, which were used to run the photon propagation simulation.

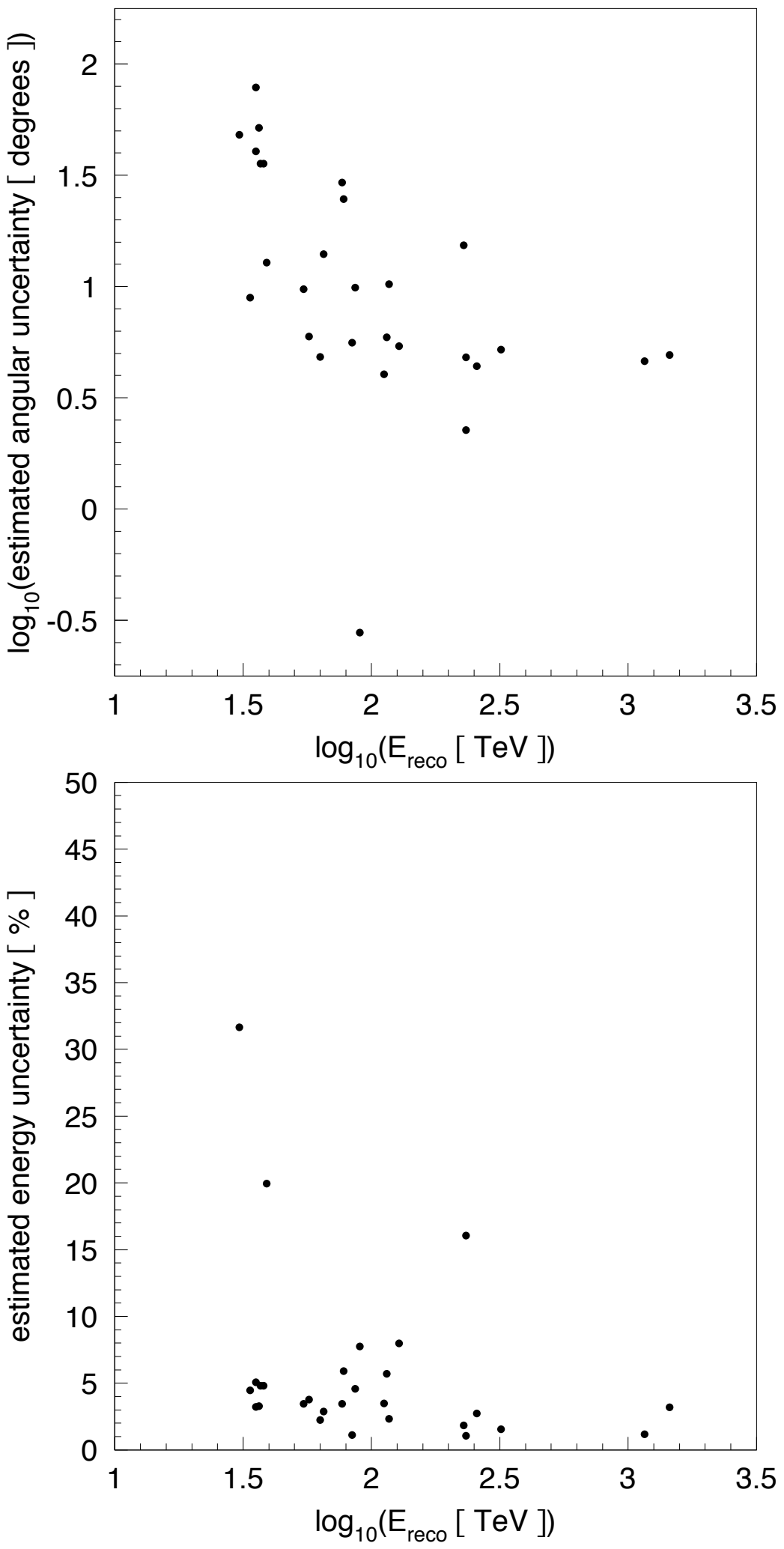

**Figure 5**: Estimated angular (top) and energy (bottom) statistical uncertainties of the 28 high energy events of [11] as a function of reconstructed energy deposited in the detector volume. The uncertainties shown were calculated as the "spread" of the reconstructed parameters, as explained in section 2.

# Robust Statistics in IceCube Initial Muon Reconstruction

THE ICECUBE COLLABORATION[1], B. RECHT [2], C. RÉ[2]

[1]*See special section in these proceedings*
[2]*Dept. of Computer Sciences, University of Wisconsin, Madison, WI 53706, USA.*

*mwellons@cs.wisc.edu*

**Abstract:** In the IceCube Neutrino Detector, muon tracks are reconstructed from the muon's light emission. The initial track "linefit" reconstruction serves as a starting point for more sophisticated track fitting, using detailed knowledge of the ice and the detector. The new approach described here leads to a substantial improvement of the accuracy in the initial track reconstruction for muons. Our approach is to couple simple physical models with robust statistical techniques. Using the metric of median angular accuracy, a standard metric for track reconstruction, this solution improves the accuracy in the reconstructed direction by 13%.

**Corresponding authors:** M. Wellons[1]
[1] *Computer Sciences Department, University of Wisconsin, Madison*

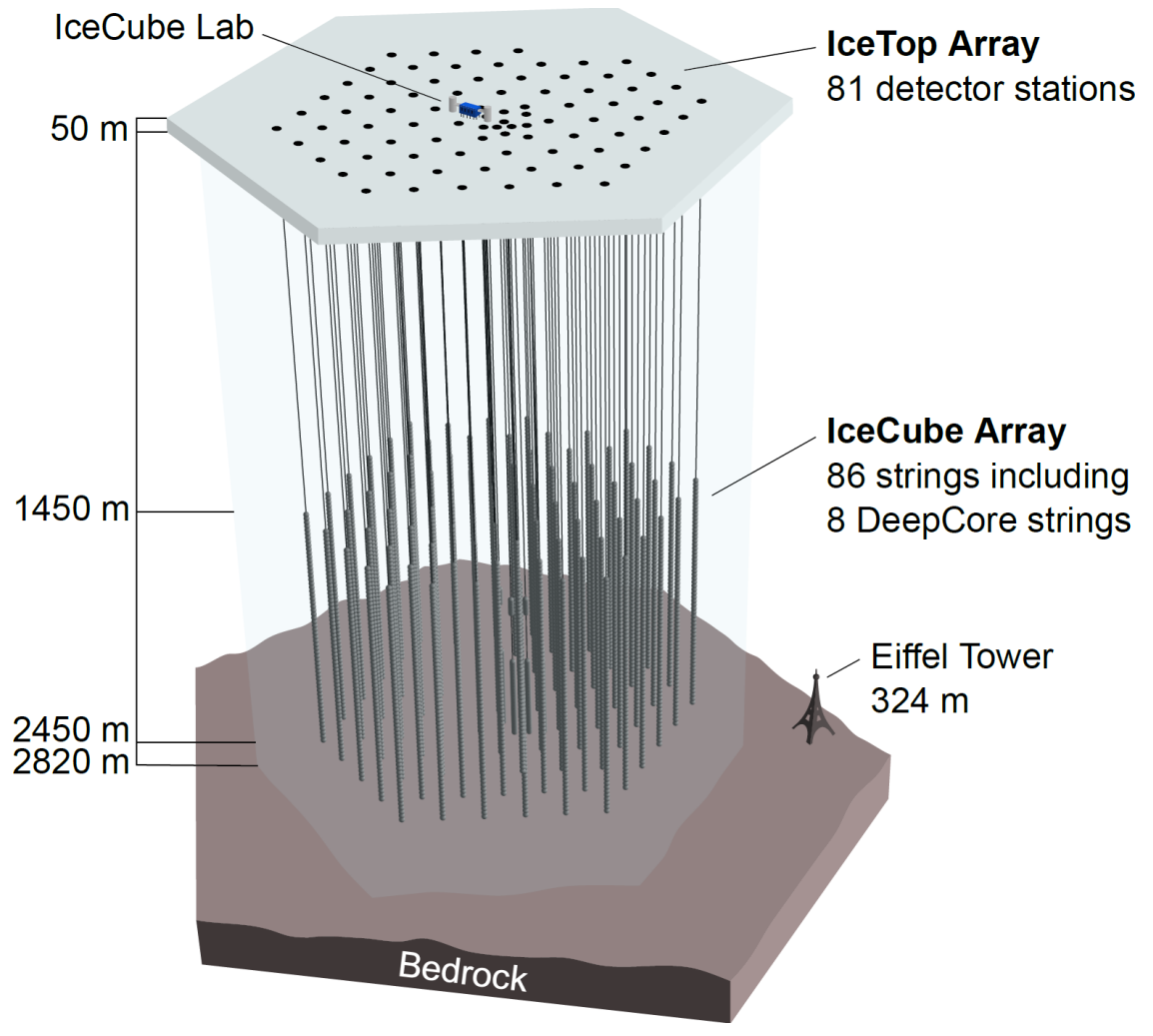

**Fig. 1**: The IceCube neutrino detector in the Antarctic ice. A picture of the Eiffel Tower is shown for scale.

## 1 Introduction

The IceCube neutrino detector searches for neutrinos that are generated by the universe's most violent astrophysical events: exploding stars, gamma ray bursts, and cataclysmic phenomena involving black holes and neutron stars [1]. These neutrinos are detected by the charged particles, often a muon, produced in their interaction with the rock or ice near the detector.

The detector, roughly one cubic kilometer in size, is located near the geographic South Pole and is buried at depths 1.5-2.5 km in the Antarctic ice [2]. The detector is illustrated in Figure 1 and a more complete description is given in Section 2.

This manuscript describes an improvement in the reconstruction algorithm used to generate the initial track position and direction of detected muons in the IceCube detector. We achieve this improvement in accuracy with the addition robust statistical techniques to the reconstruction algorithm.

## 2 Background

The IceCube detector is composed of 5,160 optical detectors, each composed of a photomultiplier tube (PMT) and onboard digitizer [3]. The PMTs are spread over 86 vertical strings arranged in a hexagonal shape, with a total instrumented volume of approximately one cubic kilometer. The PMTs on a given string are separated vertically by 17 m, and the string-to-string separation is roughly 125 m.

When a neutrino enters the telescope, it sometimes interacts with the ice and generates a muon. As the muon travels though the detector, it radiates light [4], which is observed by the PMTs and divided into discrete *hits* [5]. A collection of hits is called an *event*, and when the number of hits in an event is sufficiently large, the muon track reconstruction algorithm is triggered.

### 2.1 Cosmic Ray Muons

In addition to neutrinos, muons can also be generated by cosmic rays. One of the simpler techniques used to separate neutrino muons from cosmic ray muons is reconstructing the muon track and determining whether the muon was traveling downwards into the Earth or upwards out of the Earth. Since neutrinos can penetrate the Earth but cosmic ray muons cannot, it follows that a muon traveling out of the Earth must have been generated by a neutrino. Thus, by selecting only the muons that are reconstructed as up-going, the neutrino muons can, in principle, be identified.

While separation is possible in principle, the number of observed cosmic ray muons exceeds the number of observed neutrino muons by more than five orders of magnitude [6]. Thus, high-accuracy reconstructions are critical for preventing erroneously reconstructed cosmic ray muons from dominating the neutrino analysis. While this technique does not recover neutrinos from the top half of the sky, there are alternative techniques that attempt to recover down-going neutrinos [7].

### 2.2 Challenges in Neutrino Detection

There are several challenges for the reconstruction algorithms used in the detector.

**Modeling Difficulties** The underlying physics of the system are nontrivial to model. The muon's light is scattered by the dust impurities and air bubbles in the ice medium. This scattering cannot be analytically well-approximated, and the scattering properties of the ice vary with depth [8]. These challenges make it difficult to design a complete model of the muon's light scattering.

**Noise** The outliers inherent in the data present an additional challenge. The PMTs are so sensitive to light that they can record hits from the radioactive decay in the surrounding glass [9].

**Computational Constraints** Reconstruction algorithms need to be efficient enough to process about 3,000 muons per second with the computing resources available at the South Pole. Thus, algorithms with excessive computational demands are disfavored.

### 2.3 Prior IceCube Software

Starting with the positions and times of each hit, the detector reconstructs the muon track. Once the data is collected, it is passed through a series of filters that removes hits isolated in space and time [10].

After removing outliers, the data is processed using a simple reconstruction algorithm, *linefit*, which finds the track that minimizes the sum of the squares of the distances between the track and the hits. More formally, assume there are $N$ hits; denote the position and time of the $i^{th}$ hit as $(\vec{x}_i, t_i) \in \mathbb{R}^3 \times \mathbb{R}$. Let the muon have a reconstructed velocity of $\vec{v}$, and let $(\vec{x}_0, t_0)$ be a point on the reconstructed track. The linefit reconstruction solves the *least-squares* optimization problem

$$\min_{t_0, \vec{x}_0, \vec{v}} \sum_{i=1}^{N} \rho_i(t_0, \vec{x}_0, \vec{v})^2, \tag{1}$$

where

$$\rho_i(t_0, \vec{x}_0, \vec{v}) = \|\vec{v}(t_i - t_0) + \vec{x}_0 - \vec{x}_i\|_2, \tag{2}$$

and $\|\cdot\|_2$ is the 2-norm.

The linefit reconstruction is primarily used to generate an initial track to be used as a *seed* to a more sophisticated reconstruction.

The reconstruction algorithm for the sophisticated reconstruction is *Single-Photo-Electron-Reconstruction (SPE)* [6]. SPE takes the result of the least-squares reconstruction and event data, and uses a likelihood maximization algorithm to reconstruct the muon track. The SPE reconstruction typically takes about two orders of magnitude more time to compute than linefit. The complete reconstruction process is outlined in Figure 2.

## 3 Improvements to Muon Track Reconstruction

We now discuss the improvements we have made to the reconstruction algorithm. By augmenting the reconstruction algorithm using robust data analysis techniques, we improve the reconstruction algorithm's accuracy.

### 3.1 Algorithm Improvement

The accuracy of the SPE reconstruction is dependent on the accuracy of the seed. Given a seed that is inaccurate by 6° or more, SPE often cannot recover, and can produce a reconstruction result that is inaccurate by 6° or more. In addition, the likelihood space for the SPE reconstruction can contain multiple local maxima, so improving the accuracy of a seed already near the true solution will improve the accuracy of the SPE reconstruction . Thus, we focused our work on improving the quality of the seed.

As the muon travels through the detector, it generates hits. As indicated in Equation 1, linefit fits a line to these hits, weighting each hit quadratically in its distance from this line. This quadratic weighting makes the reconstruction result sensitive to outliers. There are two reasons why outliers may appear far from the muon track:

1. Some of the photons can scatter in the ice and get delayed by more than a microsecond. When these scattered photons are recorded by a PMT, the muon will be over 300 m away, so these photons are no longer useful indicators of the muon's position.
2. While the noise reduction filters remove most of the outlier noise, the noise hits that survive are unrelated to the muon.

To solve the outlier problem we made two changes: improve the modeling of the scattering and replace the least-squares optimization problem with a robust line-fitting algorithm.

#### 3.1.1 Improving the Scattering Model

The least-squares model does not model the scattering. Thus, hits generated by photons that scattered for a significant length of time are not useful predictors of the muon's position within this model. We found that a filter could identify these scattered hits, and improve accuracy by almost a factor of two by removing them prior to performing the fit.

A hit $(\vec{x}_i, t_i)$ is considered a scattered hit if there exists a neighboring hit $(\vec{x}_j, t_j)$ that is within a distance of $r$ and has a time coordinate, $t_j$, that is earlier than $t_i$ by an amount of time given by $t$. If $(\vec{x}_i, t_i)$ is a scattered hit, it is filtered out.

More formally, let H be the set of all hits for a particular event. Then, we define the scattered hits as

$$\left\{(\vec{x}_i, t_i) \mid \exists (\vec{x}_j, t_j) \in H : \left\|\vec{x}_i - \vec{x}_j\right\|_2 \le r \text{ and } t_i - t_j \ge t\right\}. \tag{3}$$

Optimal values of $r$ and $t$ were found to be 156 m and 778 ns, respectively.

#### 3.1.2 Adding Robustness to the Model

As described in Section 2.3, the least-squares model gives all hits quadratic weight, whereas we would like to limit the weight of the outliers. Some models in classical statistics marginalize the weight of outliers. We find that replacing the least-squares model with a Huber reconstruction [11] improves the reconstruction accuracy.

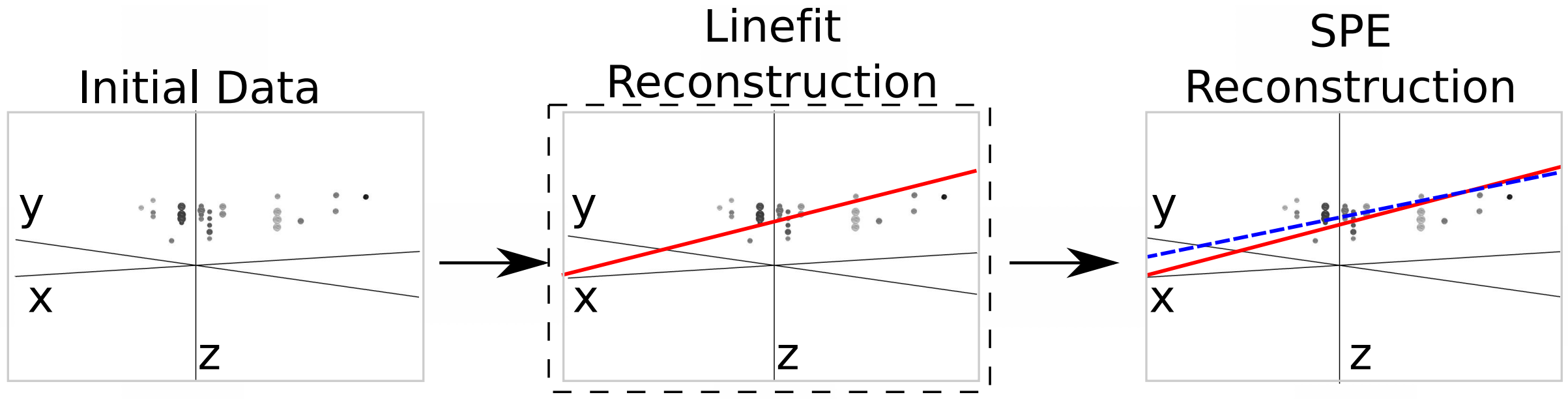

**Fig. 2**: The reconstruction pipeline used to process data in the IceCube detector. Each point indicates a PMT that recorded a photon (PMTs that recorded nothing are omitted). After initial data is collected and passed though some noise filters, the data is processed by a linefit (solid line), which is used as the seed for the SPE (dashed line). The SPE reconstruction is then evaluated as a potential neutrino. Our work on the reconstruction problem makes changes to the linefit reconstruction algorithm (indicated by the dashed box).

For the Huber reconstruction, we replace Equation 1 with the optimization problem:

$$\min_{t_0,\vec{x}_0,\vec{v}} \sum_{i=1}^{N} \phi(\rho_i(t_0,\vec{x}_0,\vec{v})), \quad (4)$$

where the Huber penalty function $\phi(\rho)$ is defined as

$$\phi(\rho) \equiv \begin{cases} \rho^2 & \text{if } \rho < \mu \\ \mu(2\rho - \mu) & \text{if } \rho \geq \mu \end{cases} . \quad (5)$$

Here, $\rho_i(t_0,\vec{x},\vec{v})$ is defined in Equation 2 and $\mu$ is a constant calibrated to the data (for this application, the optimal value of $\mu$ is 153 m).

The Huber penalty function has two regimes. In the near-hit regime ($\rho < \mu$), hits are assumed to be strongly correlated with the muon's track, and the Huber penalty function behaves like least squares, giving point quadratic weight. In the far-hit regime ($\rho \geq \mu$), the Huber penalty function gives points a weaker linear weight, as they are more likely to be noise.

In addition to its attractive robustness properties, the Huber reconstruction's weight assignment also has the added benefit that it inherently labels points as outliers (those with $\rho \geq \mu$). Thus, once the Huber reconstruction is computed, we can go one step further and simply remove the labeled outliers from the dataset. A better reconstruction is then obtained by computing the least-squares reconstruction on the data with the outliers removed.

#### 3.1.3 Implementation

Our scattering filter has a worst-case complexity that is quadratic in the number of PMTs that recorded a hit, but this is typically only between 10 and 100 PMTs. Unlike linefit, the Huber regression does not have a closed form solution, and thus must be solved iteratively. We use an alternating direction method of multipliers [12] to implement the Huber regression.

### 3.2 Results

We now present our empirical results, which validate our changes to the linefit. We also present our runtime performance results.

#### 3.2.1 Accuracy Improvement

Our goal is to improve the accuracy of the reconstruction in order, and to better separate neutrinos from cosmic rays. Thus we present three measurements: (1) the accuracy change between linefit and the new algorithm, (2) the accuracy change when SPE is seeded with the new algorithm, and (3) the improvement in separation between neutrinos and cosmic rays.

To measure the accuracy improvement, we use the metric of *median angular resolution* $\theta_{med}$, which is the arc distance between the reconstruction and the simulated true track. Our dataset is simulated neutrino data designed to be similar to that observed at the Pole. We find that we can improve the median angular resolution of the simple reconstruction by 57.6%, as shown in Table 1.

**Table 1**: Median angular resolution (degrees) for reconstruction improvements. The first line is the accuracy of the prior least-squares model, and the subsequent lines are the accuracy measurements from cumulatively adding improvements into the simple reconstruction algorithm.

| Algorithm | $\theta_{med}$ (°) |
|---|---|
| Linefit Reconstruction (Least-Squares) | 9.917 |
| With Addition of Scattering Filter | 5.205 |
| With Addition of Huber Regression | 4.672 |
| With Addition of Outlier Removal | 4.211 |

We also find that seeding the SPE reconstruction with the improved simple reconstruction generates an improvement in the angular resolution of 12.9%, and that these improvements in the reconstruction algorithm result in 10% fewer atmospheric muons erroneously reconstructed as up-going, and 1% more muons correctly reconstructed as up-going.

#### 3.2.2 Runtime Performance

We now report the runtime of our implementation, which is written in `C++`. The individual mean runtime of each component of the new algorithm is presented in Table 2. As shown, our new algorithm is more computationally demanding than linefit, but only by approximately a factor of six.

## 4 Conclusions

Muon track detection is a challenging problem in the IceCube detector. We achieve a 13% improvement in reconstruction accuracy with the addition of a scattering filter, and a more robust line-fitting algorithm. We achieve these results with a reconstruction algorithm that is only 6 times slower than the previous algorithm. Our reconstruction software runs at the South Pole in the detector, and is included in all IceCube analyses.

**Table 2**: The mean runtime for each component of the new simple reconstruction, contrasted with the mean runtime of the original linefit. As shown, the total runtime is approximately six times that of the original linefit.

| Algorithm | Runtime ($\mu s$) |
|---|---|
| Linefit Reconstruction (Least-Squares) | 24.2 |
| Scattering Filter | 56.6 |
| Huber Regression | 47.5 |
| Outlier Removal | 51.8 |

## References

[1] IceCube Collaboration, IceCube webpage http://icecube.wisc.edu/.
[2] A. Achterberg et al, Astroparticle Physics 26 (2006) 155173.
[3] R. Abbasi et al, Nuclear Instruments and Methods in Physics Research Section A 618 (2010) 139-152.
[4] R. Abbasi et al, Nuclear Instruments and Methods in Physics Research Section A 703 (2012) 190-198.
[5] R. Abbasi et al, Nuclear Instruments and Methods in Physics Research Section A 601 (2009) 294-316.
[6] J. Ahrens et al, Nuclear Instruments and Methods in Physics Research Section A 524 (2004) 169-194.
[7] IceCube Collaboration, paper 0565 these proceedings.
[8] M. Wolf, Verification of South Pole glacial ice simulations in IceCube and its relation to conventional and new, accelerated photon tracking techniques. Masters thesis University Heidelberg September 2010.
[9] R Abbasi et al, Astronomy & Astrophysics 535 (2011) 18.
[10] M. Ackermann, Searches for signals from cosmic point-like sources of high energy neutrinos in 5 years of AMANDA-II data PhD thesis Humboldt-Universita?t zu Berlin (2006).
[11] S. Boyd, and L. Vandenberghe, (2009) Convex Optimization, Cambridge University Press.
[12] S. Boyd, N. Parikh, E. Chu, B. Peleato, and J. Eckstein, Foundations and Trends in Machine Learning 3 (2011) 1-122.